\documentclass[traditabstract,longauth]{aa}

\usepackage{graphicx}
\usepackage{txfonts}
\usepackage{natbib}
\usepackage[colorlinks=true,citecolor=blue]{hyperref}
\usepackage{xspace}
\usepackage{array}
\usepackage{rotating}
\usepackage{xcolor}
\usepackage{enumitem}
\usepackage{dsfont}

\bibpunct{(}{)}{;}{a}{}{,}

\newcommand{\MJup}{\ensuremath{M_{\mathrm{Jup}}}\xspace}

\newcommand{\MSun}{\ensuremath{M_{\odot}}\xspace}

\newcommand{\as}{\hbox{$^{\prime\prime}$}\xspace}

\newcommand{\Lpf}{\ensuremath{L_{\mathrm{pf}}}\xspace}

\usepackage{pifont}
\definecolor{OliveGreen}{rgb}{0.31, 0.5, 0.0}
\newcommand{\cmark}{\textcolor{OliveGreen}{\ding{51}}}%
\definecolor{BrickRed}{rgb}{0.8, 0.25, 0.33}
\newcommand{\xmark}{\textcolor{BrickRed}{\ding{55}}}%

\newcommand{\fbdb}{\ensuremath{f_{\mathrm{BDB}}}\xspace}
\newcommand{\fppl}{\ensuremath{f_{\mathrm{PPL/LN}}}\xspace}
\newcommand{\fpmod}{\ensuremath{f_{\mathrm{BDB\,+\,PPL/LN}}}\xspace}
\newcommand{\fgi}{\ensuremath{f_{\mathrm{GI}}}\xspace}
\newcommand{\fca}{\ensuremath{f_{\mathrm{CA}}}\xspace}
\newcommand{\fsmod}{\ensuremath{f_{\mathrm{GI+CA}}}\xspace}

\defcitealias{SHINEPaperI}{Paper~I}
\defcitealias{SHINEPaperII}{Paper~II}

\begin{document}

\title{The SPHERE infrared survey for exoplanets (SHINE)}
\subtitle{III. The demographics of young giant exoplanets below 300\,au with SPHERE}

\titlerunning{The SPHERE infrared survey for exoplanets (SHINE). III.}

\author{
    A.~Vigan\inst{\ref{lam}} \and 
    C.~Fontanive\inst{\ref{bern},\ref{padova}} \and
    M.~Meyer\inst{\ref{umich},\ref{eth}} \and
    B.~Biller\inst{\ref{ifa_uoe},\ref{supa_edin},\ref{mpia}} \and
    M.~Bonavita\inst{\ref{ifa_uoe}} \and
    M.~Feldt\inst{\ref{mpia}} \and
    S.~Desidera\inst{\ref{padova}} \and
    G.-D.~Marleau\inst{\ref{tueb},\ref{bern},\ref{mpia}} \and
    A.~Emsenhuber\inst{\ref{tucson},\ref{bern}} \and
    R.~Galicher\inst{\ref{lesia}} \and
    K.~Rice\inst{\ref{supa_edin},\ref{ces_uoe}} \and
    D.~Forgan\inst{\ref{supa_stand}} \and
    C.~Mordasini\inst{\ref{bern},\ref{mpia}} \and
    R.~Gratton\inst{\ref{padova}} \and
    H.~Le Coroller\inst{\ref{lam}} \and
    A.-L.~Maire\inst{\ref{star},\ref{mpia}} \and
    F.~Cantalloube\inst{\ref{mpia}} \and
    G.~Chauvin\inst{\ref{ipag},\ref{umi}} \and
    A.~Cheetham\inst{\ref{geneva}} \and
    J.~Hagelberg\inst{\ref{geneva}} \and
    A.-M.~Lagrange\inst{\ref{ipag}} \and
    M.~Langlois\inst{\ref{cral},\ref{lam}} \and
    M.~Bonnefoy\inst{\ref{ipag}} \and
    J.-L.~Beuzit\inst{\ref{lam},\ref{ipag}} \and
    A.~Boccaletti\inst{\ref{lesia}} \and
    V.~D'Orazi\inst{\ref{padova}} \and
    P.~Delorme\inst{\ref{ipag}} \and
    C.~Dominik\inst{\ref{ams}} \and
    Th.~Henning\inst{\ref{mpia}} \and
    M.~Janson\inst{\ref{mpia},\ref{stock}} \and
    E.~Lagadec\inst{\ref{oca}} \and
    C.~Lazzoni\inst{\ref{padova}} \and
    R.~Ligi\inst{\ref{brera}} \and
    F.~Menard\inst{\ref{ipag}} \and
    D.~Mesa\inst{\ref{padova}} \and
    S.~Messina\inst{\ref{catania}} \and
    C.~Moutou\inst{\ref{irap},\ref{lam}} \and
    A.~M\"uller\inst{\ref{mpia}} \and
    C.~Perrot\inst{\ref{lesia},\ref{valpo1},\ref{valpo2}} \and
    M.~Samland\inst{\ref{mpia},\ref{stock}} \and
    H.~M.~Schmid\inst{\ref{eth}} \and
    T.~Schmidt\inst{\ref{lesia}} \and
    E.~Sissa\inst{\ref{padova}} \and
    M.~Turatto\inst{\ref{padova}} \and
    S.~Udry\inst{\ref{geneva}} \and
    A.~Zurlo\inst{\ref{lam},\ref{diegoportales1},\ref{diegoportales2}} \and
    L.~Abe\inst{\ref{oca}} \and
    J.~Antichi\inst{\ref{padova}} \and
    R.~Asensio-Torres\inst{\ref{mpia}} \and
    A.~Baruffolo\inst{\ref{padova}} \and
    P.~Baudoz\inst{\ref{lesia}} \and
    J.~Baudrand\inst{\ref{lesia}} \and
    A.~Bazzon\inst{\ref{eth}} \and
    P.~Blanchard\inst{\ref{lam}} \and
    A.~J.~Bohn\inst{\ref{leiden}} \and
    S.~Brown~Sevilla\inst{\ref{mpia}} \and
    M.~Carbillet\inst{\ref{oca}} \and
    M.~Carle\inst{\ref{lam}} \and
    E.~Cascone\inst{\ref{padova}} \and
    J.~Charton\inst{\ref{ipag}} \and
    R.~Claudi\inst{\ref{padova}} \and
    A.~Costille\inst{\ref{lam}} \and
    V.~De Caprio\inst{\ref{capodimonte}} \and
    A.~Delboulb\'e\inst{\ref{ipag}} \and
    K.~Dohlen\inst{\ref{lam}} \and
    N.~Engler\inst{\ref{eth}} \and
    D.~Fantinel\inst{\ref{padova}} \and
    P.~Feautrier\inst{\ref{ipag}} \and
    T.~Fusco\inst{\ref{onera},\ref{lam}} \and
    P.~Gigan\inst{\ref{lesia}} \and
    J.~H.~Girard\inst{\ref{stsci},\ref{ipag}} \and
    E.~Giro\inst{\ref{padova}} \and
    D.~Gisler\inst{\ref{eth}} \and
    L.~Gluck\inst{\ref{ipag}} \and
    C.~Gry\inst{\ref{lam}} \and
    N.~Hubin\inst{\ref{eso_garching}} \and
    E.~Hugot\inst{\ref{lam}} \and
    M.~Jaquet\inst{\ref{lam}} \and
    M.~Kasper\inst{\ref{eso_garching},\ref{ipag}} \and
    D.~Le Mignant\inst{\ref{lam}} \and
    M.~Llored\inst{\ref{lam}} \and
    F.~Madec\inst{\ref{lam}} \and
    Y.~Magnard\inst{\ref{ipag}} \and
    P.~Martinez\inst{\ref{oca}} \and
    D.~Maurel\inst{\ref{ipag}} \and
    O.~M\"oller-Nilsson\inst{\ref{mpia}} \and
    D.~Mouillet\inst{\ref{ipag}} \and
    T.~Moulin\inst{\ref{ipag}} \and
    A.~Orign\'e\inst{\ref{lam}} \and
    A.~Pavlov\inst{\ref{mpia}} \and
    D.~Perret\inst{\ref{lesia}} \and
    C.~Petit\inst{\ref{onera}} \and
    J.~Pragt\inst{\ref{ipag}} \and
    P.~Puget\inst{\ref{ipag}} \and
    P.~Rabou\inst{\ref{ipag}} \and
    J.~Ramos\inst{\ref{ipag}} \and
    E.~L.~Rickman\inst{\ref{geneva}} \and
    F.~Rigal\inst{\ref{ipag}} \and
    S.~Rochat\inst{\ref{ipag}} \and
    R.~Roelfsema\inst{\ref{nova}} \and
    G.~Rousset\inst{\ref{lesia}} \and
    A.~Roux\inst{\ref{ipag}} \and
    B.~Salasnich\inst{\ref{padova}} \and
    J.-F.~Sauvage\inst{\ref{onera},\ref{lam}} \and
    A.~Sevin\inst{\ref{lesia}} \and
    C.~Soenke\inst{\ref{eso_garching}} \and
    E.~Stadler\inst{\ref{ipag}} \and
    M.~Suarez\inst{\ref{eso_garching}} \and
    Z.~Wahhaj\inst{\ref{eso_chile},\ref{lam}} \and
    L.~Weber\inst{\ref{geneva}} \and
    F.~Wildi\inst{\ref{geneva}}
}

\institute{
    Aix Marseille Univ, CNRS, CNES, LAM, Marseille, France \label{lam} \\
    \email{\href{mailto:arthur.vigan@lam.fr}{arthur.vigan@lam.fr}}
    \and
    Center for Space and Habitability, University of Bern, 3012 Bern, Switzerland \label{bern}
    \and
    INAF - Osservatorio Astronomico di Padova, Vicolo della Osservatorio 5, 35122, Padova, Italy \label{padova}
    \and
    Department of Astronomy, University of Michigan, Ann Arbor, MI 48109, USA \label{umich}
    \and
    Institute for Particle Physics and Astrophysics, ETH Zurich, Wolfgang-Pauli-Strasse 27, 8093 Zurich, Switzerland \label{eth}
    \and 
    Institute for Astronomy, University of Edinburgh, EH9 3HJ, Edinburgh, UK \label{ifa_uoe}
    \and
    Scottish Universities Physics Alliance (SUPA), Institute for Astronomy, University of Edinburgh, Blackford Hill, Edinburgh EH9 3HJ, UK \label{supa_edin}
    \and
    Centre for Exoplanet Science, SUPA, School of Physics \& Astronomy, University of St Andrews, St Andrews KY16 9SS, UK \label{supa_stand}
    \and
    Max Planck Institute for Astronomy, K\"onigstuhl 17, D-69117 Heidelberg, Germany \label{mpia}
    \and
    Lunar and Planetary Laboratory, University of Arizona 1629 E. University Blvd. Tucson, AZ 85721, USA \label{tucson}
    \and
    Universit\"at T\"ubingen, Auf der Morgenstelle 10, D-72076 T\"ubingen, Germany \label{tueb}
    \and
    LESIA, Observatoire de Paris, Université PSL, Universit\'e de Paris, CNRS, Sorbonne Université, 5 place Jules Janssen, 92195 Meudon, France \label{lesia}
    \and
    Unidad Mixta Internacional Franco-Chilena de Astronom\'{i}a, CNRS/INSU UMI 3386 and Departamento de Astronom\'{i}a, Universidad de Chile, Casilla 36-D, Santiago, Chile \label{umi}
    \and
    STAR Institute, University of Li\`ege, All\'ee du Six Ao\^ut 19c, B-4000 Li\`ege, Belgium \label{star}
    \and
    Centre for Exoplanet Science, University of Edinburgh, Edinburgh EH9 3FD, UK \label{ces_uoe}
    \and
    Univ. Grenoble Alpes, CNRS, IPAG, F-38000 Grenoble, France \label{ipag}
    \and
    Department of Astronomy, Stockholm University, SE-10691 Stockholm, Sweden \label{stock}
    \and
    CRAL, CNRS, Université Lyon 1,Université de Lyon, ENS, 9 avenue Charles Andre, 69561 Saint Genis Laval, France \label{cral}
    \and
    INAF - Catania Astrophysical Observatory, via S. Sofia 78, I-95123 Catania, Italy \label{catania}
    \and
    Univ. de Toulouse, CNRS, IRAP, 14 avenue Belin, F-31400 Toulouse, France \label{irap}
    \and
    ONERA (Office National dEtudes et de Recherches Arospatiales), B.P.72, F-92322 Chatillon, France \label{onera}
    \and
    European Southern Observatory (ESO), Karl-Schwarzschild-Str. 2,85748 Garching, German \label{eso_garching}
    \and
    Geneva Observatory, University of Geneva, Chemin des Mailettes 51, 1290 Versoix, Switzerland \label{geneva}
    \and
    Universit\'e C\^ote d’Azur, OCA, CNRS, Lagrange, France \label{oca}
    \and
    Anton Pannekoek Institute for Astronomy, Science Park 9, NL-1098 XH Amsterdam, The Netherlands \label{ams}
    \and
    Leiden Observatory, Leiden University, PO Box 9513, 2300 RA Leiden, The Netherlands \label{leiden}
    \and 
    INAF - Osservatorio Astronomico di Brera, Via E. Bianchi 46, 23807 Merate, Italy \label{brera}
    \and
    N\'ucleo de Astronom\'ia, Facultad de Ingenier\'ia y Ciencias, Universidad Diego Portales, Av. Ejercito 441, Santiago, Chile \label{diegoportales1}
    \and
    Escuela de Ingenier\'ia Industrial, Facultad de Ingenier\'ia y Ciencias, Universidad Diego Portales, Av. Ejercito 441, Santiago, Chile \label{diegoportales2}
    \and
    INAF - Osservatorio Astronomico di Capodimonte, Salita Moiariello 16, 80131 Napoli, Italy \label{capodimonte}
    \and
    European Southern Observatory, Alonso de C\`ordova 3107, Vitacura, Casilla 19001, Santiago, Chile \label{eso_chile}
    \and 
    Space Telescope Science Institute, 3700 San Martin Drive, Baltimore, MD, 21218, USA \label{stsci}
    Instituto de F\'isica y Astronom\'ia, Facultad de Ciencias, Universidad de Valpara\'iso, Av. Gran Breta\~na 1111, Valpara\'iso, Chile \label{valpo1}
    \and
    N\'ucleo Milenio Formaci\'on Planetaria - NPF, Universidad de Valpara\'iso, Av. Gran Breta\~na 1111, Valpara\'iso, Chile  \label{valpo2}
    \and
    NOVA/UVA \label{nova}
}

\date{Received 7 April 2020 / Accepted 6 July 2020}

\abstract{
    The SpHere INfrared Exoplanet (SHINE) project is a 500-star survey performed with SPHERE on the Very Large Telescope for the purpose of directly detecting new substellar companions and understanding their formation and early evolution. Here we present an initial statistical analysis for a subsample of 150 stars spanning spectral types from B to M that are representative of the full SHINE sample. Our goal is to constrain the frequency of substellar companions with masses between 1 and 75\,\MJup and semimajor axes between 5 and 300\,au. For this purpose, we adopt detection limits as a function of angular separation from the survey data for all stars converted into mass and projected orbital separation using the BEX-COND-hot evolutionary tracks and known distance to each system. Based on the results obtained for each star and on the 13 detections in the sample, we use a Markov chain Monte Carlo tool to compare our observations to two different types of models. The first is a parametric model based on observational constraints, and the second type are numerical models that combine advanced core accretion and gravitational instability planet population synthesis. Using the parametric model, we show that the frequencies of systems with at least one substellar companion are $23.0_{-9.7}^{+13.5}\%$, $5.8_{-2.8}^{+4.7}\%$, and $12.6_{-7.1}^{+12.9}\%$ for BA, FGK, and M stars, respectively. We also demonstrate that a planet-like formation pathway probably dominates the mass range from 1--75\,\MJup for companions around BA stars, while for M dwarfs, brown dwarf binaries dominate detections. In contrast, a combination of binary star-like and planet-like formation is required to best fit the observations for FGK stars. Using our population model and restricting our sample to FGK stars, we derive a frequency of $5.7_{-2.8}^{+3.8}\%$, consistent with predictions from the parametric model. More generally, the frequency values that we derive are in excellent agreement with values obtained in previous studies.
}


\maketitle

\section{Introduction}

In the past 20 years, large-scale direct-imaging surveys for exoplanets have discovered approximately 60 substellar and planetary-mass companions around young nearby stars \citep[see, e.g.,][]{Wagner2019}. Early surveys were relatively modest in size, with samples of 50--100 stars, while ongoing surveys target 500--600 stars. The largest projects to date are the SpHere INfrared Exoplanets (SHINE) project conducted with SPHERE \citep{Chauvin2017} and the Gemini Planet Imager (GPI) Exoplanet Survey \citep[GPIES;][]{Macintosh2015}. SHINE and GPIES have added three new exoplanet detections to the growing cohort of directly imaged objects \citep{Macintosh2015,Chauvin2017,Keppler2018} and several additional higher mass brown dwarfs \citep{Konopacky2016,Cheetham2018}. However, new exoplanet detections are just one goal of large-scale direct-imaging surveys. These surveys also provide key spectral and orbital characterization data for known exoplanets \citep[e.g.,][]{DeRosa2016,Samland2017,Chauvin2018,Wang2018,Mueller2018,Cheetham2019,Lagrange2019,Maire2019}, and statistical constraints on the distribution of such objects at star--planet separations $>20$\,au \citep[e.g.,][]{Kasper2007,Nielsen2010,Heinze2010,Janson2011,Vigan2012,Biller2013,Rameau2013,Brandt2014,Galicher2016,Lannier2016,Vigan2017,Stone2018,Baron2019,Nielsen2019}.

In particular, the current generation of surveys strongly constrains the distribution of wide ($>10$\,au) giant exoplanets and substellar companions to young stars \citep{Reggiani2016,Nielsen2019}. The distribution of gas giant planets and brown dwarf companions as a function of mass and orbital separation can provide insight into formation mechanisms because different formation channels (e.g., planet formation in a disk versus brown dwarf binary formation in a protostellar core) may dominate in different circumstances (mass ratio of companion to host, orbital separation, and total system mass). Contrast limits from these surveys, and the cohort of detected objects, can be used with a Bayesian approach to constrain the fraction of systems hosting planetary or substellar companions and to explore distribution functions of their architectures (semimajor axis, mass, or eccentricity distributions). Diverse demographic models can be tested: a)~parametric models based on a wide range of point estimates of frequency over fixed ranges of mass and orbital separation \citep[e.g.,][]{Reggiani2016,Meyer2018}, extrapolated to the mass and separation ranges probed by direct-imaging surveys; and b)~population synthesis models \citep[e.g.,][]{Mordasini2009,Forgan2013}, which use numerical estimates based on physical theories of various formation mechanisms to predict a population of exoplanets, which can be compared to our observations. 

Imaging surveys have yielded significantly fewer exoplanet detections than expected using extrapolations of radial velocity (RV) planet populations to larger semimajor axes \citep[e.g.,][]{Cumming2008}. These extrapolations predicted dozens of detections with optimistic assumptions. While this is disappointing from the perspective of detection, these results constrain the distribution of giant exoplanets and brown dwarf companions at separations $>10$\,au from host stars. SHINE achieves typical contrasts of $10^{-5}$--$10^{-6}$ at separations of 0.4--0.5\as (\citealt{SHINEPaperII}, hereafter \citetalias{SHINEPaperII}). Based on evolutionary models used for mass--luminosity conversion and on the ages and distances of targets in our sample (\citealt{SHINEPaperI}, hereafter \citetalias{SHINEPaperI}), we expect that the SHINE survey will be sensitive to 1--75 Jupiter mass (\MJup) companions at separations 5--300\,au. We note, however, that as for all direct-imaging surveys, the sensitivity to the lowest masses improves for larger semimajor axes and is expected to reach a minimum only at a few dozen astronomical units. Predictions of what SHINE will see depend on the planet mass function, the orbital distribution, any correlations between the two, and perhaps on host star properties. Because only a small number of companions are detected (typically a few in a given large-scale survey), we must simplify the models to a few free parameters, preferably based on measured populations of substellar companions and extrasolar planets obtained by other methods \citep[e.g.,][]{Meyer2018}.

Several formation mechanisms can lead to the formation of 1--75\,\MJup companions that are detected in these surveys. In addition to formation channels for very low-mass binary companions \citep[e.g.,][]{Kratter2010,Offner2010}, companions can be formed in multiple modes as planets in circumstellar disks as well. The core accretion (CA) scenario is a bottom-up framework where a solid core of a few Earth masses forms first \citep{Mizuno1980,Pollack1996,Alibert2004}, and then the rapid accretion of gas builds up gas giant planets \citep{Piso2015a,Venturini2015}. In contrast, the disk instability, or gravitational instability (GI), is a top-down binary star-like framework where planets form very quickly in the outer parts of disks from clumps that detach from the rest of the disk, become gravitationally bound, and contract into a giant planet \citep{Boss1998,Vorobyov2013}. Multiple theoretical approaches provide simulated populations of planets that formed through these mechanisms, which can then be compared to planets that are detected through direct imaging. The comparison of direct-imaging observations with theoretical predictions was pioneered by \citet{Janson2011} and \citet{Rameau2013}. \citet{Vigan2017} were then the first to compare observations to the outputs of population synthesis models based on the GI scenario. The authors found that these models can describe a fraction of the population, wide-orbit giant companions. With the improved sensitivity in mass and semimajor axis of new surveys, it becomes realistic to compare observations to predictions of both CA models \citep[e.g.,][]{Mordasini2017} and GI models \citep[e.g.,][]{Forgan2013,Forgan2015}.

In this paper we present a first statistical analysis of the properties of the population of 1--75\,\MJup companions at orbital separations from 5--300\,au based on the first 150 stars observed in the SHINE survey. In Sect.~\ref{sec:stat_sample_detection_limits} we present the target sample considered in our analysis, the detections that are taken into account, how the detection limits were derived and converted into mass, and finally, the survey sensitivity derived from the observations. In Sect.~\ref{sec:exoplanet_models} we introduce the exoplanet population models to which we compare our observations, and in Sect.~\ref{sec:stat_tools} we present the simulation tools we used for the comparison. In Sect.~\ref{sec:results} we present all of our results, and finally, in Sect.~\ref{sec:discussion} we discuss them in a broader context and compare the SHINE results to previously published surveys.

\section{Statistical sample and detection limits}
\label{sec:stat_sample_detection_limits}

This section provides information regarding the sample of targets we considered (Sect.~\ref{sec:stat_sample}), the observations and data analysis (Sect.~\ref{sec:obs_data_analysis}), how the planetary candidates were treated (Sect.~\ref{sec:planetary_candidates}), the statistical weight attributed to the detections (Sect.~\ref{sec:stat_weight}), and finally, the mass conversion of the detection limits (Sect.~\ref{sec:mass_conv}). The detailed properties of the statistical sample are separately treated in a companion paper \citepalias{SHINEPaperI}, and the observations and data analysis are discussed in a second companion paper \citepalias{SHINEPaperII}.

\subsection{Statistical sample}
\label{sec:stat_sample}

The SHINE survey is being performed by the SPHERE consortium and exploits 200 nights of guaranteed time of observation (GTO). The main goal of SHINE is to observe a sample of 500 stars out of a larger sample of 800 nearby young stars to search for new substellar companions. The sample is oversized with respect to the available telescope time by a factor of approximately two on the basis of the adopted observing strategy, which assumes observations across meridian passage in order to achieve the maximum field-of-view (FoV) rotation for optimal angular differential imaging. This requires some flexibility in the target list in order to optimize the scheduling. Moreover, a sample of at least a few hundred objects is mandatory to achieve robust inference on the frequency of planets because the expected frequency of substellar companions is likely low \citep[e.g.,][]{Vigan2017}.

The sample includes a broad range of stellar masses to explore the effect of this parameter on planet frequency. The stellar masses in the sample range from $\sim$3.0~\MSun  to 0.3--0.5~\MSun, and the faint end is determined by the working limit of the adaptive optics system of SPHERE \citep{Sauvage2016,Beuzit2019}. The 800 targets were divided into four priority bins, called P1, P2, P3, and P4 in decreasing order of priority, which were used to schedule the observations. Roughly a dozen targets of special interest were added to the sample for scientific reasons (presence of known substellar objects, disks, RV planets, etc.) and classified as P0 (highest priority). Some of these special objects were originally drawn from the 800-star sample built for the statistical analysis, but some were also added at a later stage. Both the selection of the targets from a wide database of nearby young stars and the priority ranking were based on simulations of planet detectability with SPHERE that was performed before the start of the survey \citepalias{SHINEPaperI}. These simulations adopted two different expected distributions and dependences on the stellar mass for the planet population in order to avoid biasing our results by relying on a single assumed distribution.

In addition to obvious magnitude and declination limits, the selection criteria exclude known spectroscopic and close visual binaries within the radial FoV of the SPHERE/IRDIS science camera (6\as). No selection was made in favor of stars with known disks or IR excess.

A total of 150 targets with first-epoch observations until February 2017 were included in the present analysis, and second-epoch observations extended until 2019. At this stage, the sample was not complete in any aspect, considering that the scheduling was optimized for the whole survey, but is fully representative of the whole sample. The sample includes 53 BA stars, 77 FGK stars, and 20 M stars. The median stellar age of the sample used in this early statistical analysis is 45\,Myr (90\% between 11 and 450\,Myr), the median stellar mass is 1.15~\MSun (90\% between 0.57 and 2.37~\MSun), and the median distance is 48\,pc (90\% limits of 11 and 137\,pc). 

Most of the stars in the present sample belong to young moving groups. The age range considered for group members corresponds to the typically accepted spread of the mean age of the group. Age spread within the groups is not included, although mild kinematic outliers are considered individually, and their age uncertainties are typically larger than those of bona fide members. Therefore we expect that the effect is negligible for most groups, which show no clear evidence of a large age spread, and that it may be present only for targets in the Scorpius-Centaurus OB2 association, which are known to have an age spread of up to 50\% \citep{Pecaut2016}. However, the targets in this association constitute only 20\% of our present sample, therefore the overall effect should be small. The general design of the survey, the sample selection, and the simulations performed for building it, and the parameters of the individual targets in this series of papers are fully described in \citetalias{SHINEPaperI}.

\subsection{Observations and data analysis}
\label{sec:obs_data_analysis}

The complete observing strategy, data analysis, and detection performance for the targets in the sample are described in \citetalias{SHINEPaperII}. All observations were performed with the SPHERE instrument at the Very Large Telescope (VLT) \citep{Beuzit2019} in either \texttt{IRDIFS} or \texttt{IRDIFS-EXT} mode, that is, the two near-IR (NIR) subsystems, IFS and IRDIS, observed in parallel. The IFS covers a 1.7\as$\times$1.7\as FoV and IRDIS covers a circular unvignetted FoV of diameter $\sim$9\as. Some targets were observed multiple times because of known companions and/or the detection of (new) candidate companions. This varied the observations for each target.

All the data were downloaded at the SPHERE data center \citep{Delorme2017b} and processed with the SpeCal software \citep{Galicher2018} for speckle suppression, derivation of detection limits, and astro-photometry of the detected candidates. The final data products were then transferred to a private part of the DIVA+ database\footnote{\url{http://cesam.lam.fr/diva/}} \citep{Vigan2017} from where they were retrieved for our analysis.

More specifically, we used the 5$\sigma$ IRDIS and IFS detection limits of each observation for all the targets in the sample. These detection limits are derived based on the noise in the speckle-subtracted image, compensated for the throughput of the algorithm (calibrated with simulated planet injections), the transmission of the coronagraph (calibrated from measurements in SPHERE), and the small sample statistics \citep{Mawet2014}. More details are provided in \citet{Galicher2018} and \citetalias{SHINEPaperII}.

\subsection{Planetary candidates}
\label{sec:planetary_candidates}

For 91 of the 150 targets in the sample, candidates were identified in the first-epoch observations. Valid follow-up observations were obtained for 45 targets, complemented by archival or published data for a total of 39 targets. The use of archival data enabled us to recover the position of a significant number of candidates on a very long temporal baseline and minimize the need for follow-up observations (see \citetalias{SHINEPaperII}). Multi-epoch data were therefore obtained for 66 targets, which left only 25 targets without follow-up. In practice, follow-up observations were attempted for all of these 25 targets but could not be performed because of scheduling problems or poor weather during the observing runs.

The stellar proper motion of the targets with candidates in the sample is 81$\pm$64\,mas/yr, with a minimum and maximum of 13 and 454\,mas/yr, respectively. Follow-up observations were only scheduled after a time span that would unambiguously enable us to distinguish between a bound companion and background source, resulting in temporal baselines of 1.94$\pm$1.22\,yr. The motion of the stars with candidates over their respective time baselines is 156$\pm$145\,mas for the SHINE data, with a minimum and maximum of 10 and 500\,mas/yr, respectively (even extending beyond a few arcseconds when the archival data are considered). With a typical astrometric accuracy of a few mas, the follow-up observations were therefore distant enough in time to reliably assess the status of candidates.

We were not always able to obtain a clear confirmation of companion or background status for the 66 targets with follow-up observations, sometimes despite multi-epoch follow-up of some candidates. This is in most cases due to the non-redetection of some candidates because the observing conditions between epochs varied. Follow-up observations of all remaining candidates is foreseen in a future dedicated programme.

Of the 1454 individual candidates, 16 were confirmed as companions or were already known to be companions (Table~\ref{tab:detections}), 1134 were confirmed as background using either relative astrometry or classification based on color-magnitude diagrams \citepalias[see][]{SHINEPaperII}, but 304 remain unconfirmed, sometimes despite multiple observations. This count is largely dominated by one target close to the galactic plane (TYC\,7879-0980-1) that has only one epoch and more than 100 candidates in the IRDIS FoV, and a handful of other targets with a few dozen candidates. Based on the distance of the targets in the sample, the projected separation of unconfirmed candidates ranges from 3 to~1300\,au. In Appendix~\ref{sec:candidates_histo} we provide a cumulative histogram of the number of candidates as a function of projected semimajor axis to illustrate that a small number of targets largely contributes to the total number of undefined candidates.

The statistics of young substellar companions beyond 300\,au has been well established in the past decade by numerous direct-imaging surveys that used various instruments and targeted stars with a wide range of properties \citep[e.g.,][]{Nielsen2010,Heinze2010,Janson2011,Vigan2012,Rameau2013,Galicher2016,Lannier2016,Vigan2017,Stone2018,Baron2019}. One of the main goals of the new generation of exoplanet imagers such as SPHERE is to set the first meaningful constraints below 100\,au. To further this goal, we restricted our analysis to projected semimajor axes $\le$\,300\,au. With this upper limit in terms of physical separation, the number of unconfirmed candidates decreases to only 187, again mostly clustered around a handful of targets.

For our targets with incomplete follow-up and unconfirmed candidates, two approaches can be followed in the statistical analysis. Either the detection limits for individual targets can be raised above the level of the brightest unconfirmed candidate, regardless of its position in the FoV, or the limit can be cut at the separation of the closest unconfirmed candidate. For this analysis we chose the latter approach in order to retain the best possible sensitivity at the closest separations. 

\subsection{Statistical weight of detections}
\label{sec:stat_weight}

\begin{table*}
    \caption[]{Substellar companions detected around targets within the current sample}
    \label{tab:detections}
    \centering
    \begin{tabular}{lccccccccc}
    \hline \hline
    Companion                 & SpT & $M_{\star}$ & Semimajor axis & Mass    & $q$           & Original & Updated  & Statistical & References \\
                              &     &                     &                 &         & $M_p/M_\star$ & priority & priority & weight & \\
                              &     & [$M_{\odot}$]       & [au]            & [\MJup] & [\%]          &          &          &           & \\
    \hline
    \multicolumn{10}{c}{New SHINE detections} \\
    \hline
    \object{HIP\,64892} B     & B9  & 2.09                & 147--171         & 29--37   & 1.3--1.7\%     & P1       &          & 1.00      & 1 \\
    \object{HIP\,65426} b     & A2  & 1.96                & 80--210          & 7--9     & 0.3--0.4\%     & P1       &          & 1.00      & 2, 3 \\
    \hline
    \multicolumn{10}{c}{Previously known detections -- no priority update} \\
    \hline
    \object{$\eta$\,Tel} B    & A0  & 2.00                & 125--432         & 20--50   & 1.0--2.4\%     & P1       &          & 1.00      & 4, 5 \\
    \object{CD\,-35\,2722} B  & M1  & 0.56                & 74--216          & 23--39   & 3.9--6.6\%     & P1       &          & 1.00      & 6, 5 \\
    \hline
    \multicolumn{10}{c}{Previously known detections -- updated priority} \\
    \hline
    \object{HIP\,78530 B}\tablefootmark{a}     & B9  & 1.99                & $\sim$620        & 19--26   & 0.9--1.2\%     & P1       & P0       & 0.60      & 7 \\
    \object{$\beta$\,Pic} b   & A3  & 1.61                & 8.5--9.2         & 9--16    & 0.5--0.9\%     & P1       & P0       & 0.60      & 8, 9 \\
    \object{HR\,8799} b       & A5  & 1.42                & 62--72           & 5.3--6.3 & 0.3--0.4\%     & P1       & P0       & 0.60      & 10 \\
    HR\,8799 c                & A5  & 1.42                & 39--45           & 6.5--7.8 & 0.4--0.5\%     & P1       & P0       & 0.60      & 10 \\
    HR\,8799 d                & A5  & 1.42                & 24--27           & 6.5--7.8 & 0.4--0.5\%     & P1       & P0       & 0.60      & 10 \\
    HR\,8799 e                & A5  & 1.42                & 14--17           & 6.5--7.8 & 0.4--0.5\%     & P1       & P0       & 0.60      & 10 \\
    \object{HD\,95086} b      & A8  & 1.55                & 28--64           & 2--9     & 0.1--0.6\%     & P1       & P0       & 0.60      & 11, 12 \\
    \object{51\,Eri} b        & F0  & 1.45                & 10--16           & 6--14    & 0.4--0.9\%     & P1       & P0       & 0.60      & 13, 14 \\
    \object{HIP\,107412} B    & F5  & 1.32                & 6.2--7.1         & 15--30   & 1.1--2.2\%     & P4       & P0       & 0.01      & 15, 16 \\
    \object{PZ\,Tel} B        & G9  & 1.07                & 19--30           & 38--54   & 3.4--4.8\%     & P1       & P0       & 0.60      & 17, 18 \\
    \object{AB\,Pic} B        & K1  & 0.97                & $\sim$250        & 13--30   & 1.3--3.0\%     & P1       & P0       & 0.60      & 19, 20 \\
    \object{GSC\,8047-0232} B & K2  & 0.89                & 190--880         & 15--35   & 1.6--3.8\%     & P2       & P0       & 0.35      & 21, 22 \\
    \hline 
    \end{tabular} 
    \tablefoot{\tablefoottext{a}{With a semimajor axis of $\sim$620\,au and no additional published constraints, HIP\,78530 is not taken into account into most of our simulations, which use a cutoff at 300\,au.}}
    \tablebib{(1) \citet{Cheetham2018}; (2) \citet{Chauvin2017}; (3) \citet{Cheetham2019}; (4) \citealt{Neuhauser2011}; (5) \citet{Blunt2017}; (6) \cite{Wahhaj2011}; (7) \citet{Lafreniere2011}; (8) \citet{Dupuy2019}; (9) \citet{Lagrange2019}; (10) \citet{Wang2018}; (11) \citet{DeRosa2016}; (12) \citet{Chauvin2018}; (13) \citet{Samland2017}; (14) \citet{Maire2019}; (15) \citet{Delorme2017a}; (16) \citet{Grandjean2019}; (17) \citet{Maire2016}; (18) \citet{Bowler2020}; (19) \citet{Chauvin2005}; (20) \citet{Bonnefoy2010}; (21) \citet{Chauvin2005b}; (22) \citet{Ginski2014}.
    }
\end{table*}

In the construction of the complete SHINE sample, each individual target was attributed a scientific priority from P1 to P4 based on planet detectability simulations \citepalias{SHINEPaperI}. The additional very high priority bin (P0) was also created to enforce the observation of specific targets, such as those with already known companions or with companions detected in parallel with SHINE, but by other teams (e.g., GPIES or open-time programs with SPHERE). The original scientific priority of stars with detections are listed in Table~\ref{tab:detections}, along with the reassignment to the P0 priority when applicable.

The reassignment of P0 priority to some targets based on a priori knowledge of the presence of companions necessarily introduces a statistical bias. Without a priori knowledge, these targets may not have been immediately observed in the course of the survey, or may even have had a very low probability of being observed (e.g., HIP\,107412, \citealt{Milli2017}). To properly take into account the previously known detections in the sample, we introduced a statistical weight related to the probability that the target would have been observed if the companion had not been known before. The value of this weight is between zero and one.

The most straightforward case is for completely new detections around HIP\,65426 \citep{Chauvin2017} and HIP\,64892 \citep{Cheetham2018}. These two targets were in the P1 category and were observed as part of the normal course of the survey. Each of these detections are therefore counted as full detections (statistical weight of 1.0).

Then, there are cases of targets that were known to have a companion, but were not given a higher priority based on this knowledge. Only two such targets are included in the current sample: $\eta$\,Tel and CD\,-35\,2722. These objects were observed independently of the fact that there was knowledge about a substellar companions, and we can safely assume that if their companions had not been known, they would certainly have been detected. The latter statement is not a strong assumption because of the relatively low contrast and large angular separation of $\eta$\,Tel\,B \citep{Lowrance2000} and CD\,-35\,2722\,B \citep{Wahhaj2011}. Each of these detections are therefore also counted as full detections (statistical weight of 1.0).

Finally, there are cases of targets for which the priority was boosted to P0 because of previously known companions (HIP\,78530, $\beta$\,Pic, HR\,8799, HD\,95086, PZ\,Tel, AB\,Pic, and GSC\,8047-0232) or because of the discovery of a companion by another team after the start of the SHINE survey (51\,Eri and HIP\,107412). For these stars, the assigned statistical weight is equal to the probability that a star from the same priority bin (P1 to P4) would have been observed by a fixed date, specifically, the date where the early SHINE statistical sample was frozen. Because the current sample was frozen in the course of the survey, this date also corresponds to the time where 100\% of first-epoch observations were obtained for the stars in the sample. Following this analysis, detections around targets that originally were in the P1, P2, and P4 priority bins\footnote{No companions are detected around P3 targets in the current SHINE subsample.} were attributed a statistical weight of 0.60, 0.35, and 0.01 respectively. The weight values were computed numerically a posteriori, based on the definition of the sample and on the dates of all the SHINE observations. For example, a weight of 0.6 implies that 60\% of the stars within the original priority class of that particular star were observed at the point at which the survey was frozen for the analysis, independently of the stellar types.

The statistical weight of each detection considered in the analysis is taken into account in the Markov chain Monte Carlo (MCMC) simulations. These are described in Sect.~\ref{sec:stat_tools}.

\subsection{Mass conversion of the detection limits}
\label{sec:mass_conv}

To convert the detection limits obtained in luminosity space into mass detection limits, it is necessary to use a mass-luminosity relationship, $L(M)$. Whereas for old ($\gtrsim1$~Gyr) systems this relationship is essentially unique for gas giants, at young ages, the value of the post-formation luminosity still remains uncertain \citep{Marley2007,Spiegel2012,Marleau2014,Bonnefoy2014kap,Bonnefoy2014bet}. In recent years, first steps toward predicting the post-formation luminosity of planets have been taken \citep{Berardo2017,BerardoCumming2017,Cumming2018,Marleau2017,Marleau2019shock}. While detailed predictions are not quite available yet, these theoretical studies suggest that warm or hot starts are more likely (see also the discussion in Sect.~\ref{sec:impact_assumptions}). This agrees with observational results that cold starts are disfavored for massive companions (see review in e.g.,~\citealt{Nielsen2019}). For lower-mass companions, the question remains open from the observational side. For example, a cold start is \textit{\textup{allowed}} by the data for 51\,Eri\,b, but they do not exclude a hot start either (e.g., \citealp{Rajan2017,Samland2017}).

When luminosity was converted into mass, we used hot starts as the fiducial model, but also consider warm starts in Sect.~\ref{sec:impact_assumptions} and Appendix~\ref{sec:depth_of_search}. Specifically, we took the Bern EXoplanet cooling tracks (BEX) coupled with the COND atmospheric models \citep{Allard2001} and assumed hot-start or warm-start initial conditions \citep{Marleau2019}. Extending the fits of \citet[][Equations~(1b) and~(1c), respectively]{Marleau2019}, we took the post-formation (i.e., initial) luminosity $\Lpf$ of the BEX-hot and BEX-warm tracks as a function of planet mass $M_p$ as
\begin{subequations}
    \label{eq:LpfBEX}
    \begin{align}
    \tilde{M}_n &\equiv \frac{M_p}{n~\MJup},\\
    \Lpf^{\textrm{BEX-hot}} & =
      2.62  \times 10^{-5} \,\tilde{M}_1^{1.4}~L_{\odot},\\
    \Lpf^{\textrm{BEX-warm}} & = \begin{cases}
        4.35 \times 10^{-6} \,\tilde{M}_1^{0.5}~L_{\odot}, &
        M_p \leqslant 10~\MJup \\
        1.39 \times 10^{-5} \,\tilde{M}_{10}^{7}~L_{\odot}, &
        10 < M_p/\MJup \leqslant 20 \\
        1.74 \times 10^{-3} \,\tilde{M}_{20}^{1.4}~L_{\odot}, &
        20~\MJup \leqslant M_p.
        \end{cases}
    \end{align}
\end{subequations}
Finally, we also considered the COND-2003 cooling tracks \citep{Baraffe2003}, which have even higher $\Lpf(M_p)$ that is not based on any formation model.

\begin{figure}[t]
    \centering 
    \includegraphics[width=0.5\textwidth]{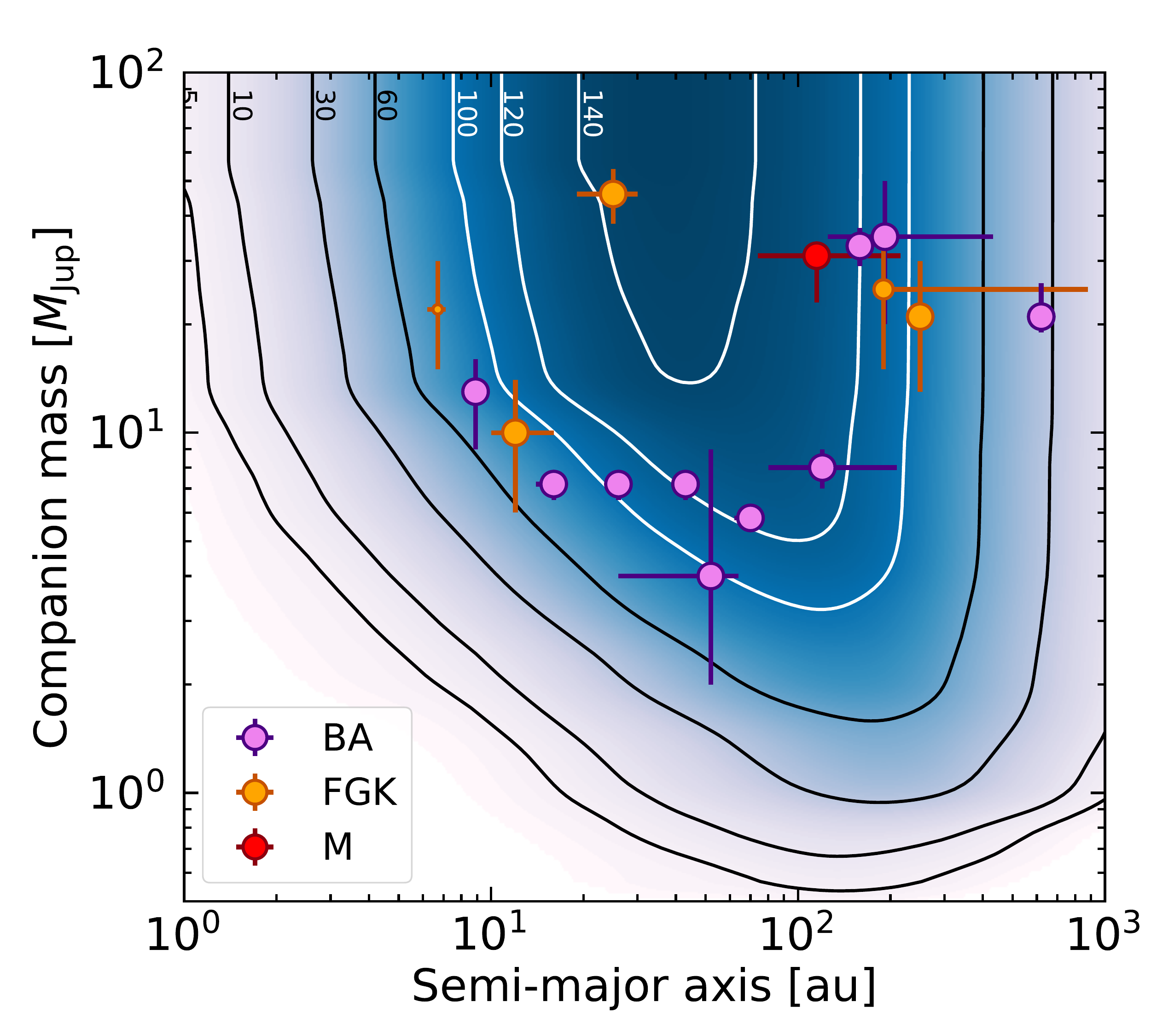}
    \caption{Depth of search of the SHINE survey for the 150 stars in the sample. The black and white contour lines give the numbers of stars around which the survey is sensitive to substellar companions as a function of mass and semimajor axis. The mass conversion of the detection limits is based on the nominal stellar ages and on the BEX-COND-hot evolutionary models \citep{Marleau2019}. The colored circles represent the detected substellar companions in the sample. The color indicates the spectral type of the primary star (BA, FGK, or M). The size of the symbol is proportional to the weight of the detection in the statistical analysis (see Sect.~\ref{sec:stat_weight} and Table~\ref{tab:detections} for details).}
    \label{fig:shine_sensitivity}
\end{figure}

\begin{figure}[t]
    \centering 
    \includegraphics[width=0.5\textwidth]{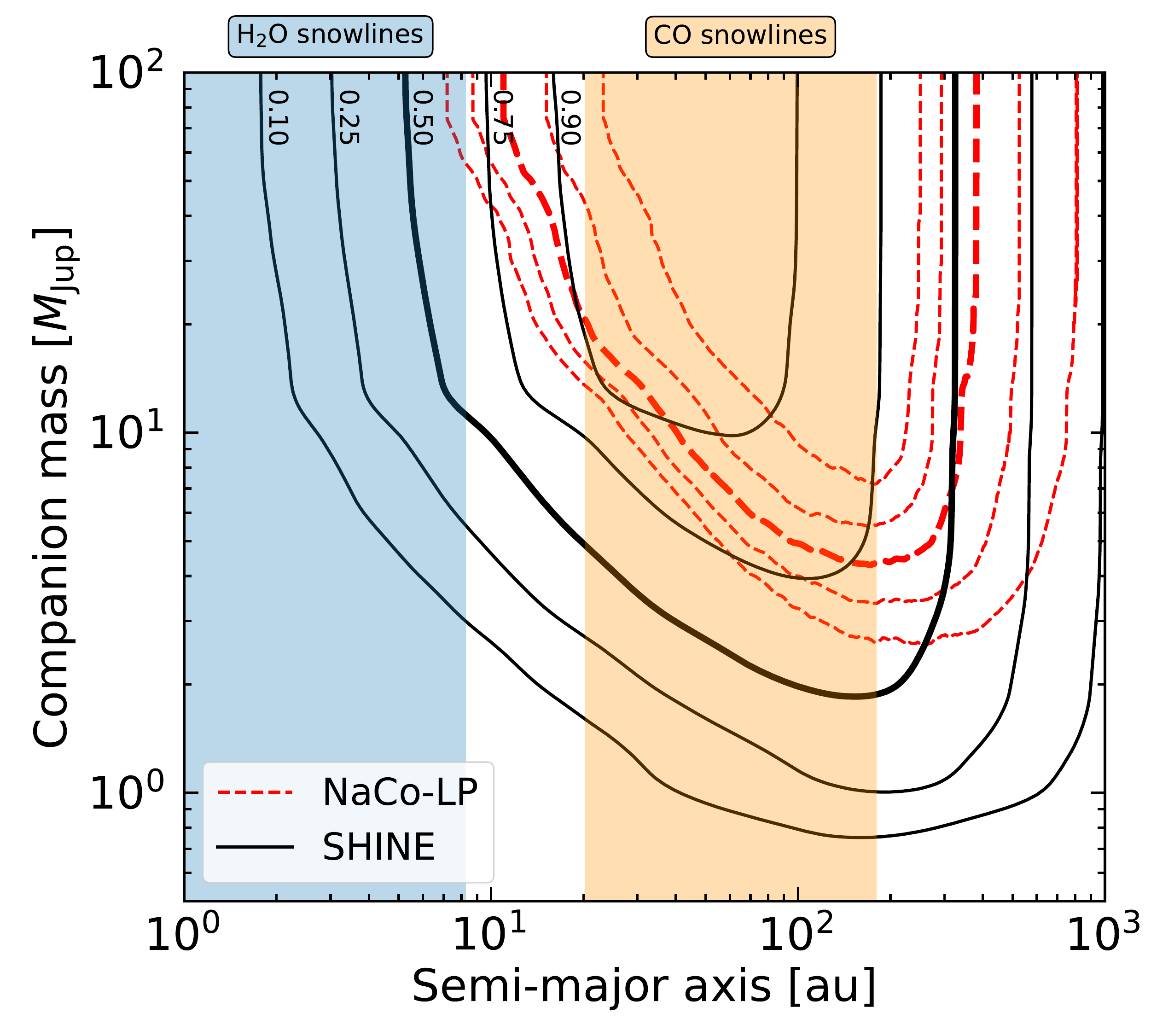}
    \caption{Comparison of the sensitivities of the NaCo-LP (\citealt{Vigan2017}; dashed red lines) and SHINE (this work; solid black lines) surveys (with the current sample), based on the average probability of detecting a companion as a function of its mass and semimajor axis. The analysis is based on detection limits that were converted using the COND-2003 evolutionary tracks for both surveys. The contours for the NaCo-LP are not labeled but are the same as for SHINE, and correspond to equal levels of detection probability. The range of semimajor axes spanning the H$_2$O and CO snow lines for the stars in the sample are overplotted (see Sect.~\ref{sec:survey_sensitivity} for details).}
    \label{fig:shine_vs_nacolp_sensitivity}
\end{figure}

\subsection{Survey sensitivity}
\label{sec:survey_sensitivity}

In order to constrain the statistical properties of our observed sample, we first converted the observed detection limits into the same parameter space as the models, that is, from projected separation to semimajor axis and from detection contrast to companion mass, so as to determine the completeness of the survey in terms of semimajor axis $a$ and companion mass $M_p$. For each star, we defined a grid of semimajor axis and mass values uniformly distributed in log space, with 500 values ranging from 0.1 to 10\,000\,au in $a$, and 200 values between 0.1 to 100\,\MJup in $M_p$. For each cell in the grid, we generated $10^4$ companions with arguments of periastron and orbital phases drawn from uniform distributions, taking into account the orbital velocities along the orbit (i.e., considering the fact that an eccentric companion spends more time near apastron). We used a uniform distribution in inclination in order to simulate random orientations of orbits in space. For the eccentricity distribution, we considered the recent results derived by \citet{Bowler2020} for directly imaged exoplanets and brown dwarf companions. For this parameter we adopted a Beta distribution with parameters [$\alpha=0.95$, $\beta=1.30$], which corresponds to the best fit to the full sample of wide substellar companions studied in \citet{Bowler2020}.

For each simulated companion, we then computed the corresponding projected separation from the drawn orbital elements and the semimajor axis $a$ of that grid point. We finally determined whether the companion is detectable in our observations by verifying that the mass value $M_p$ of that cell lies above the contrast curve converted into mass of the considered star at the obtained projected separation (see Sect.~\ref{sec:mass_conv}), and that this projected separation value lies within the FoV for that star. The fraction of detectable companions in each grid cell provides the fractional completeness as a function of mass and semimajor axis for each star in our sample. Summing all derived completeness maps and dividing by the number of targets, we obtained the average 2D completeness of the survey. This task was repeated using the mass limits obtained with the various evolutionary models described in Sect.~\ref{sec:mass_conv}, and considering the nominal, minimum, and maximum ages of the stellar primaries. This provided a separate completeness map for each specific analysis to be performed. 

Using the completeness maps for each of the targets in the sample, we computed the depth of search of the complete survey, which provides the number of stars around which the survey is sensitive for a given substellar companion mass and semimajor axis. This metric is useful for estimating the statistical strength of the results presented later. The depth of search for the 150 stars of our sample is presented in Fig.~\ref{fig:shine_sensitivity}, based on the nominal stellar ages and the BEX-COND-hot models (see Sect.~\ref{sec:mass_conv}). The core of the sensitivity (>100 stars) reaches 7--9\,au for objects >10\,\MJup. At lower masses, the sensitivity to the lowest masses around at least 100 targets is reached at $\sim$100\,au with a mass of $\sim$3\,\MJup. Sensitivity to 1\,\MJup planets is only reached around $\sim$30 stars at separations of 100--200\,au.

The mean completeness map for the whole sample provides the average sensitivity of the survey, that is, the average probability of detecting an object of given mass and semimajor axis. This metric enables a direct comparison of the SHINE survey to surveys performed using the previous generations of instruments. In Fig.~\ref{fig:shine_vs_nacolp_sensitivity} we compare the sensitivity of SHINE with that of the NaCo-LP survey \citep{Chauvin2015,Vigan2017}, which were both computed using detection limits converted into mass using the COND-2003 evolutionary tracks \citep{Baraffe2003}. While the two surveys do not share strictly identical samples, they both target a large pool of relatively young nearby stars, so that the probability of detection in the mass versus semimajor axis space averaged over all targets is a good metric for comparison. Clearly, the new generation of instruments such as SPHERE provides a significant boost in sensitivity for 1--10\,\MJup planets in the 5--50\,au range. However, the core of the sensitivity (probability >50\%) still remains beyond 10\,au, even for the most massive substellar companions (10--300\,au for companions $>$10\,MJup).

We also plot in Fig.~\ref{fig:shine_vs_nacolp_sensitivity} an estimate of the range of H$_2$O and CO snow lines for the stars in the SHINE sample. The snow lines are estimated based on a parametric disk temperature profile as derived from the composition of Solar System bodies \citep{Lewis1974} and on observations of a large sample of protoplanetary disks \citep{Andrews2005,Andrews2007a,Andrews2007b}. The average evaporation temperatures for H$_2$O and CO have been reported in \citet{Oberg2011}, specifically, they are 135\,K and 20\,K, respectively. Because protoplanetary disk physics and chemistry are complex, these estimates of the snow lines locations are approximate, but they enable a first-order comparison of the sensitivity of SHINE in locations that are important for giant planet formation. It is interesting to note that SHINE has some sensitivity to massive objects at the level of the water snow-line, which might constitute a turnover point in the giant planet occurrence rate \citep{Fernandes2019}, although the core of the sensitivity is shifted toward larger orbital separations. If the water snow-line is indeed a turnover point, the low detection rate of new planetary companions in the SHINE and GPIES surveys might qualitatively indicate that this turnover might apply to low masses where SHINE (this work) and GPIES \citep{Nielsen2019} have little sensitivity.

\section{Exoplanet population modeling}
\label{sec:exoplanet_models}

We here compare our observations to two different types of exoplanet population models. The first type is a parametric model based on inputs from both theoretical and observational work (Sect.~\ref{sec:p-model}), which aims at being a better representation than the simple power-law distributions in mass and semimajor axis used previously \citep[e.g.,][]{Lafreniere2007, Kasper2007,Nielsen2010,Vigan2012}. Although relatively straightforward, this remains a simplified parametric approach to describing the giant exoplanet population. The second type of model is based on exoplanet population synthesis models, which by definition rely on very detailed (although often simplified) physical modeling of the planet formation, interactions, and evolution (Sect.~\ref{sec:s-model}). The parametric and population model types both include a top-down binary star-like formation component and a bottom-up planet-like formation component in an attempt to capture different formation pathways for the observed detections in the SHINE sample.

\subsection{Parametric model}
\label{sec:p-model}

\begin{figure}
    \centering 
    \includegraphics[width=0.5\textwidth]{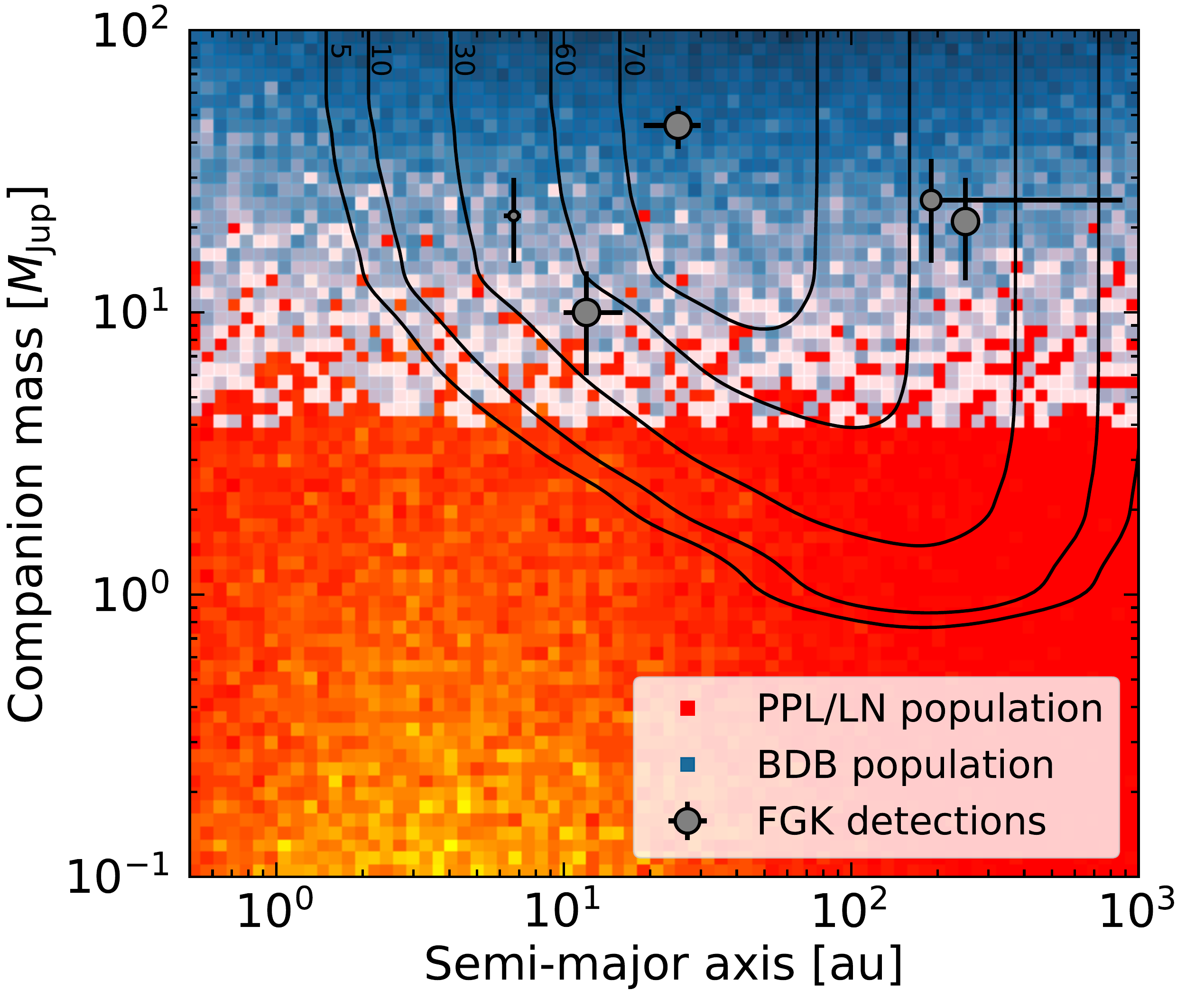}
    \caption{Comparison of the depth of search of the SHINE survey for the 77 FGK stars in the sample with a population of 20000 draws from our parametric model presented in Sect.~\ref{sec:p-model}. The contour lines give the numbers of stars around which the survey is sensitive to substellar companions as a function of mass and semimajor axis. The PPL/LN part of the model is represented with shades of red (low density of companions) to yellow (high density of companions), and the BDB part of the model is represented with shades of white (low density of companions) to blue (high density of companions). Only the detections around FGK stars are plotted.}
    \label{fig:shine_pop_param_comparison}
\end{figure}

\begin{figure}[t]
    \centering 
    \includegraphics[width=0.5\textwidth]{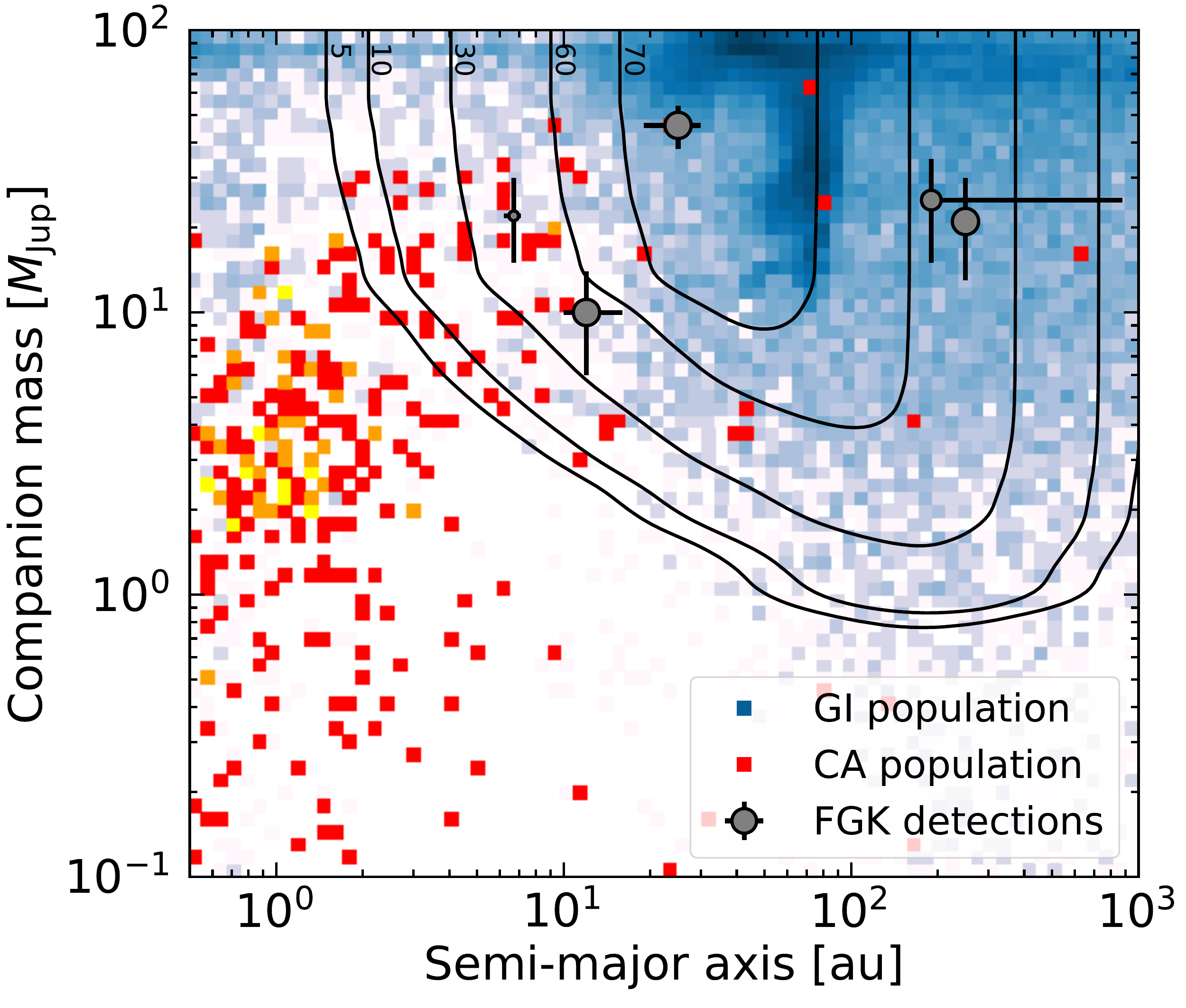}
    \caption{Comparison of the depth of search of the SHINE survey for the 77 FGK stars in the sample with the population synthesis models based on the CA and GI formation scenarios presented in Sect.~\ref{sec:GI_pop} and Sect.~\ref{sec:CA_pop}, respectively. The contour lines give the numbers of stars around which the survey is sensitive to substellar companions as a function of mass and semimajor axis. The CA companions are represented with shades of red (low density of companions) to yellow (high density of companions), and the GI companions are represented with shades of white (low density of companions) to blue (high density of companions). The apparent lower density of CA objects arises because the vast majority of the CA population is located outside the range of mass and semimajor axis considered in this plot. Only the detections around FGK stars are plotted.}
    \label{fig:shine_pop_synth_comparison}
\end{figure}

We compared our results to a parametric model that was developed to explain a wide range of observations. Details of this model are presented in \citet{Meyer2020}. We provide here an overview of the key features of the model for our needs in the context of the SHINE survey.

For each bin of stellar spectral type (BA, FGK, and M), the model comprises two parts that represent two different populations of substellar companions: one is a planet-like population, and the other is a binary star-like population. For each of these two parts, we considered different distributions of objects as a function of mass and semimajor axis: a distribution of planets as a function of orbital separation ($a$) and  a planet mass function ($b$) for the planet-like population, and an orbital distribution of low-mass binary companions ($c$) and a companion mass ratio distribution ($d$) for the binary star-like population. The planet part of the model ($a$ and $b$) and the binary star part of the model ($c$ and $d$) require a different normalization. In principle, all parameters of the model can be fit to the data. However, because our survey includes a limited number of observations, we only fit the normalization of the planet part and binary part separately (two free parameters). These normalization factors represent the amplitudes, or the relative frequencies, of having a very low-mass binary-like companion or a planet-like companion. Combined, the resulting fit represents the total probability for a star to have one or more substellar companions. 

For part $(a)$, the orbital distribution of gas giant planets, we assumed a Gaussian distribution in $\log a$, $a$ being the semimajor axis, with fixed mean and sigma. These properties likely depend on host star mass, and based on results to date, we adopted a log-normal distribution with mean $\log a = 0.45$ and $\sigma = 0.52$ for M stars \citep{Meyer2018,Fernandes2019}, $\log a = 0.58$ and $\sigma = 0.69$ for FGK stars, and $\log a = 0.79$ and $\sigma = 0.77$ for BA stars \citep{Meyer2020}. For part $(b),$ the planet mass function, we assumed a power law where the frequency $f$ depends on the ratio of the companion to the host star mass, $q = M_p/M_{\mathrm{\star}}$, that is, $f \propto q^{\beta}$, with $\beta = -1.31$ for all stellar types \citep{Cumming2008,Wagner2019}. We furthermore assumed that the planet mass function does not depend on orbital separation. The amplitude factor associated with the product of these two functions, \fppl, is the first of our two fit variables. The planet part of our phenomenological model, which combines $(a)$ and $(b)$, is abbreviated PPL/LN (planet power-law, log-normal) from here on.

For part $(c)$ we assumed a log-normal surface density of binary companions, as measured for stellar masses (e.g., \citealt{Raghavan2010} for FGK stars and \citealt{Winters2019} for M dwarfs) with mean $\log a = 1.30$ and $\sigma = 1.16$ for M dwarfs, $\log a = 1.70$ and $\sigma = 1.68$ for FGK stars, and $\log a = 2.59$ and $\sigma = 0.79$ for BA stars \citep{DeRosa2014}. For part $(d)$ we assumed a universal companion mass ratio distribution, which is roughly flat with the mass ratio (power-law slope of 0.25; \citealt{Reggiani2013}). We assumed that the companion mass ratio distribution extends to the minimum mass for fragmentation \citep[cf.][]{Reggiani2016} and that the companion mass ratio distribution does not depend on orbital separation. The other amplitude factor associated with the product of these two functions, \fbdb, is our second fit variable. The binary part of our phenomenological model, which combines $(c)$ and $(d)$, is abbreviated BDB (brown dwarf binary) from here on.

An illustrative comparison of the output populations with the survey sensitivity around FGK stars is provided in Fig.~\ref{fig:shine_pop_param_comparison}. The BDB and PPL/LN parts of the model are clearly visible: the density of planetary companions (PPL/LN) is highest at low masses, with a peak at a few astronomical units and a density decreasing toward higher masses and larger orbital separations, while the density of binary companions (BDB) is highest for higher masses and then slowly decreases toward planetary masses.

In our analysis, we fit only the relative frequencies \fbdb and \fppl for the parametric model and for each bin of stellar spectral type (BA, FGK, and M). We also computed the total frequency for the sum of the planetary and binary parts of the model, \fpmod. 

\subsection{Population model}
\label{sec:s-model}

The population model consists of two different population synthesis models based on the GI scenario and the CA scenario, which are described in Sect.~\ref{sec:GI_pop} and~\ref{sec:CA_pop}, respectively. Combined, they comprise the full population model. These models are currently computed only for solar-mass stars, therefore we compare them only to the observations of FGK stars in the sample (see Sect.~\ref{sec:full_pop_model} and~\ref{sec:results}). Comparison with higher and lower mass stars will be the subject of future work.

\subsubsection{GI population}
\label{sec:GI_pop}

The synthetic GI populations are based on those first presented by \citet{Forgan2013} and then updated by \citet{Forgan2018}. These models involved running, in advance, a suite of 1D disk models that smoothly proceed from an epoch in which the GI dominates their evolution \citep{Rice2009} to an epoch in which it is dominated by an alternative angular momentum transfer mechanism, such as the magnetorotational instability \citep{Balbus91}. These models also include photoevaporation, which plays an important role in disk dispersal \citep{Owen2011}. The outer radius of each disk was taken to be 100\,au, which optimizes the likelihood of the disk to undergo fragmentation, after which dynamical interactions can then sculpt the semimajor axis distribution. The disk-to-star mass ratios varied from 0.125 to 0.375, and the host star masses varied from 0.8 to 1.2\,\MSun.

To generate the synthetic populations, a disk model was selected and fragments were then placed in this disk. The innermost fragment was placed at the smallest radius where fragmentation is possible, typically beyond $\sim$50\,au \citep{Rafikov2005,Clarke2009}, and the subsequent fragments were then placed at separations that were initially a random number of Hill radii (uniform distribution between 1.5 and 3 Hill radii). The fragment masses were set by the local Jeans mass, their radii were set using the assumption that they are equivalent to the initial radii of star-forming cores, and their initial temperatures were set to be the virial temperature \citep{Nayakshin2010}. 

The fragments then followed a tidal downsizing process where they contracted and cooled, and evolved through disk migration and $n$-body interactions. Grains within the fragment can grow and sediment, potentially forming a solid core. When the radius of an embryo exceeds its Hill radius, it can be tidally disrupted, potentially allowing for the emergence of a terrestrial-mass protoplanetary core. Each system was evolved for a duration of 1\,Myr to ascertain the effect of object--object scattering on the planetary orbital parameters \citep{Forgan2015}. Although each system was evolved for a time that is shorter than the observed ages of the objects to which we would like to compare to disk fragment models, this relatively short simulation time was used partly to reduce computational expense and partly because systems that produce scattering events express this instability within a few ten thousand years \citep{Chambers1996,Chatterjee2008}.

This process was repeated many times to produce a large population of planetary systems that have formed via GI. These systems were then used as input for the SHINE simulations for comparison with our observational results. The relative frequency of systems with at least one companion associated with the GI model of formation is noted \fgi from here on.

\subsubsection{CA population}
\label{sec:CA_pop}

The synthetic CA populations were obtained using the new Bern generation 3 model of planetary formation and evolution described in \citet{Emsenhuber2020A}, which corresponds to an update of the model presented in \citet{Mordasini2018}. This model in turn has evolved out of earlier versions of the Bern model described in \citet{Alibert2004}, \citet{Mordasini2012}, and \citet{Benz2014}. The model self-consistently evolves a 1D gas disk, the dynamical state of the solids, the accretion by the protoplanets, gas-driven migration of the protoplanets, the interiors of the planets, and their dynamical interactions. The specific population we used is population \texttt{NG76} from the new-generation planetary population synthesis (NGPPS) series.

For the gas disk, the model assumes that it is viscously evolving \citep{LyndenBellPringle1974} and the macroscopic viscosity is given by the standard $\alpha$ parameterization \citep{ShakuraSunyaev1973}. The vertical structure was computed using a vertically integrated approach \citep{NakamotoNakagawa1994} that includes the effect of stellar irradiation. We included additional sink terms for the accretion by the planets, and both internal and external photoevaporation, following \citet{Clarke2001} and \citet{Matsuyama2003}, respectively.

The model assumes planetesimal accretion in the oligarchic regime \citep{IdaMakino1993,Ohtsuki2002,Thommes2003}. The model solves the internal structure equations \citep{BodenheimerPollack1986} for the gas envelope. In the initial (or attached) phase, the envelope is in equilibrium with the surrounding disk gas, and accretion is governed by the ability of the planet to radiate the gravitational energy released from the accretion of both solids and gas. When the accretion rate exceeds the supply from the disk, the envelope is no longer in equilibrium with the disk and contracts \citep{Bodenheimer2000}. Planets undergo gas-driven migration, and the dynamical interactions are followed by means of an \textit{n}-body simulation.

After 20 Myr, the model transitions into the evolution stage, where the planets are followed individually up to 10 Gyr. In this stage, the model computes the thermodynamical evolution of the envelope, atmospheric escape, and tidal migration, but the gravitational interactions with other planets in the system are not considered.

To obtain a synthetic population, we followed the procedure outlined in \citet{Mordasini2009} and \citet{Emsenhuber2020B}. The distributions for the disk masses follow \citet{Tychoniec2018}, and we used the relationship described by \citet{Andrews2010} to determine the characteristic radius that defines the radial distribution of the gas. The inner edge of the disk is based on the work of \citet{Venuti2017}, with a log-normal distribution in period with a mean of 4.7\,d. The dust-to-gas ratio was obtained as described in \citet{Mordasini2009} from the observed stellar [Fe/H], but we used the primordial solar metallicity as a reference \citep{Lodders2003} without an enhancement factor. The initial slope of the surface density of solids is steeper than the slope of the gas disk, following \citet{Ansdell2018}.

The population used here consists of 1000 systems with 1\,\MSun stars. Each disk started with 100 planetary embryos of lunar mass ($10^{-2}\,M_\oplus$), whose initial positions were randomly selected between the inner edge of the disk up to 40 au, with a uniform probability in the logarithm of the semimajor axis. 

The generated systems were then used as input for the SHINE simulations. The relative frequency of systems in which at least one companion is associated with the CA mode of formation is noted \fca from here on.

\subsubsection{Full population model}
\label{sec:full_pop_model}

The two population synthesis models described above were combined to form the full population model. Because the population synthesis models were computed only for solar-mass stars, we restricted our analysis with this model to the 77 FGK stars that are part of the present SHINE sample. In our analysis we fit the relative frequencies \fca and \fgi that are associated with the CA and GI parts of the model, respectively, and the total frequency for the sum of the two parts, \fsmod.

An illustrative comparison of the output populations with the survey sensitivity is provided in Fig.~\ref{fig:shine_pop_synth_comparison}. Similarly to what has been described in \citet{Vigan2017} for the GI population, a large cluster of massive objects (>10\,\MJup) is located at separations of 50--100\,au where the SHINE survey is the most sensitive. In contrast, the CA population only shows a rather small population of 1--30\,\MJup objects that are scattered at separations ranging from a few up to a few dozen astronomical units.

\section{Statistical tools}
\label{sec:stat_tools}

We used a statistical tool based on the MCMC sampling method described in \citet{Fontanive2018,Fontanive2019} to constrain the companion fractions of our observed sample. The tool was built using the {\tt emcee} \citep{Foreman-Mackey2013} python implementation of the affine-invariant ensemble sampler for MCMC \citep{Goodman2010}.  The code was adapted to use two separate exoplanet population models, each made of two parts: the parametric model presented in Sect.~\ref{sec:p-model}, and the population model described in Sect.~\ref{sec:s-model}. In all simulations, the shapes of the underlying companion distributions in mass and semimajor axis were fixed to those of the models, leaving as only MCMC parameters the relative companion frequencies of the two model populations considered. The companion fractions $f_1$ and $f_2$ of populations 1 and 2, respectively, are defined over fixed semimajor axis and companion mass ranges, [$a_\mathrm{min}$, $a_\mathrm{max}$] and [$M_{p,\mathrm{min}}$, $M_{p,\mathrm{max}}$]. We sought the posterior distributions of $f_1$ and $f_2$ given our observed data, where 1 and 2 designate the two parts of our models, either BDB and PPL/LN for the parametric model, or GI and CA for the population model.

In order to take the uncertainties on the measured masses and semimajor axes of the detected planets and brown dwarfs around the observed targets into account, we followed the method of sampling approximation to the marginalized likelihood from \citet{Hogg2010}. This offers a powerful approach in the framework of Bayesian statistics to inform a population-level likelihood using the posterior distributions of individual systems. At each step in the MCMC, $K = 10^3$ sets of semimajor axes and masses are generated for each of the $N_\mathrm{comp}$ detected companions. Values are drawn from Gaussian distributions centered on the measured masses and semimajor axes, with Gaussian widths set to the uncertainties of the measurements (Table~\ref{tab:detections}). When no uncertainties are available, the measured value is always chosen. When the 1$\sigma$ interval is asymmetric around the most likely value, we defined a two-piece Gaussian (see, e.g., \citealp{Wallis2014}) from which values were randomly selected. When the drawn values are between [$a_\mathrm{min}$, $a_\mathrm{max}$] and [$M_{p,\mathrm{min}}$, $M_{p,\mathrm{max}}$], a companion was counted towards the detections in that region of the parameter space. For each iteration $k$, we thus obtained a number $N_{\mathrm{sys},k}$ of systems with at least one companion in the probed range, which might vary when the drawn parameters occasionally fell outside the ranges of interest. We note that $N_{\mathrm{sys},k}$ may be smaller than the total number of detected companions when multiple planets or brown dwarfs are found around the same star. 

For all iterations, we started by estimating the total number of companions expected to be detected in our observations. This was done by drawing simulated companions between [$a_\mathrm{min}$, $a_\mathrm{max}$] and [$M_{p,\mathrm{min}}$, $M_{p,\mathrm{max}}$] from the two model distributions, and injecting them into the combined completeness map defined in Sect.~\ref{sec:survey_sensitivity}, using only the targets considered in a specific analysis (e.g., the BA, FGK, or M stars) with the selected evolutionary models and stellar ages. For the parametric model, $N_1 = N_2 = 10^4$ companions were drawn from the continuous separation and mass ratio distributions describing the BDB and PPL/LN populations. We used the mean stellar mass of the studied subset to convert the mass ratios of the model companions into corresponding companion masses. When we worked with the synthetic population models (FGK stars only), we injected all companions found in each model within the considered semimajor axis and mass limits, adding up to totals of $N_1$ and $N_2$ companions, respectively. The expected total number of detections $\lambda$ around the observed targets is then given by
\begin{equation}
        \lambda = \left[ \frac{f_1}{N_1} \sum_{i=1}^{N_1} p_i + \frac{f_2}{N_2} \sum_{j=1}^{N_2} p_j \right] \times N_\star ,
    \label{eq:expected_detections}
\end{equation}
where $N_\star$ is the number of stars in the studied subsample, and $p_i$ and $p_j$ are the probabilities of detecting simulated companions $i$ and $j$ from model populations 1 and 2, given the survey sensitivity. The first term in the square brackets thus provides the fraction of detectable companions from population 1 with companion frequency $f_1$, and the second term the fraction from population 2 with companion frequency $f_2$. The sum of these two terms gives the total fraction of companions that can be detected in the survey from the injected populations. This value was then multiplied by $N_\star$ to obtain the total number of companions expected to be detected for respective companion frequencies $f_1$ and $f_2$ for the two parts of the model population.

The number $\lambda$ of expected substellar detections may then be compared to the observed number of systems $N_{\mathrm{sys},k}$ using Poisson statistics, as was done in \citet{Fontanive2018}, providing a value $\mathcal{L}_{\mathrm{P},k}$ at each step $k$. Averaged over the $K$ iterations, this provides the first part of the likelihood function, which allows us to constrain the overall companion fraction. As detailed in Sect.~\ref{sec:stat_weight}, some of the detections are weighted to correct for observational biases due to the presence of previously known companions. The total number $N_{\mathrm{sys},k}$ of detected systems is thus given by the sum of the effective detection rates for the companions to retain, listed in Table~\ref{tab:detections} (counting the HR\,8799 system only once).

The second part of the likelihood compares the position of the companions in the mass--semimajor axis space to the model distributions in order to scale the relative companion frequencies of the two populations. This was done by defining a joint 2D probability density describing the semimajor axis--mass distributions of the combined model populations, weighting each part of the model by taking into account the relative companion fractions of each population, $f_1$ and $f_2$. Following the approach from \citet{Fontanive2018}, we were then able to compute the probabilities of the detected companions being drawn from this overall model distribution. The full model probability density was convolved with the completeness map for the targets we investigated, so as to represent the distribution of companions that could be observed given the survey sensitivity. When the semimajor axis and mass log spaces are divided into bins of width 0.2\,dex, the probability of observing a companion in a specific mass--semimajor axis bin is given by the volume below the probability density function delimited by the edges of that bin. For each of the $N_\mathrm{comp}$ detections in the considered subsample, we thus computed at each of the $K$ steps the integral within the bin enclosing the drawn mass and semimajor axis of the companion. When the drawn values for a companion fall outside the considered parameter space, the integral was set to 0. For each detection $n$, the probability that this companion is drawn from the joint model distribution is given by averaging the integrals obtained for each iteration $k$ ($\mathcal{L}_\mathrm{nk}$).

The final likelihood $\mathcal{L}$ is then computed as
\begin{equation}
        \mathcal{L} = \left( \prod_{n=1}^{N_\mathrm{comp}} \frac{1}{K} \sum_{k=1}^{K} \mathcal{L}_{nk} \right) \times \left( \frac{1}{K} \sum_{k=1}^{K} \mathcal{L}_{\mathrm{P},k} \right),
\label{eq: likelihood}
\end{equation}
where $\mathcal{L}_\mathrm{nk}$ is the integral computed above for the $n^\mathrm{th}$ observed giant planet or brown dwarf companion at the $k^\mathrm{th}$ iteration, and $\mathcal{L}_{\mathrm{P},k}$ is the Poisson likelihood of the total number of detected systems at the $k^\mathrm{th}$ step ($N_{\mathrm{sys},k}$) for an expected number of systems $\lambda$ given by Eq.~\ref{eq:expected_detections}. The term in the left set of brackets hence gives the probability that the detected companions are drawn from the overall model distribution, and the probability for each companion is calculated as the average over the $K$ iterations, and the final value for this term given by the product of the value for each detection. The term on the right provides the probability for detecting the number of companions that the survey yielded, taken again as the average of the Poisson likelihoods over the $K$ iterations.

We adopted uniform priors between 0 and 1 for the two companion fractions, $f_1$ and $f_2$. The combination of the prior distributions and likelihood function according to Bayes' theorem allows for the calculation of the posterior distribution for our two model parameters $f_1$ and $f_2$. In each simulation presented below, the code was run with $10^3$ walkers taking $10^4$ steps each. The initial 500 steps were discarded to remove the so-called burn-in phase, as a mean acceptance fraction was reached after some hundred steps.

\section{Results}
\label{sec:results}

\begin{figure*}
    \centering 
    \includegraphics[width=1\textwidth]{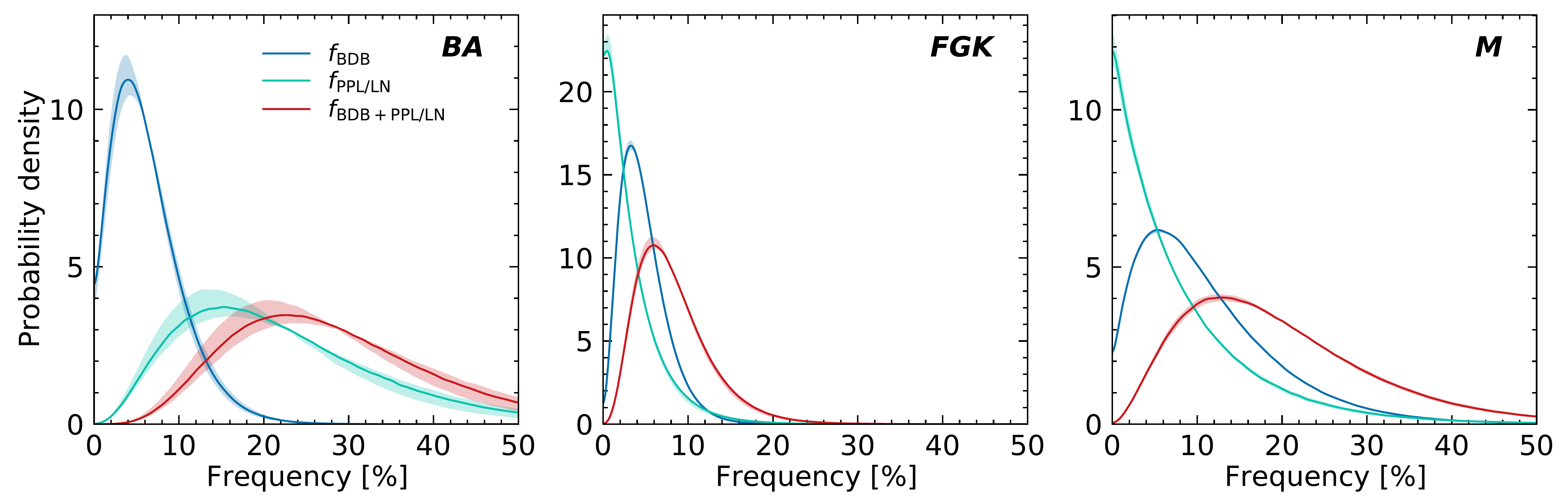}
    \caption{Probability density functions of the frequencies of substellar companions around BA (left), FGK (center), and M stars (right) based on the parametric model, computed for companions with masses in the range $M_p = 1$--75\,\MJup and semimajor axes in the range $a = 5$--300\,au, and using the BEX-COND-hot evolutionary tracks for the mass conversion of the detection limits. Each plot shows the PDFs for the relative frequencies of the two components of the model (\fbdb and \fppl), and for the total frequency for the full model (\fpmod). The plain lines show the PDFs for the nominal stellar ages, while the shaded envelopes show the variation of these PDFs for the maximum and minimum stellar ages. The median values and 68\% confidence intervals are provided in Table~\ref{tab:frequencies}.}
    \label{fig:p-model_SpT}
\end{figure*}

\begin{table*}
    \caption[]{Constraints on the frequency of substellar companions}
    \label{tab:frequencies}
    \centering
    \begin{tabular}{cccccccc}
    \hline\hline
    Mass range & s.m.a. range & Evol. model   & Ages & SpT    & Planet model & Median & 68\% CI      \\
    $[\MJup]$  & [au]      &               &      &        &              & [\%]   & [\%]         \\
    \hline
    \multicolumn{8}{c}{Parametric model} \\
    \hline
    1--75      & 5--300    & BEX-COND-hot  & Nominal & BA     & Full      & 23.0  & 13.3--36.5 \\
               &           &               &         &        & BDB       & 4.1   & 1.1--8.3   \\
               &           &               &         &        & PPL/LN    & 14.8  & 6.9--28.6  \\
    \hline
    1--75      & 5--300    & BEX-COND-hot  & Nominal & FGK    & Full      & 5.8   & 3.0--10.5  \\
               &           &               &         &        & BDB       & 3.2   & 1.4--6.2   \\
               &           &               &         &        & PPL/LN    & 0.4   & 0.0--4.0   \\
    \hline
    1--75      & 5--300    & BEX-COND-hot  & Nominal & M      & Full      & 12.6  & 5.5--25.5  \\
               &           &               &         &        & BDB       & 5.4   & 1.0--14.1  \\
               &           &               &         &        & PPL/LN    &       & $<9.7$     \\
    \hline
    \multicolumn{8}{c}{Parametric model -- impact of input assumptions} \\
    \hline
    1--75      & 5--300    & BEX-COND-hot  & Nominal & FGK    & Full      & 5.8   & 3.0--10.5 \\
               &           &               & Minimum &        &           & 5.7   & 3.0--10.1 \\
               &           &               & Maximum &        &           & 6.0   & 3.1--10.8 \\
    1--75      & 5--300    & BEX-COND-warm & Nominal & FGK    & Full      & 5.9   & 3.1--10.6 \\
    1--75      & 5--300    & COND-2003     & Nominal & FGK    & Full      & 6.0   & 3.1--10.7 \\
    1--75      & 10--300   & BEX-COND-hot  & Nominal & FGK    & Full      & 5.5   & 2.8--9.5  \\
    \hline
    \multicolumn{8}{c}{Synthetic population model} \\
    \hline
    1--75      & 5--300    & BEX-COND-hot  & Nominal & FGK    & Full      & 5.7   & 2.9--9.5  \\
               &           &               &         &        & GI        & 1.5   & 0.5--3.6  \\
               &           &               &         &        & CA        & 2.5   & 0.8--5.6  \\
    \hline
    \end{tabular} 
\end{table*}

\begin{figure*}
    \centering
    \includegraphics[width=0.33\textwidth]{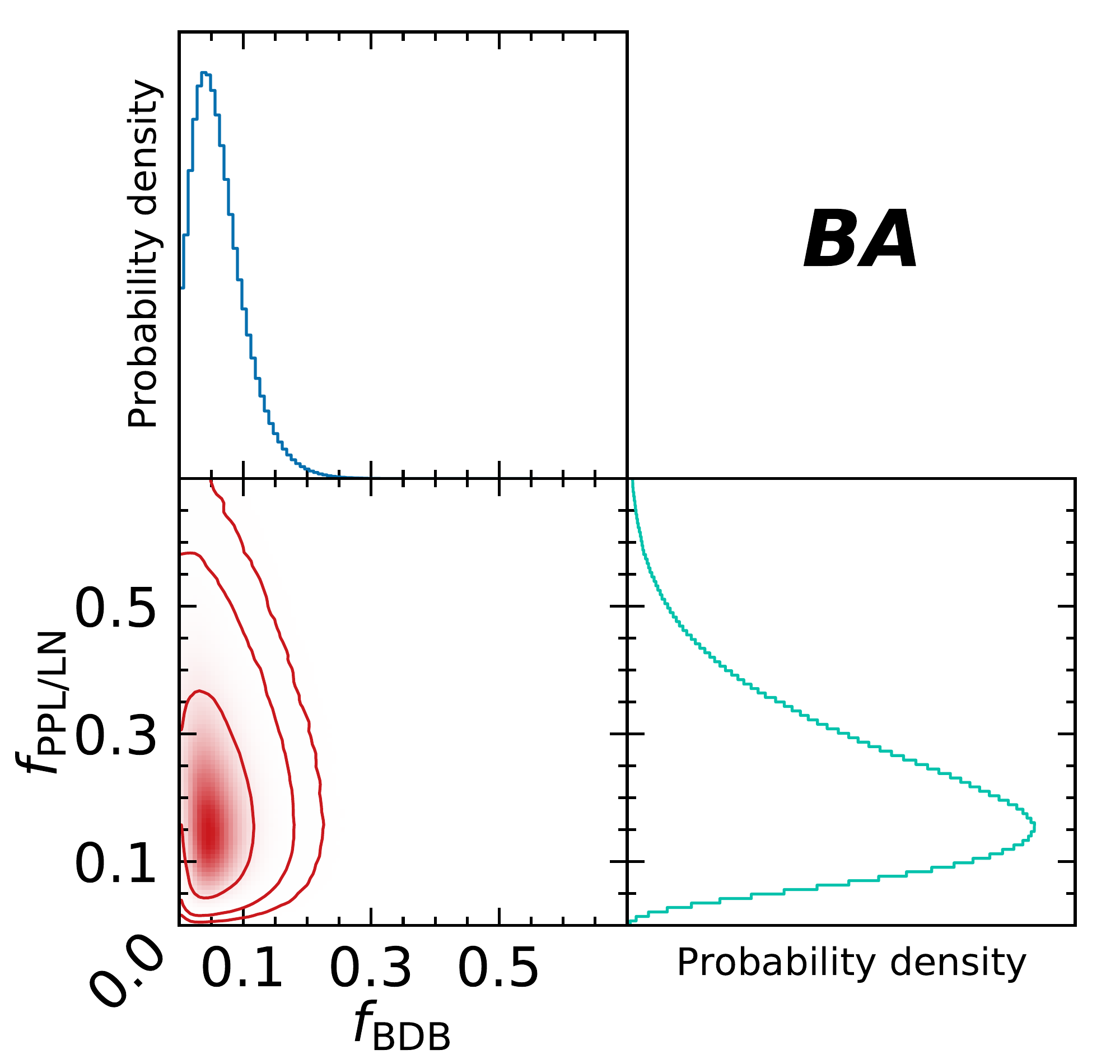}
    \includegraphics[width=0.33\textwidth]{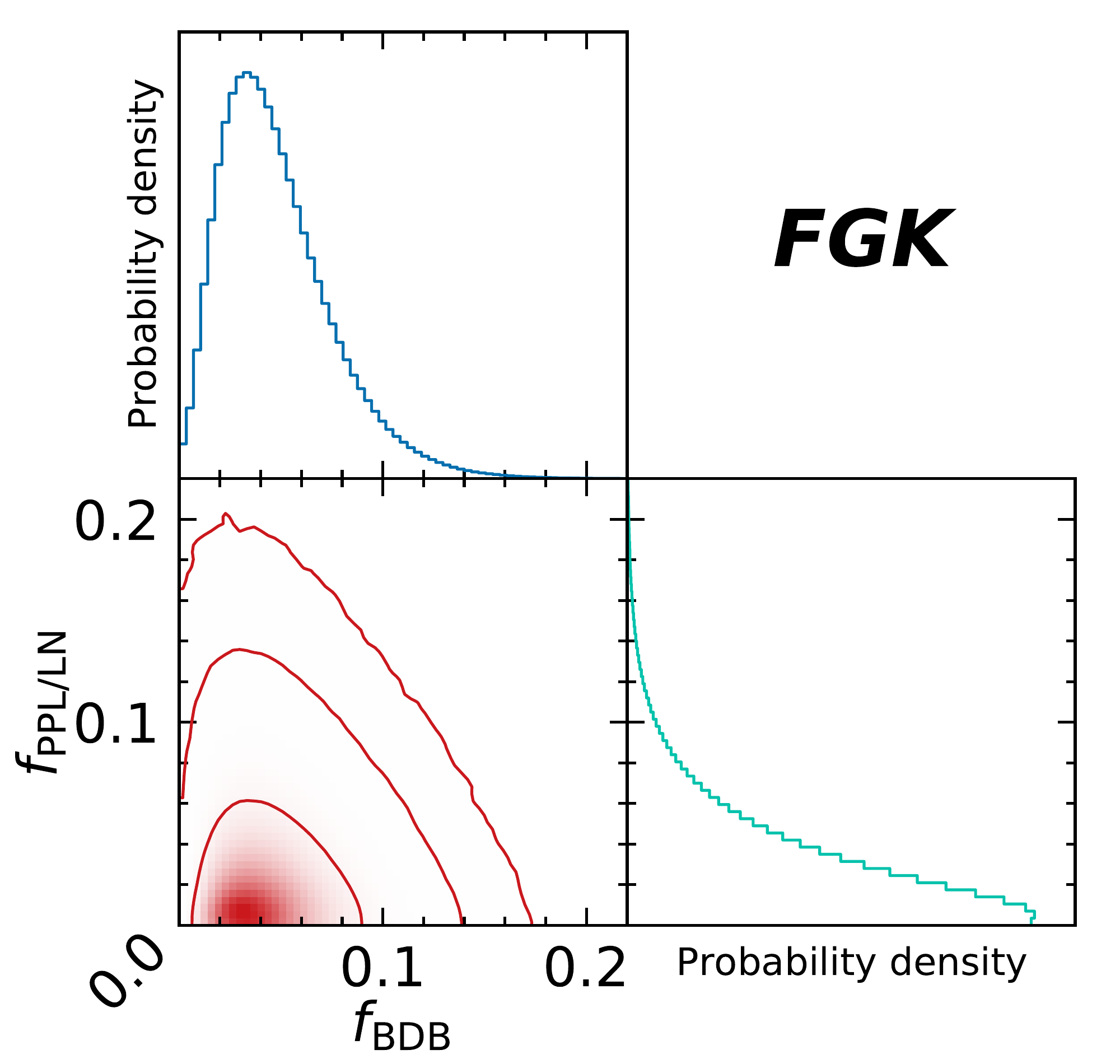}
    \includegraphics[width=0.33\textwidth]{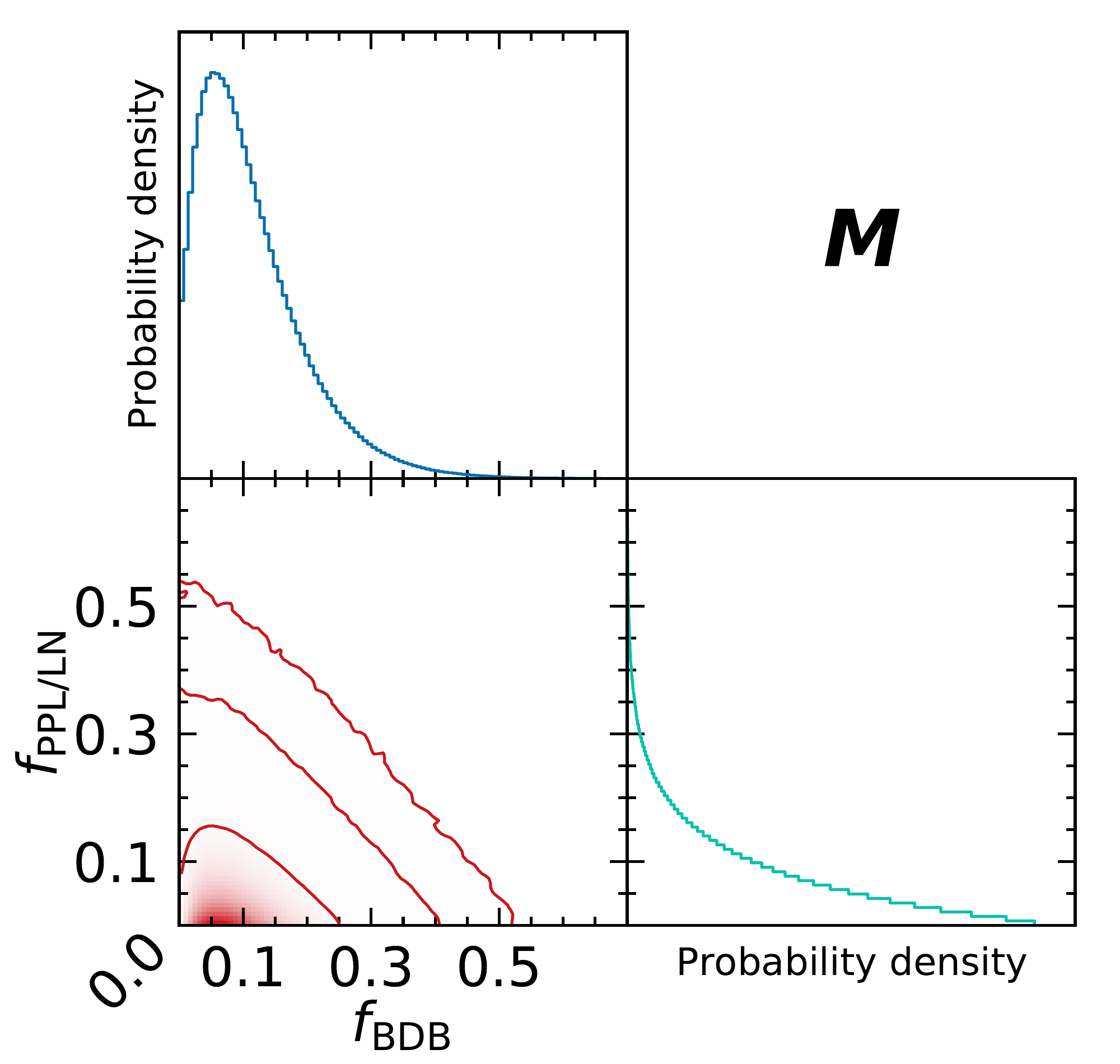}
    \caption{Correlation plots and marginalized probability density functions for \fbdb and \fppl in the parametric model around BA (left), FGK (center), and M (right) stars computed for companions with masses in the range $M_p = 1$--75\,\MJup and semimajor axes in the range $a = 5$--300\,au, and using the BEX-COND-hot evolutionary tracks at the optimal stellar ages. Contour lines in the correlation plots correspond to regions containing 68\%, 95\%, and 99\% of the posterior, respectively. For the FGK subsample, the scale of the axes is different from that of the two other subsamples.}
    \label{fig:p-models_cornerplot}
\end{figure*}

In this section we present the results from our comparison of the two models described above to our observations. We begin with results from our base parametric model (Sect.~\ref{sec:companion_freq_vs_stellar_mass}), then explore variations of the results as a function of input assumptions (Sect.~\ref{sec:impact_assumptions}), and conclude with tests of specific planet formation models using our population model (Sect.~\ref{sec:testing_planet_formation}).

\subsection{Frequency of substellar companions from parametric models versus host star mass}
\label{sec:companion_freq_vs_stellar_mass}

Our basic parametric model explores the companion frequency as a function of the companion mass ratio distribution and orbital distribution. For each host star, the mass range over which we are sensitive (1--75\,\MJup) results in a unique range of mass ratio $q$ (approximately 0.0005--0.2). We explored the companion frequency from 5 to~300\,au and used the age-dependent mass-luminosity conversions from the BEX-COND-hot model (see Sect.~\ref{sec:mass_conv}). We performed the mass-luminosity conversion for the nominal age, and the minimum and maximum ages, presented in \citetalias{SHINEPaperI}. Our fitting explores the best-fit combinations of relative frequencies for the brown dwarf binary companion model (BDB) and the planet distributions (PPL/LN). All the results are presented in Fig.~\ref{fig:p-model_SpT}, and the corresponding point estimates of frequency (probability that one or more companions lie the quoted ranges)  are reported in Table~\ref{tab:frequencies}.

Figure~\ref{fig:p-model_SpT} presents the probability density function (PDF) of the integrated frequency that one or more companions lies within 1--75\,\MJup and 5--300\,au derived for the BDB and PPL/LN parts of our parametric model, and for the combined model (BDB + PPL/LN), as a function of spectral type. The frequency estimate for the planet contribution is significantly higher for the higher mass BA stars in our sample than for the lower mass M dwarfs. This is consistent with the idea that for 1--75\,\MJup the range of $q$ probed for higher mass stars (0.0005--0.036 for 2\,\MSun) is lower for the planet mass function ($dN_p \propto q^{-1.3}$) than in lower mass stars ($q$ = 0.0015--0.228 for a 0.3\,\MSun star). Similarly, the frequency of brown dwarf companions is much higher for low-mass M dwarfs than for higher mass BA stars. This reflects the fact that the binary brown dwarf companion mass range probed in our survey (1--75\,\MJup) is at higher $q$ for lower mass stars than for higher mass stars ($dN_{BD} \propto q^{0.25}$). This is qualitatively similar to results from the GPIES survey \citep{Nielsen2019}. 

In this framework, where we combine two components in the model that represent planet-like and star-like formation pathways, it is interesting to study the degeneracy between the two components of the model. In Fig.~\ref{fig:p-models_cornerplot} we show the degeneracy between the relative frequencies derived for the individual parts of the model for the BA, FGK, and M stars in the sample. For the BA and M stars, the observations are well explained by only a single part of the model, either PPL/LN or BDB. This is extremely clear for BA stars, where the likelihood is clustered close to zero for \fbdb and significantly elongated for \fppl. While this is slightly less pronounced for M stars, the fact that only an upper limit can be derived for \fppl is an indication that the contribution of the BDB part to the model is higher than that of the PPL/LN part. We note, however, that this result may be due to the small size of the M sample (20 stars) and the single detection we have in that subset \citep[cf.][]{Lannier2016}. At higher confidence levels, similar probabilities are found in the correlation plot for roughly equal contributions from both parts of the model, and for either part being the predominant underlying population.

The result is more nuanced for the FGK stars, which appear as a transition between BA stars, dominated by planet-like formation over this range of $q$ for companions, and M stars, dominated by binary star-like formation. FGK stars have a comparable contribution from both parts of the model in this range of $q$, but the total together is a lower frequency overall than either the BA or the M dwarf subsample. The contribution of the BDB and PPL/LN parts of the model is clearly inverted with respect to the BA stars: the the BDB part dominates and the PPL/LN part has a small contribution, but the two parts are still required to fit the data best. While it is difficult to determine the formation scenario of individual objects, at the population level, the observed companions around FGK stars are therefore most likely explained by a combination of planet-like and star-like formation scenarios. To fully understand the transition from BA to M, it would be interesting to work in smaller bins of stellar spectral types, but the current data do not allow this because the overall number of detections is small.

Finally, our results appear to show a local minimum in the frequency of substellar companions around FGK stars. Because our sample contains only 20 M stars, we caution that this result should not be overinterpreted. The analysis of the full SHINE sample at the end of the survey will provide much stronger constraints based on a subsample of M stars that is two to three times larger than the current one.

\subsection{Effect of input assumptions}
\label{sec:impact_assumptions}

Our results are based on some important assumptions and parameters that need to be evaluated and tested with additional simulations. For these tests, we used as a reference the FGK sample and converted the detection limits into mass using the BEX-COND-hot models. All the results are summarized in Table~\ref{tab:frequencies}.

\subsubsection{Stellar ages}

Some important parameters are the stellar ages and time of planet formation because giant gaseous exoplanets are expected to slowly cool down and therefore eventually decrease in overall luminosity \citep{Burrows1997,Baraffe2002,Fortney2011,Linder2019}. We assumed that the planet age is equal to that of the star. The full age derivation for the sample is presented in \citetalias{SHINEPaperI}. We study here the variation in PDFs of the frequency of substellar companions when the minimum and maximum ages for all the stars of the sample are compared to the nominal PDF. The corresponding depths of search plots at the different ages are provided in Appendix~\ref{sec:depth_of_search}. In Fig.~\ref{fig:p-model_SpT} the plain lines show the PDFs for the nominal age of the stars, while the shaded regions around each curve show the PDFs computed assuming the minimum and maximum ages for the stars. Generally speaking, the effect of the ages on our results can be considered negligible, with changes of less than 2\% in the peak frequency of companions (or upper limit) between the nominal ages and the minimum or maximum ages, mainly because most of the stars in the sample ($\sim$80\%) are members of young nearby moving groups for which the allowed age range is narrow and well established. For targets that are not part of such moving groups, the age range is generally much larger, but because these targets form only a small fraction of the targets, the overall effect on the PDFs remains low. 

\begin{figure}
  \centering
  \includegraphics[width=0.5\textwidth]{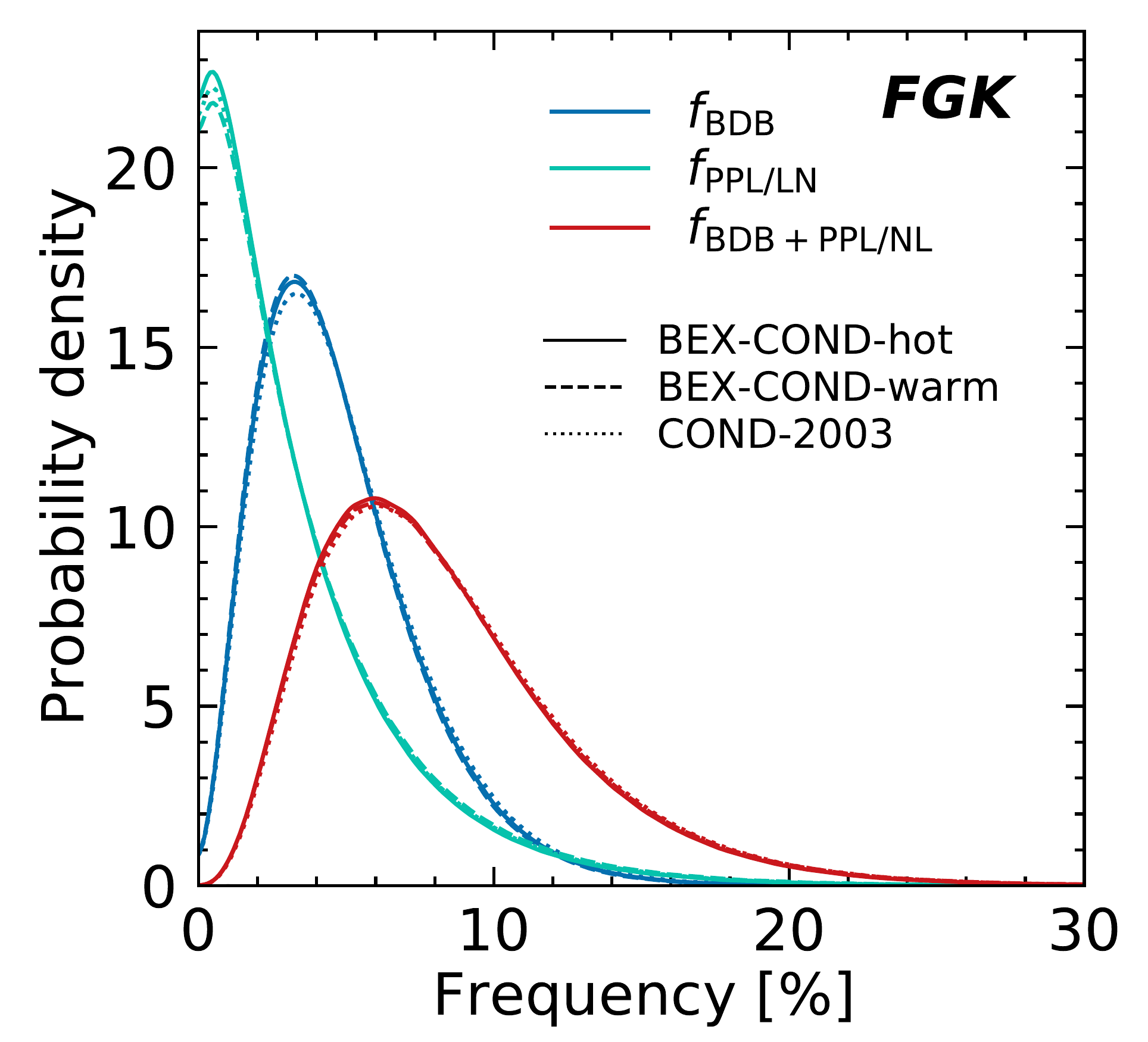}
  \caption{Probability density functions of the frequencies of substellar companions around FGK stars based on the parametric model, computed for companions with masses in the range $M_p = 1$--75\,\MJup and semimajor axes in the range $a = 5$--300\,au, and using the BEX-COND-hot (plain line), BEX-COND-warm (dashed line) or COND-2003 (dotted line) evolutionary tracks for the mass conversion of the detection limits. The median values and 68\% confidence intervals are provided in Table~\ref{tab:frequencies}.}
  \label{fig:p-model_test_evol_model}
\end{figure}

\subsubsection{Initial entropy}

Another major astrophysical assumption is the initial entropy that is used as an input for the evolutionary models. Several studies in the past decade have demonstrated the significant effect of the assumed energy transfer method during the gas accretion phase onto the protoplanetary embryos \citep{Marley2007,Fortney2008,Spiegel2012,Marleau2014}. An extreme outcome is that the entire energy of the infalling gas is transformed into thermal energy by the accretion shock front without radiative losses, so that the entropy remains high; this leads to a bright planet post-formation. If conversely, the entire energy is radiated away at the shock, the entropy of the postshock gas is much lower, which leads to a faint planet at the end of its formation. These two extreme scenarios are generally known as ``hot start'' and ``cold start'' , respectively \citep{Marley2007}, and in reality, a whole range of intermediate initial entropy levels exists that are known as ``warm starts'' \citep{Spiegel2012,Marleau2014}. The most recent advanced models \citep{Marleau2017,Berardo2017,BerardoCumming2017,Cumming2018,Marleau2019shock} and global formation calculations \citep{Mordasini2013,Mordasini2017} clearly suggest that the classical (very) cold starts first proposed by \citet{Marley2007} are unlikely and would occur for a very small fraction of planets that are formed by CA. The initial luminosity of young Jupiters may strongly correlate with the size of their core \citep{Mordasini2013}, with a realistic core mass associated with high entropy even within nominally cold gas accretion \citep{Mordasini2017}.

In order to test the effect of the post-formation entropy and luminosity, we converted our detection limits into mass using three sets of evolutionary tracks: BEX-COND-warm and BEX-COND-hot, which are described in Sect.~\ref{sec:mass_conv}, and COND-2003 \citep{Baraffe2003}. These are the standard tracks that have been used by most studies in the past. They correspond to an even hotter start than BEX-COND-hot. The corresponding depths of search plots for these various models are presented in Appendix~\ref{sec:depth_of_search}, and a comparison of the PDFs of the frequency of substellar companions is presented in Fig.~\ref{fig:p-model_test_evol_model}. The effect of varying the model and/or initial entropy for the evolutionary tracks is negligible. This is perfectly in line with the results presented in \citet{Mordasini2017}, who showed that the luminosity distributions of planets for the BEX-COND-hot and BEX-COND-warm models are extremely similar, and that the BEX-COND-hot tracks are equivalent to those of the COND-2003 tracks after a few~million years (see also Figure~1 of \citealp{Marleau2019}). One should finally note that the COND-2003 models assume initial conditions that are arbitrary and not based on a formation model, in contrast to the BEX models. 

The results are presented only for FGK stars, but the same conclusion applies for BA and M stars. We conclude that our results are independent of the choice of initial entropy in the evolutionary models.

\begin{figure}
    \centering 
    \includegraphics[width=0.5\textwidth]{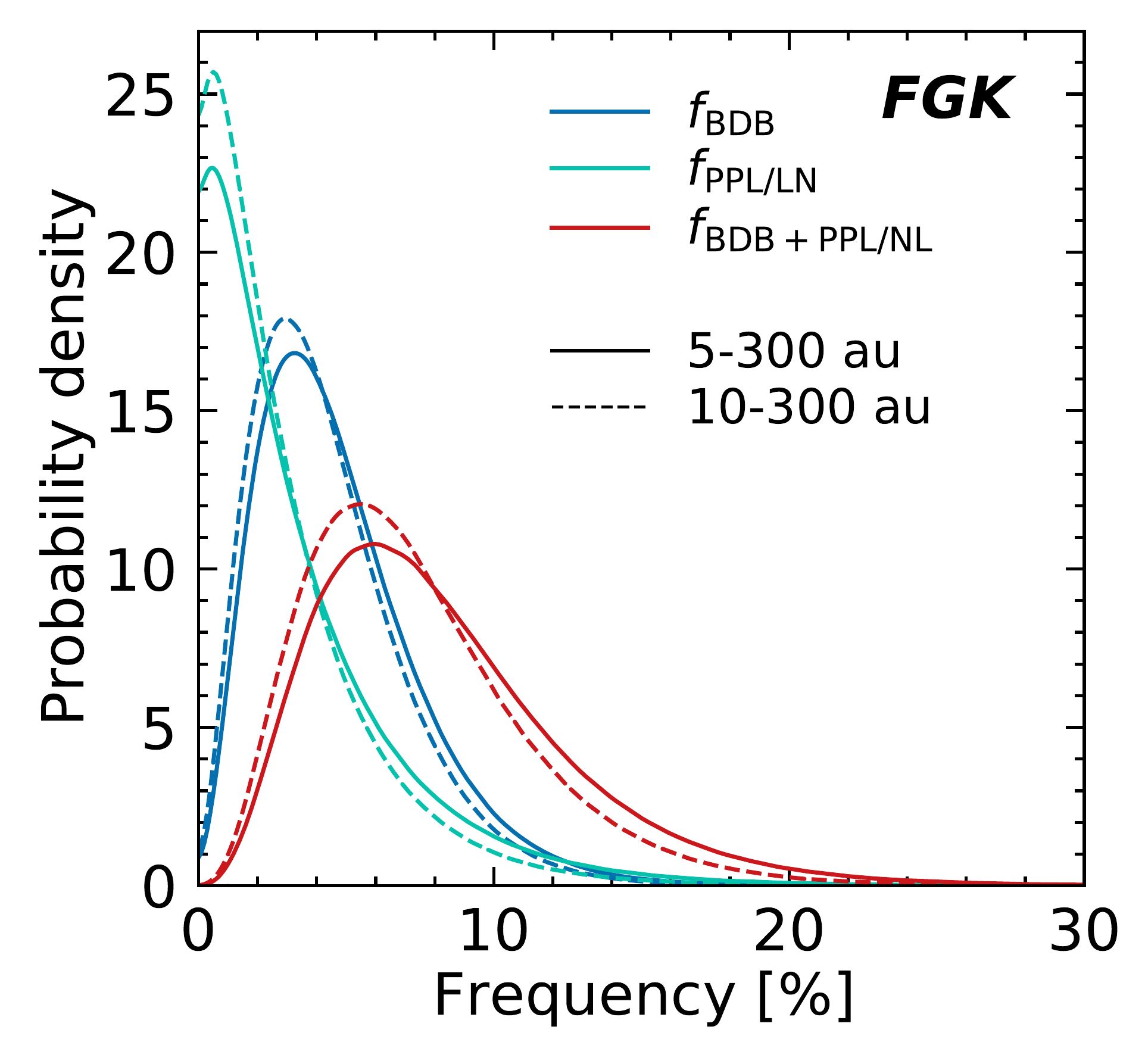}
    \caption{Probability density functions of the frequencies of substellar companions around FGK stars based on the parametric model, computed for companions with semimajor axes in the range $a = 5$--300\,au (plain line) or $a = 10$--300\,au (dashed line), and using the BEX-COND-hot evolutionary tracks for the mass conversion of the detection limits. The median values and 68\% confidence intervals are provided in Table~\ref{tab:frequencies}. The same plots for BA and M stars are shown in Fig.~\ref{fig:p-model_test_sma_BA+M}.}
    \label{fig:p-model_test_sma_FGK}
\end{figure}

\subsubsection{Semimajor axis cutoff}

Finally, another important parameter is the range of semimajor axes that we used to estimate the companion frequency. Our baseline uses a range extending from 5 to 300\,au. As explained in Sect.~\ref{sec:planetary_candidates}, the outer limit of this range is primarily driven by the result of previous direct-imaging surveys, which have already constrained the frequency of companions well down to 300\,au. However, the inner limit of the range is more arbitrary and should be driven by the sensitivity of the survey. A lower limit that is too high (e.g., 50\,au) will provide reliable constraints because in 50--300\,au the sensitivity of SHINE is excellent: more than 100 out of 150 targets are sensitive down to 3\,\MJup (Fig.~\ref{fig:shine_sensitivity}). It will include only about half of the detected companions, however, and will therefore have less statistical significance in a regime that is already dominated by small-number statistics. In contrast, a limit that is too low (e.g., 1\,au) will provide looser constraints because the sensitivity at small separations is lower:~only five targets provide sensitivity at $\sim$1\,au. It is only applicable for masses higher than $\sim$40\,\MJup.

Our final choice of 5\,au is set by the detections around $\beta$\,Pic (companion at $a = 9$\,au) and HIP\,107412 (companion at $a = 6.7$\,au), but in terms of sensitivity to low masses, fewer than five observations are sensitive to the lowest estimated mass for HD\,95086\,b (2\,\MJup) at this separation. Starting at $\sim$10\,au, about a dozen of our observations have a sensitivity down to 2\,\MJup, and at $\sim$30\,au, about a dozen have a sensitivity down to 1\,\MJup. In Fig.~\ref{fig:p-model_test_sma_FGK} we show the variation in PDF of the frequency of substellar companions around FGK stars when the lower limit cutoff in semimajor axis is changed from 5 to 10\,au. In the latter case, where one detection is removed from the analysis (HIP\,107412\,B, around an F5 star), the PDF is slightly modified; the peak frequency for the full parametric model shifts from 5.8\% to 5.5\%. The two peak frequencies remain fully compatible within their respective 68\% confidence intervals: 3.0--10.5\% for the 5--300\,au analysis and 2.8--9.5\% for the 10--300\,au analysis. This result demonstrates that our conclusions are reliable in the selected range of semimajor axes. The plots for BA and M stars are also provided in Appendix~\ref{sec:sma_cutoff_test}.

Although the effect of the semimajor axis cutoff appears to be stronger than that of the stellar ages and the initial entropy, with variations of the peak frequencies up to 5\%, it does not change the conclusions we drew in Sect.~\ref{sec:companion_freq_vs_stellar_mass}. The observed trends remain the same even though detections are removed when a higher cutoff is chosen. This is a strong confirmation of our conclusions.

\subsection{Frequency of substellar companions from formation models}
\label{sec:testing_planet_formation}

\begin{figure}
    \centering 
    \includegraphics[width=0.5\textwidth]{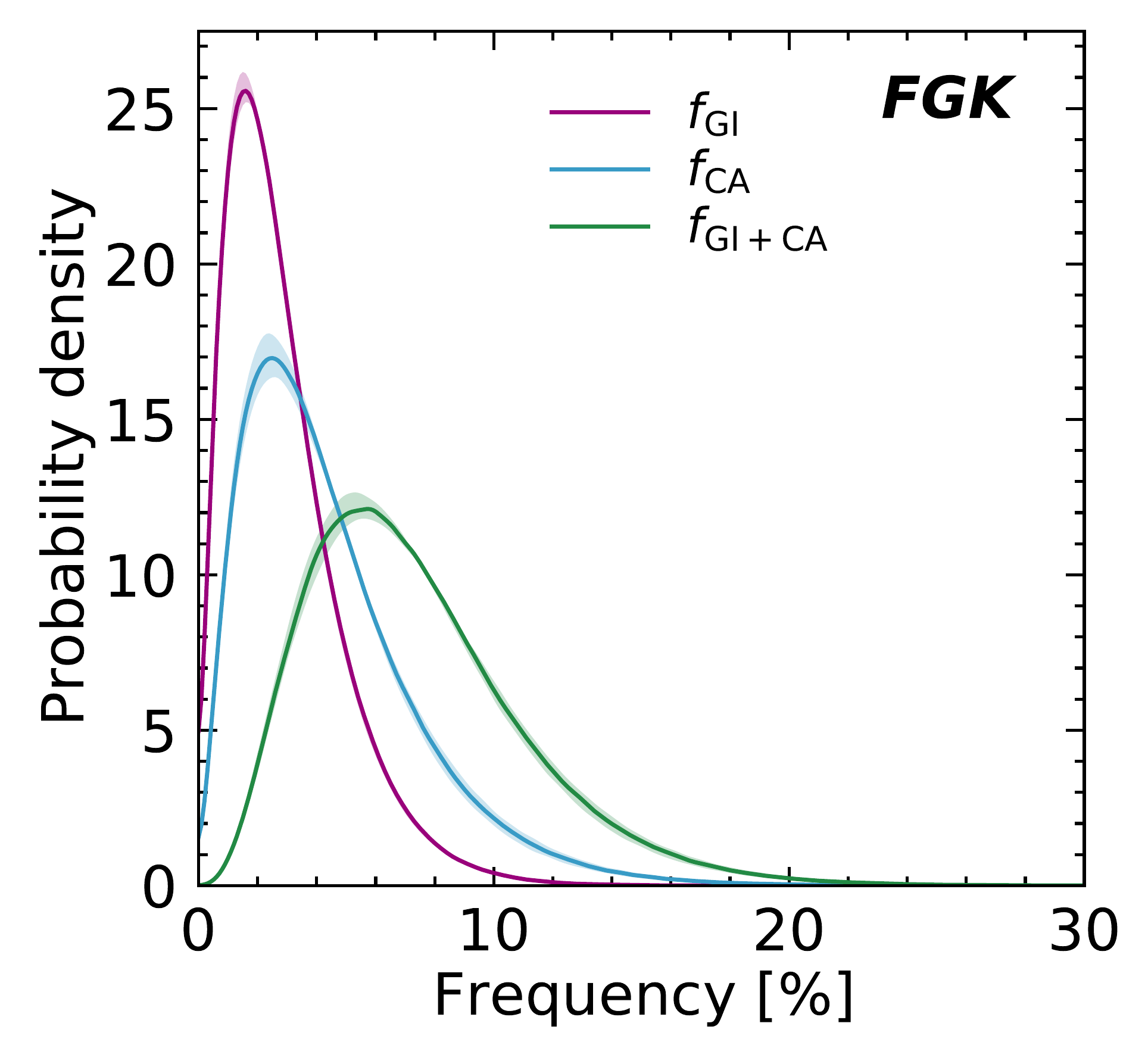}
    \caption{Probability density functions of the frequencies of substellar companions around FGK stars based on the population model, computed for companions with masses in the range $M_p = 1$--75\,\MJup and semimajor axes in the range $a = 5$--300\,au, and using the BEX-COND-hot evolutionary tracks for the mass conversion of the detection limits. Each plot shows the PDFs for the relative frequencies of the two components of the model (\fgi and \fca), and for the total frequency for the full model (\fsmod). The plain lines show the PDFs for the nominal stellar ages, and the shaded envelopes show the variation of these PDFs for the maximum and minimum stellar ages. The median values and 68\% confidence intervals are provided in Table~\ref{tab:frequencies}.}
    \label{fig:s-model}
\end{figure}

\begin{figure}
    \centering
    \includegraphics[width=0.5\textwidth]{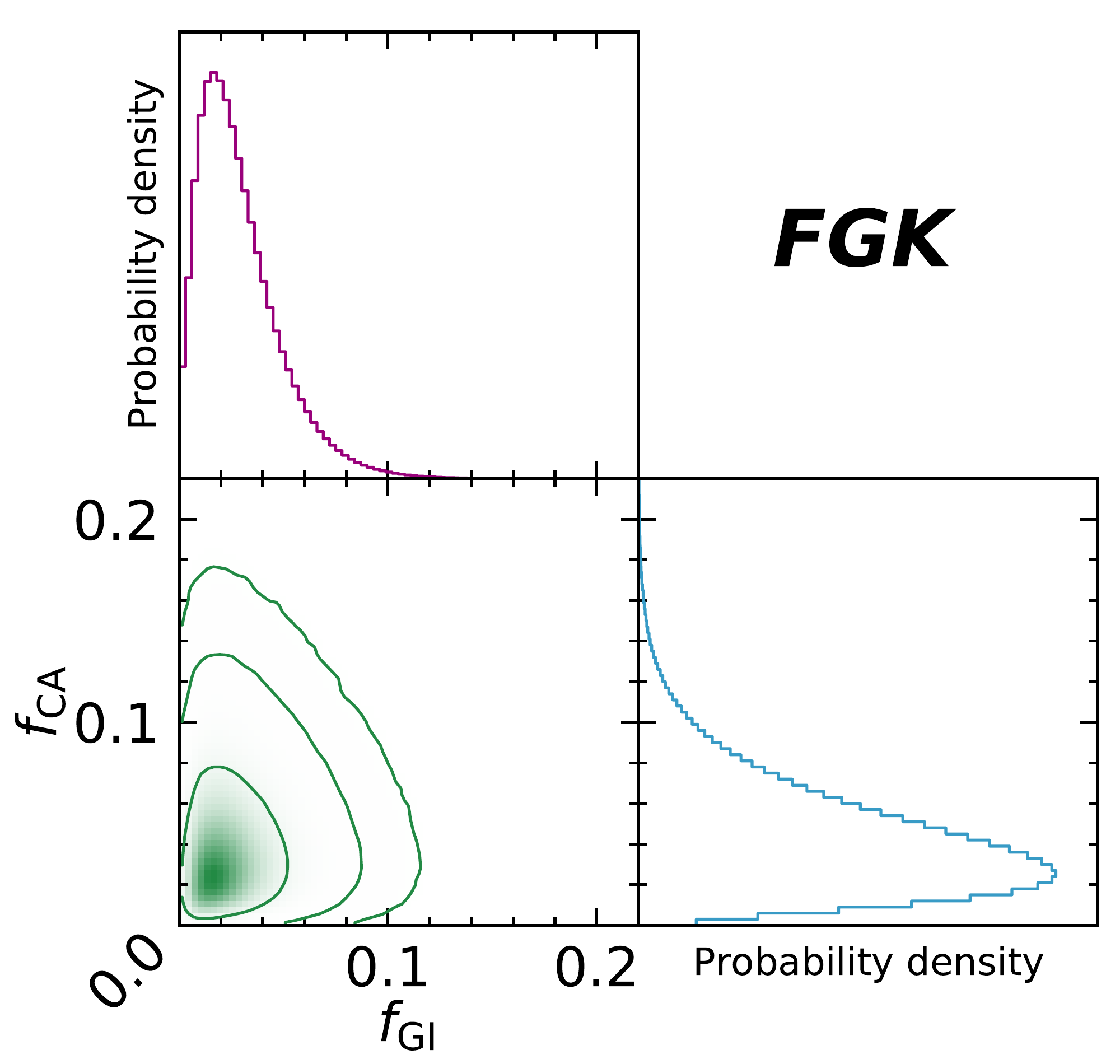}
    \caption{Correlation plots and marginalized PDFs for \fgi and \fca in the population model around FGK stars, computed for companions with masses in the range $M_p = 1$--75\,\MJup and semimajor axes in the range $a = 5$--300\,au, and using the BEX-COND-hot evolutionary tracks at the optimal stellar ages. Contour lines in the correlation plots correspond to regions containing 68\%, 95\%, and 99\% of the posterior, respectively.}
    \label{fig:FGK_s-models_cornerplot}
\end{figure}

\begin{figure}
    \centering
    \includegraphics[width=0.5\textwidth]{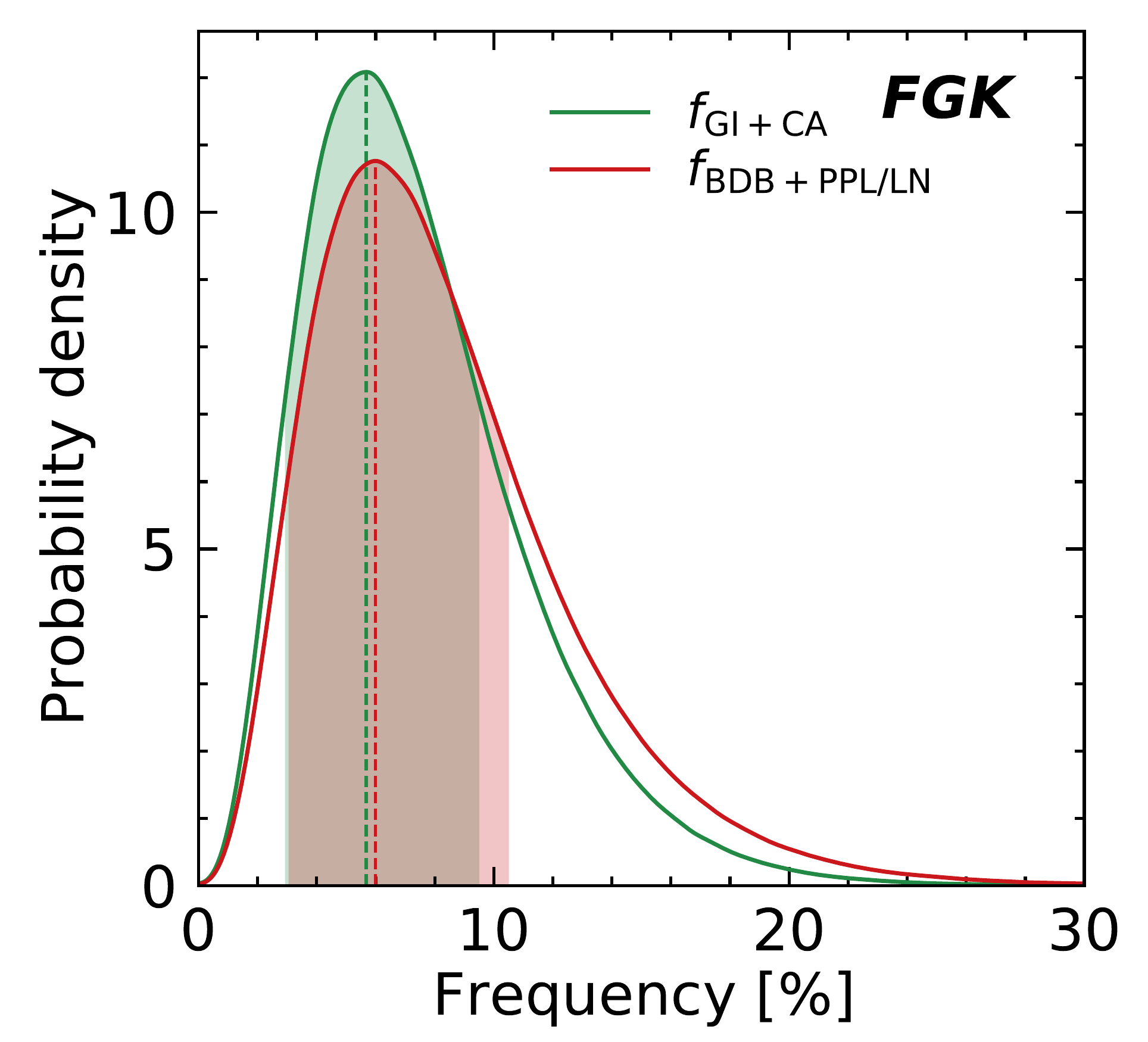}
    \caption{Comparison of the PDF of the frequency of systems with at least one companion for the full parametric and population models, \fpmod and \fsmod , respectively.}
    \label{fig:FGK_ftot_s-vs-p_models}
\end{figure}

In addition to the detection and study of substellar companions, one of the main goals of the SHINE and GPIES campaigns has always been to provide meaningful constraints for planet formation models, or at least to distinguish between different formation scenarios for different categories of objects. In a previous work based on a sample of 200 FGK stars \citep[the NaCo-LP;][]{Vigan2017}, we compared our direct-imaging observations with population synthesis models based on the GI formation scenario. The sensitivity of the NaCo-LP observations was, however, not sufficient to reach a regime of mass and semimajor axis where CA would have been a viable formation scenario (Fig.~\ref{fig:shine_vs_nacolp_sensitivity}), hence the focus on GI at the time. With the improved SHINE sensitivity at small semimajor axes and low masses, combined with new-generation CA models, it is now possible to compare our observations with outcomes of both GI and CA population synthesis models, as is qualitatively illustrated in Fig.~\ref{fig:shine_pop_synth_comparison}.

Here we compare our observations with a combination of CA and GI population synthesis models. From the theoretical point of view, using a superposition of these two formation scenarios is a reasonable assumption. The bottom-up CA formation pathway is very powerful in explaining properties of the exoplanet population within 5--10\,au, but it faces great difficulties in explaining the formation of giant planets farther out than 10--20\,au because the formation timescales that would be involved are prohibitively long \citep{alibert2005,kennedy2008}. Although the pebble accretion process has been proposed as a way to solve the problem \citep{Lambrechts2012,Levison2015}, simulations show that this mechanism does not really form giant planets (several \MJup) at large orbital distances \citep[][their Figs. 4 and 5]{Bitsch2015}. Perhaps the only viable scenario to place CA-formed planets on very wide orbits is to invoke scattering between multiple planets in the systems that originate from different initial embryos \citep{Veras2009,Dawson2013,Marleau2019}. The populations described in Sect.~\ref{sec:CA_pop} include multiple embryos and their subsequent interactions during the early evolution of the protoplanetary disk (20\,Myr). Despite the scattering, very few objects are scattered out to distances of several dozen or hundreds of au where some detections are observed (see Fig.~\ref{fig:shine_pop_synth_comparison}). The top-down GI formation pathway more readily explains the existence of gas giants on wide orbits, but the conditions that can lead to disk fragmentation are still not fully understood \citep{Meru2011,Paardekooper2012,Rice2012,Rice2014,Young2016}. Disk-planet interactions \citep{Kley2012} or planet-planet scattering certainly affect the original semimajor axis distribution of protoplanets and result in exoplanets that cover a wide range of possible masses, sizes, locations, and compositions. These effects are also taken into account in the populations described in Sect.~\ref{sec:GI_pop}.

\begin{table*}
    \caption[]{Comparison of SHINE results based on our parametric model with previously published work}
    \label{tab:frequencies_comparison}
    \centering
    \begin{tabular}{lcccccccccc}
    \hline\hline
                         &            &                          &               &                     & \multicolumn{2}{c}{Published study} & \multicolumn{2}{c}{SHINE} & Compatible\tablefootmark{b} \\
    Study                & Mass       & S.m.a.\tablefootmark{d}  & Distribution  & SpT                 & Median    & 68\% CI\tablefootmark{a} & Median  & 68\% CI & \\
                         & $[\MJup]$  & [au]                     &               &                     & [\%]    & [\%]      &  [\%]          & [\%]            &     \\
    \hline
    \citet{Vigan2012}    & 3--14      & 5--320                   & Uniform       & AF\tablefootmark{c} & 8.7     & 5.9--18.8 & 6.1      & 3.2--11.3 & \cmark \\
                         & 15--75     & 5--320                   & Uniform       & AF\tablefootmark{c} & 2.8     & 2.0--8.9  & 9.0      & 5.6--14.0 & \cmark \\
    \hline
    \citet{Galicher2016} & 4--14      & 25--940                  & Uniform       & BA                  & 1.9     & 0.5--10.1 & 2.7      & 1.7--4.4  & \cmark \\
                         & 4--14      & 25--940                  & Power law     & BA                  & 2.1     & 0.5--11.1 & 2.7      & 1.7--4.4  & \cmark \\
                         & 4--14      & 25--856                  & Uniform       & FGK                 & 1.2     & 0.6--6.6  & 0.5      & 0.3--0.9  & \cmark \\
                         & 4--14      & 25--856                  & Power law     & FGK                 & 1.1     & 0.3--6.1  & 0.5      & 0.3--0.9  & \cmark \\
                         & 1--13      & 10--200                  & Uniform       & M                   &         & $<9.2$    & 1.6      & 0.5--4.5  & \cmark \\
                         & 1--13      & 10--200                  & Power law     & M                   &         & $<11.9$   & 1.6      & 0.5--4.5  & \cmark \\
    \hline
    \citet{Lannier2016}  & 2--14      & 8--400                   & Uniform       & M                   & 2.3     & 1.6--8.1  & 2.0      & 0.1-4.5   & \cmark \\
    \hline
    \citet{Bowler2016}   & 5--13      & 10--100                  & Uniform       & BA                  & 7.7     & 1.7--16.7 & 2.2      & 1.2--4.1  & \xmark \\
                         & 5--13      & 10--100                  & Uniform       & FGK                 &         & $<6.8$    & 0.3      & 0.1--0.8  & \cmark \\
                         & 5--13      & 10--100                  & Uniform       & M                   &         & $<4.2$    & 0.8      & 0.3--1.7  & \cmark \\
    \hline                              
    \citet{Vigan2017}    & 0.5--75    & 20--300                  & Uniform       & FGK                 & 2.1     & 1.5--4.5  & 3.5      & 1.9--6.2  & \cmark \\
    \hline                              
    \citet{Nielsen2019}  & 2--13      & 3--100                   & Uniform       & BA                  & 24      & 14--37    & 8.6      & 4.1--15.9 & \xmark \\
                         & 2--13      & 3--100                   & Power law     & BA                  & 8.9     & 5.3--13.9 & 8.6      & 4.1--15.9 & \cmark \\
                         & 2--13      & 3--100                   & Uniform       & FGK                 &         & $<6.9$    & 0.7      & 0.3--2.9  & \cmark \\
    \hline
    \end{tabular}
    \tablefoot{The ``Mass'' and ``S.m.a.'' columns give the ranges of companion masses and semimajor axes, respectively. 
    \tablefoottext{a}{In contrast to confidence intervals that are expressed at 68\% confidence level, all upper limits are expressed at 95\% confidence level.} \tablefoottext{b}{Compatibility between the results from SHINE and from the previous work. We assumed one asymmetric normal distribution for each measurement, and we tested the null hypothesis that the two measurements are equal with a~5\,\% risk, as described in Appendix~\ref{app:asym_distrib}. A check mark indicates that the null hypothesis is accepted, and a cross mark that it is not.} \tablefoottext{c}{In \citet{Vigan2012} the sample included only 4 F-stars, therefore we consider that the results are only marginally biased compared to SHINE BA results.} \tablefoottext{d}{The SHINE analysis is always truncated at 300\,au.}}
\end{table*}

Because the population synthesis models described in Sect.~\ref{sec:s-model} are currently computed only for solar-mass stars, we based our analysis on the 77 FGK stars from the sample, and the five detections of substellar companions around such stars. Figure~\ref{fig:s-model} shows the PDF of the frequency of substellar companions based on the synthetic population models. The peak \fsmod model is located at 5.7\%, with a 68\% confidence interval of 2.9--9.5\%. Interestingly, the corner plot showing the correlation between the two components of the model in Fig.~\ref{fig:FGK_s-models_cornerplot} looks different from the plot for the parametric model in Fig.~\ref{fig:p-models_cornerplot}. While it is not possible to draw quantitative conclusions here, the CA contribution appears to be greater than the GI part. Both parts of the model are still required to explain the observations, as is visible from the roughly triangular shape of the 2D posterior, but the shape is narrower and more elongated in the direction of CA. The CA part of the model therefore contributes a slightly larger fraction of the full companion population that is required to explain the data. At this stage, it would certainly be necessary to extend our analysis to BA and M stars to confirm if the trends identified in Sect.~\ref{sec:companion_freq_vs_stellar_mass} hold when based on physical population models rather than empirical parametric models. This will be explored in future work using the full SHINE sample, with population synthesis models computed for higher and lower mass stars.

As a cross-check with our parametric model we overplot in Fig.~\ref{fig:FGK_ftot_s-vs-p_models} the PDFs of the full models. With peak frequencies at 5.8\% and 5.7\%, and 68\% confidence intervals of 3.0--10.5\% and 2.9--9.5\% for \fpmod and \fsmod , respectively, the results appear to be fully consistent between the two modeling approaches. This is expected because each model includes a combination of planet-like (i.e., bottom-up) and binary star-like (i.e., top-down) formation components.

\section{Discussion and conclusion}
\label{sec:discussion}

\subsection{Comparison to previous works}
\label{sec:comparison_previous_works}

The SHINE survey is certainly one of the deepest, and it is one of the first to open the low-mass regime at semimajor axes 5--50\,au. This enables us to obtain quantitative statistical constraints in that range. It also offers some overlap with the parameter space that has been explored by previous works, which have placed strong statistical constraints on the population of young giant planets on wide orbits. In this section we assess the compatibility of some of these previous works with the SHINE results. The comparison is not completely straightforward because the considered ranges of mass, semimajor axes, and stellar spectral types vary from one study to the next. In addition, numerous studies have so far either made no physical assumptions on the underlying population of planets, often considering flat distributions in mass and semimajor axes, or used power-law parametric models that sometimes were simply extrapolated from RV surveys and truncated to avoid a continuous growth of the number of planets at wide orbits. The latter approach is contradicted by the latest observational results, which show indications for a turnover in the frequency of companions at the snow line \citep[e.g.,][]{Fernandes2019} and a negative power-law distribution in semimajor axis at wide orbital separations \citep[e.g.,][]{Nielsen2019}. In most cases, our comparison is therefore more a consistency check than a quantitative comparison.

The comparison of our SHINE results with previous studies is presented in Table~\ref{tab:frequencies_comparison}. For this comparison we have recomputed the frequency of systems based on our parametric model while trying to match the other parameters as closely as possible: mass range, semimajor axis range, or stellar spectral type bins. This is not always possible, and some caveats are inevitable. Nonetheless, the estimates from previous surveys are generally compatible with the new values derived for SHINE for the different stellar spectral types. To estimate the compatibility, we tested the null hypothesis that the two measurements are equal with a 5\% risk, as described in Appendix~\ref{app:asym_distrib}. In most cases we find that measurements are compatible with each other within this 5\% risk. In only two cases are the values not compatible with SHINE for BA stars: the GPIES analysis from \citet{Nielsen2019}, which we discuss in more detail in the next paragraph, and the meta-analysis from \citet{Bowler2016}. For the latter, it might be argued that their sensitivity in the quoted ranges of mass and semimajor axes around BA stars is marginal at best, which has a strong effect on the frequency that they derive. Because of the sensitivity of SHINE at small semimajor axes, our results can be considered far more robust.

In contrast to \citet{Bowler2016}, the sensitivity of the GPIES survey \citep{Nielsen2019} below 100\,au is comparable to that of SHINE. Their frequency estimation for BA stars using a uniform distribution as a prior clearly contradicts our own estimation, but uniform distributions as a prior are not physically realistic. It is more reasonable to compare our derived value with the value they derived using their power-law parametric model. Similarly to our parametric model, their model aims at realistically modeling the underlying population of substellar companions with just a few parameters. They modeled the planet population with a distribution of the form
\begin{equation*}
    \frac{dN^2}{dm\,da} = f C_1 m^\alpha a^\beta,
\end{equation*}
\noindent where $f$ is the frequency of planetary systems, and $m$ and $a$ the mass and semimajor axis of planets, respectively. Based on the GPIES data acquired for stars in the range 1.5--5\,\MSun, which corresponds to our BA sample, their best fit is obtained for $f = 8.9_{-3.6}^{+5.0}\%$, $\alpha = -2.37$ and $\beta = -1.99$ for planets in the range 3--100\,au and 2--13\,\MJup. Within the error bars, their value of $f$ is indeed almost exactly equal to the value of $\fpmod = 8.6_{-4.5}^{+7.3}\%$ that we derive based on the SHINE data using our own parametric model. We therefore consider that the two surveys agree excellently for early-type stars within the assumptions of our respective parametric models. The agreement can partly be explained by the overlap in terms of targets (67 targets) and of detections in the two surveys \citep{SHINEPaperI}, but this partial overlap cannot by itself fully explain the agreement. We instead consider the agreement as confirmation that the frequency of substellar systems around young BA stars is indeed around 8\%, regardless of the sample that is considered.

\subsection{Implications for formation theory}
\label{sec:implications_formation_theory}

Based on our observation of 150 stars that are part of the full SHINE sample, we can already conclude that gas giant companions are more commonly found around higher mass stars, and brown dwarf binary companions are observed more frequently around lower mass stars. The same conclusion is reached by \citet{Nielsen2019} for the GPIES survey. This can be interpreted in the context of our parametric model, which treats the observed companion frequency in terms of mass ratio distributions that are the same for all stellar masses. Within the range of companion masses to which we are sensitive (1--75\,\MJup), the planet part of the companion mass function for higher mass stars samples planets at smaller $q$, predicting higher frequencies. Conversely, at larger $q$ for lower mass primaries (e.g., >0.1 for 30\,\MJup companions to 0.3\,\MSun stars), the brown dwarf companion part of the parametric model is expected to dominate, as observed. This suggests a formation framework that is independent of stellar mass, but for which different pathways dominate as a function of $q$.

We recall that the deuterium burning limit at 13\,\MJup likely plays no specific role in the physical evolution of young very low-mass companions as both GI and CA can produce objects that are above and below that threshold \citep{Molliere2012,Chabrier2014}. It is likely that binary star-like formation and planet-like formation pathways both contribute to the substellar mass distribution, without strict physical limits between the two. In this framework, brown dwarfs constitute the low-end tail of the stellar companion mass ratio distribution extended to the very low-mass regime, as was initially suggested by \citet{Metchev2009}. This universality of the companion mass distribution has since then received some strong observational support \citep{Reggiani2011,Reggiani2013}. On the other hand, more planet-like formation models such as GI or CA (with or without pebble accretion) clearly suggest that massive companions at wide orbits can be formed, constituting the high-end tail of their mass distribution \citep{Forgan2013,Forgan2015,Mordasini2012,Lambrechts2012}.

Our parametric analysis predicts only the relative probability that a given mass ratio $q$ may have formed from binary star-like or planet-like processes without discontinuities. Ultimately, the characterization of planetary-mass companions might reveal their formation pathway, for instance, if the comparison of volatile abundances shows differences relative to the star, perhaps indicating a disk formation process \citep{Oberg2011,Piso2015b,Mordasini2016}. The models also suggest that the overall efficiency of gas giant planet formation and low-$q$ binary formation does not depend strongly on stellar mass, although the estimates of companion frequency strongly depend on the range of $q$ that is considered in the analysis, which in turn depends on the host star mass: while the stellar mass scales the planet mass function in our model (self-similar in $q$), the normalization constants are roughly consistent throughout the range of stellar masses we studied. In addition, we note that our original sample selection for SHINE \citepalias{SHINEPaperI} has purposely removed all known visual binaries, which means that our observations are far from complete in some ranges of $q$. This caveat might be circumvented in the future using results from the ESA/Gaia \citep{Gaia2016,Gaia2018} survey to correct the completeness in $q$.

A comparison of our results to specific realizations of GI population synthesis suggests that this is probably not the dominant channel of planet formation around FGK stars. This conclusion has been reached by \citet{Vigan2017}, and our new analysis based on deeper SHINE data only strengthens this result. Perhaps more importantly, the SHINE results show that GI may not be the dominant formation scenario even for the most massive companions at large distances (5--10\,\MJup companions at $\sim$50\,au). However, additional input physics and more sophisticated simulations might alter the comparison in the future. Furthermore, some variants of GI such as early fragmentation within infalling disks that are only partially supported by rotation \citep{Stamatellos2008,Stamatellos2011,Forgan2012} or rapid inward migration might contribute significantly to low-$q$ binary populations. However, overall, the CA population synthesis models appear to be more promising. Alternative initial conditions such as disk lifetimes that depend on host star mass, and other relevant processes such as core formation based on pebble accretion \citep{Alibert2017,Ndugu2018} will illuminate the robustness of these predictions.

Like many studies in the past, our results are affected by the choice of evolutionary tracks for the conversion of the detection limits in the luminosity space into mass limits in a physical space. Significant progress is currently made in this field, which provides alternatives to the canonical evolutionary tracks \citep[e.g.,][]{Burrows1997,Baraffe2003} that usually give good estimates at old ages and high masses, but require further validation at young ages or low masses \citep[e.g.,][]{Konopacky2010,Dupuy2017}. The current and next-generation population synthesis models are gaining the ability to quantitatively predict the post-formation luminosity of young planets, which in turn drives the early evolution of the planet \citep{Mordasini2012,Mordasini2013,Emsenhuber2020A,Emsenhuber2020B}. The predictions of these models provide a robust view of the range of post-formation luminosities that planets can take \citep{Mordasini2017}. Much work still remains to be done to understand and accurately model the physics during the accretion phase, including the accretion geometry onto the planet \citep[e.g.,][]{Gressel2013,Szulagyi2017b,Batygin2018,Bethune2019,Schulik2019} or the radiative properties of the accretion shock \citep{Marleau2017,Marleau2019shock}, but the discovery and study of accreting protoplanets such as PDS\,70\,b and~c \citep{Keppler2018,Haffert2019,Thanathibodee2019,Aoyama2019,Christiaens2019b,Hashimoto2020} will certainly help. More generally, the continued direct detection of spatially resolved substellar companions that already have dynamical mass estimates either through RV and/or astrometry (e.g., \citealt{Crepp2012,Bowler2018,Peretti2019,Dupuy2019,Brandt2019}; ESA/Gaia~DR4), will help calibrate the mass-luminosity relationships even more precisely and will give greater confidence in our survey results.

Finally, the combination of results from direct imaging, RV, transit, microlensing, astrometry, and timing variations will provide a more complete picture of exoplanet demographics as a function of stellar mass. Trends in the companion mass ratio distribution as a function of orbital radius might reveal important discontinuities. Future work should also consider more carefully how planet populations depend on stellar multiplicity, and how global planet architectures affect our statistical results, such as the ratio of planet masses and orbital radii in multiplanet systems.

\subsection{Summary and perspectives}
\label{sec:summary}

We have presented the first statistical analysis of the properties of the population of substellar companions at wide orbital separation based on a subset of 150 stars from the SHINE survey. The full details of the sample, the observations, and the data analysis are presented in two companion papers \citepalias{SHINEPaperI,SHINEPaperII}. Although the size of the current sample is only a fraction of the full SHINE sample, we can already derive some important conclusions.
Based on our parametric model presented in Sect.~\ref{sec:p-model}, we draw the conclusions listed below.

\begin{enumerate}
    \item \label{item1} We determine the frequency of systems in which at least one companion has a mass in the range $M_p = 1$--75\,\MJup and a semimajor axis in the range $a = 5$--300\,au, \fpmod, to be $23.0_{-9.7}^{+13.5}\%$, $5.8_{-2.8}^{+4.7}\%$, and $12.6_{-7.1}^{+12.9}\%$ for BA, FGK, and M stars, respectively. These values were derived using a conversion of the detection limits into mass using the BEX-COND-hot evolutionary tracks and the nominal age for all the stars in the sample. These values are average estimates over the stated ranges, but the sensitivity at the lowest masses and shorter separations is limited. This means that the uncertainties increase significantly when we focus on low masses and short separations.
    \item The frequency of substellar companions is significantly higher around BA stars than around FGK and M stars, by factors of approximately 4 and 2, respectively. The apparent local minimum in frequency around FGK stars is suggestive and will be examined in detail with our full survey sample in the future.
    \item \label{item3} Our two-component parametric model shows a clear inversion between BA and M stars. While in the case of BA stars the likelihood of the \fppl part of the model dominates \fbdb, this former part only becomes an upper limit for M stars. This can be translated physically into a predominance of the planet-like formation pathway for companions detected around early-type stars over the binary star-like formation pathway for the mass ratio range that is sampled as a function of host star type.
    \item The FGK stars are a transition range of spectral types, where observations are better explained by a combination of the two parts of the model, although the contribution of the PPL part is small. While it would be extremely interesting to perform an analysis in smaller bins of spectral type, the current data do not allow this because we have only a few detections.
    \item The input assumptions such as the stellar ages, cutoff in semimajor axis, or evolutionary tracks, have little effect on the frequencies of planetary systems derived from the observations.
    \item We find that the frequency of systems in which at least one companion has a mass in the range $M_p = 2$--13\,\MJup and a semimajor axis in the range $a = 3$--100\,au around BA stars is $8.6_{-4.5}^{+7.3}\%$. This value is fully compatible with the value derived by GPIES \citep{Nielsen2019} in a survey with similar sensitivity as SHINE, but with a slightly different sample \citep{SHINEPaperI}. This confirms the reliability of the estimation.
\end{enumerate}

Based on the population model presented in Sect.~\ref{sec:s-model}, we can also draw the following conclusions for FGK stars:

\begin{enumerate}[resume]
    \item We determine the frequency of systems in which at least one companion has a mass in the range $M_p = 1$--75\,\MJup and a semimajor axis in the range $a = 5$--300\,au, \fsmod, to be $5.7_{-2.8}^{+3.8}\%$ for FGK stars. This value was derived using a conversion of the detection limits into mass using the BEX-COND-hot evolutionary tracks and the nominal age for all the stars in the sample, but again this result is not very sensitive to the input parameters. The same words of caution as in item~\ref{item1} apply here.
    \item Qualitatively, the contribution of the CA part of the model appears to be larger than the GI part, which means that CA contributes a higher fraction of the full companion population required to explain the data. Simulations extended to BA and M stars are required, however, to determine whether the general trend highlighted in item~\ref{item3} above holds when we consider our population model instead of the parametric model.
    \item The values of \fsmod and \fpmod  perfectly agree. Although the underlying model is different, the overall frequency values required to explain the observations are almost the same with the two approaches.
\end{enumerate}

The SHINE survey is due to be completed in 2020, but will certainly extend over a few more years to become complete in terms of follow-up for all candidates within at least a 300\,au and possibly even farther away. The final sample will include over 600 stars, which will make SHINE the largest high-contrast imaging survey to date, covering from B to M stars in the solar neighborhood. Beyond the reanalysis of the complete SHINE data with advanced post-processing techniques \citep{Cantalloube2015,Ruffio2017,Flasseur2018}, which will hopefully provide improved detection limits at small separations, the full power of the survey will be in the statistical conclusions based on a sample that is almost four times larger than the sample we used here. Some of the prospects for future statistical inference work include an extension of our analysis based on population synthesis models to BA and M stars, the analysis of subsamples such as stars with disks or know infrared excess \citep[e.g.,][]{Wahhaj2013} or stars that belong to nearby young moving groups \citep[e.g.,][]{Biller2013}, or an extension of the completeness of the sample in $q$ space using the ESA/Gaia~DR4 results.

\begin{acknowledgements}
    SPHERE is an instrument designed and built by a consortium consisting of IPAG (Grenoble, France), MPIA (Heidelberg, Germany), LAM (Marseille, France), LESIA (Paris, France), Laboratoire Lagrange (Nice, France), INAF - Osservatorio di Padova (Italy), Observatoire de Gen\`eve (Switzerland), ETH Z\"urich (Switzerland), NOVA (Netherlands), ONERA (France) and ASTRON (Netherlands) in collaboration with ESO. SPHERE was funded by ESO, with additional contributions from CNRS (France), MPIA (Germany), INAF (Italy), FINES (Switzerland) and NOVA (Netherlands). SPHERE also received funding from the European Commission Sixth and Seventh Framework Programmes as part of the Optical Infrared Coordination Network for Astronomy (OPTICON) under grant number RII3-Ct-2004-001566 for FP6 (2004--2008), grant number 226604 for FP7 (2009--2012) and grant number 312430 for FP7 (2013--2016). 
    
    This work has made use of the SPHERE Data Centre, jointly operated by OSUG/IPAG (Grenoble), PYTHEAS/LAM/CeSAM (Marseille), OCA/Lagrange (Nice), Observatoire de Paris/LESIA (Paris), and Observatoire de Lyon/CRAL, and supported by a grant from Labex OSUG@2020 (Investissements d’avenir -- ANR10 LABX56).

    This research has made use of the Direct Imaging Virtual Archive (DIVA), operated at CeSAM/LAM, Marseille, France.

    AV acknowledges funding from the European Research Council (ERC) under the European Union's Horizon 2020 research and innovation programme (grant agreement No.~757561).

    G-DM acknowledges the support of the DFG priority program SPP~1992 ``Exploring the Diversity of Extrasolar Planets'' (KU~2849/7-1).
    
    CM, AE, and G-DM acknowledge the support from the Swiss National Science Foundation under grant BSSGI0$\_$155816 ``PlanetsInTime''. Parts of this work have been carried out within the framework of the NCCR PlanetS supported by the Swiss National Science Foundation.
    
    A-ML acknowledges funding from \emph{Agence Nationale de la Recherche} (France) under contract number ANR-14-CE33-0018.
    
    TH and RAT acknowledge support from the European Research Council (ERC) under the European Union's Horizon 2020 research and innovation programme (grant agreement No.~832428).
    
    FMe acknowledges funding from \emph{Agence Nationale de la Recherche} (France) under contract number ANR-16-CE31-0013.
    
    CP acknowledge financial support from Fondecyt (grant~3190691) and financial support from the ICM (Iniciativa Cient\'ifica Milenio) via the N\'ucleo Milenio  de  Formaci\'on Planetaria grant, from the Universidad de Valpara\'iso.
\end{acknowledgements}

\bibliographystyle{aa}
\bibliography{paper}

\begin{thebibliography}{182}
\expandafter\ifx\csname natexlab\endcsname\relax\def\natexlab#1{#1}\fi

\bibitem[{{Alibert}(2017)}]{Alibert2017}
{Alibert}, Y. 2017, \aap, 606, A69

\bibitem[{{Alibert} {et~al.}(2004){Alibert}, {Mordasini}, \&
  {Benz}}]{Alibert2004}
{Alibert}, Y., {Mordasini}, C., \& {Benz}, W. 2004, \aap, 417, L25

\bibitem[{{Alibert} {et~al.}(2005){Alibert}, {Mordasini}, {Benz}, \&
  {Winisdoerffer}}]{alibert2005}
{Alibert}, Y., {Mordasini}, C., {Benz}, W., \& {Winisdoerffer}, C. 2005, \aap,
  434, 343

\bibitem[{{Allard} {et~al.}(2001){Allard}, {Hauschildt}, {Alexander},
  {Tamanai}, \& {Schweitzer}}]{Allard2001}
{Allard}, F., {Hauschildt}, P.~H., {Alexander}, D.~R., {Tamanai}, A., \&
  {Schweitzer}, A. 2001, \apj, 556, 357

\bibitem[{{Andrews} \& {Williams}(2005)}]{Andrews2005}
{Andrews}, S.~M. \& {Williams}, J.~P. 2005, \apj, 631, 1134

\bibitem[{{Andrews} \& {Williams}(2007{\natexlab{a}})}]{Andrews2007b}
{Andrews}, S.~M. \& {Williams}, J.~P. 2007{\natexlab{a}}, \apj, 671, 1800

\bibitem[{{Andrews} \& {Williams}(2007{\natexlab{b}})}]{Andrews2007a}
{Andrews}, S.~M. \& {Williams}, J.~P. 2007{\natexlab{b}}, \apj, 659, 705

\bibitem[{{Andrews} {et~al.}(2010){Andrews}, {Wilner}, {Hughes}, {Qi}, \&
  {Dullemond}}]{Andrews2010}
{Andrews}, S.~M., {Wilner}, D.~J., {Hughes}, A.~M., {Qi}, C., \& {Dullemond},
  C.~P. 2010, \apj, 723, 1241

\bibitem[{{Ansdell} {et~al.}(2018){Ansdell}, {Williams}, {Trapman}, {van
  Terwisga}, {Facchini}, {Manara}, {van der Marel}, {Miotello}, {Tazzari},
  {Hogerheijde}, {Guidi}, {Testi}, \& {van Dishoeck}}]{Ansdell2018}
{Ansdell}, M., {Williams}, J.~P., {Trapman}, L., {et~al.} 2018, \apj, 859, 21

\bibitem[{{Aoyama} \& {Ikoma}(2019)}]{Aoyama2019}
{Aoyama}, Y. \& {Ikoma}, M. 2019, \apjl, 885, L29

\bibitem[{{Balbus} \& {Hawley}(1991)}]{Balbus91}
{Balbus}, S.~A. \& {Hawley}, J.~F. 1991, \apj, 376, 214

\bibitem[{{Baraffe} {et~al.}(2002){Baraffe}, {Chabrier}, {Allard}, \&
  {Hauschildt}}]{Baraffe2002}
{Baraffe}, I., {Chabrier}, G., {Allard}, F., \& {Hauschildt}, P.~H. 2002, \aap,
  382, 563

\bibitem[{{Baraffe} {et~al.}(2003){Baraffe}, {Chabrier}, {Barman}, {Allard}, \&
  {Hauschildt}}]{Baraffe2003}
{Baraffe}, I., {Chabrier}, G., {Barman}, T.~S., {Allard}, F., \& {Hauschildt},
  P.~H. 2003, \aap, 402, 701

\bibitem[{{Baron} {et~al.}(2019){Baron}, {Lafreni{\`e}re}, {Artigau},
  {Gagn{\'e}}, {Rameau}, {Delorme}, \& {Naud}}]{Baron2019}
{Baron}, F., {Lafreni{\`e}re}, D., {Artigau}, {\'E}., {et~al.} 2019, \aj, 158,
  187

\bibitem[{{Batygin}(2018)}]{Batygin2018}
{Batygin}, K. 2018, \aj, 155, 178

\bibitem[{{Benz} {et~al.}(2014){Benz}, {Ida}, {Alibert}, {Lin}, \&
  {Mordasini}}]{Benz2014}
{Benz}, W., {Ida}, S., {Alibert}, Y., {Lin}, D., \& {Mordasini}, C. 2014, in
  Protostars and Planets VI, ed. H.~{Beuther}, R.~S. {Klessen}, C.~P.
  {Dullemond}, \& T.~{Henning}, 691

\bibitem[{{Berardo} \& {Cumming}(2017)}]{BerardoCumming2017}
{Berardo}, D. \& {Cumming}, A. 2017, \apjl, 846, L17

\bibitem[{{Berardo} {et~al.}(2017){Berardo}, {Cumming}, \&
  {Marleau}}]{Berardo2017}
{Berardo}, D., {Cumming}, A., \& {Marleau}, G.-D. 2017, \apj, 834, 149

\bibitem[{{B{\'e}thune}(2019)}]{Bethune2019}
{B{\'e}thune}, W. 2019, \mnras, 490, 3144

\bibitem[{{Beuzit} {et~al.}(2019){Beuzit}, {Vigan}, {Mouillet}, {Dohlen},
  {Gratton}, {Boccaletti}, {Sauvage}, {Schmid}, {Langlois}, {Petit},
  {Baruffolo}, {Feldt}, {Milli}, {Wahhaj}, {Abe}, {Anselmi}, {Antichi},
  {Barette}, {Baudrand}, {Baudoz}, {Bazzon}, {Bernardi}, {Blanchard}, {Brast},
  {Bruno}, {Buey}, {Carbillet}, {Carle}, {Cascone}, {Chapron}, {Charton},
  {Chauvin}, {Claudi}, {Costille}, {De Caprio}, {de Boer}, {Delboulb{\'e}},
  {Desidera}, {Dominik}, {Downing}, {Dupuis}, {Fabron}, {Fantinel}, {Farisato},
  {Feautrier}, {Fedrigo}, {Fusco}, {Gigan}, {Ginski}, {Girard}, {Giro},
  {Gisler}, {Gluck}, {Gry}, {Henning}, {Hubin}, {Hugot}, {Incorvaia}, {Jaquet},
  {Kasper}, {Lagadec}, {Lagrange}, {Le Coroller}, {Le Mignant}, {Le Ruyet},
  {Lessio}, {Lizon}, {Llored}, {Lundin}, {Madec}, {Magnard}, {Marteaud},
  {Martinez}, {Maurel}, {M{\'e}nard}, {Mesa}, {M{\"o}ller-Nilsson}, {Moulin},
  {Moutou}, {Orign{\'e}}, {Parisot}, {Pavlov}, {Perret}, {Pragt}, {Puget},
  {Rabou}, {Ramos}, {Reess}, {Rigal}, {Rochat}, {Roelfsema}, {Rousset}, {Roux},
  {Saisse}, {Salasnich}, {Santambrogio}, {Scuderi}, {Segransan}, {Sevin},
  {Siebenmorgen}, {Soenke}, {Stadler}, {Suarez}, {Tiph{\`e}ne}, {Turatto},
  {Udry}, {Vakili}, {Waters}, {Weber}, {Wildi}, {Zins}, \&
  {Zurlo}}]{Beuzit2019}
{Beuzit}, J.~L., {Vigan}, A., {Mouillet}, D., {et~al.} 2019, \aap, 631, A155

\bibitem[{{Biller} {et~al.}(2013){Biller}, {Liu}, {Wahhaj}, {Nielsen},
  {Hayward}, {Males}, {Skemer}, {Close}, {Chun}, {Ftaclas}, {Clarke}, {Thatte},
  {Shkolnik}, {Reid}, {Hartung}, {Boss}, {Lin}, {Alencar}, {de Gouveia Dal
  Pino}, {Gregorio-Hetem}, \& {Toomey}}]{Biller2013}
{Biller}, B.~A., {Liu}, M.~C., {Wahhaj}, Z., {et~al.} 2013, \apj, 777, 160

\bibitem[{{Bitsch} {et~al.}(2015){Bitsch}, {Lambrechts}, \&
  {Johansen}}]{Bitsch2015}
{Bitsch}, B., {Lambrechts}, M., \& {Johansen}, A. 2015, \aap, 582, A112

\bibitem[{{Blunt} {et~al.}(2017){Blunt}, {Nielsen}, {De Rosa}, {Konopacky},
  {Ryan}, {Wang}, {Pueyo}, {Rameau}, {Marois}, {Marchis}, {Macintosh},
  {Graham}, {Duch{\^e}ne}, \& {Schneider}}]{Blunt2017}
{Blunt}, S., {Nielsen}, E.~L., {De Rosa}, R.~J., {et~al.} 2017, \aj, 153, 229

\bibitem[{{Bodenheimer} {et~al.}(2000){Bodenheimer}, {Hubickyj}, \&
  {Lissauer}}]{Bodenheimer2000}
{Bodenheimer}, P., {Hubickyj}, O., \& {Lissauer}, J.~J. 2000, \icarus, 143, 2

\bibitem[{{Bodenheimer} \& {Pollack}(1986)}]{BodenheimerPollack1986}
{Bodenheimer}, P. \& {Pollack}, J.~B. 1986, \icarus, 67, 391

\bibitem[{{Bonnefoy} {et~al.}(2010){Bonnefoy}, {Chauvin}, {Rojo}, {Allard},
  {Lagrange}, {Homeier}, {Dumas}, \& {Beuzit}}]{Bonnefoy2010}
{Bonnefoy}, M., {Chauvin}, G., {Rojo}, P., {et~al.} 2010, \aap, 512, A52

\bibitem[{{Bonnefoy} {et~al.}(2014{\natexlab{a}}){Bonnefoy}, {Currie},
  {Marleau}, {Schlieder}, {Wisniewski}, {Carson}, {Covey}, {Henning}, {Biller},
  {Hinz}, {Klahr}, {Marsh Boyer}, {Zimmerman}, {Janson}, {McElwain},
  {Mordasini}, {Skemer}, {Bailey}, {Defr{\`e}re}, {Thalmann}, {Skrutskie},
  {Allard}, {Homeier}, {Tamura}, {Feldt}, {Cumming}, {Grady}, {Brandner},
  {Helling}, {Witte}, {Hauschildt}, {Kandori}, {Kuzuhara}, {Fukagawa}, {Kwon},
  {Kudo}, {Hashimoto}, {Kusakabe}, {Abe}, {Brandt}, {Egner}, {Guyon}, {Hayano},
  {Hayashi}, {Hayashi}, {Hodapp}, {Ishii}, {Iye}, {Knapp}, {Matsuo}, {Mede},
  {Miyama}, {Morino}, {Moro-Martin}, {Nishimura}, {Pyo}, {Serabyn}, {Suenaga},
  {Suto}, {Suzuki}, {Takahashi}, {Takami}, {Takato}, {Terada}, {Tomono},
  {Turner}, {Watanabe}, {Yamada}, {Takami}, \& {Usuda}}]{Bonnefoy2014kap}
{Bonnefoy}, M., {Currie}, T., {Marleau}, G.-D., {et~al.} 2014{\natexlab{a}},
  \aap, 562, A111

\bibitem[{{Bonnefoy} {et~al.}(2014{\natexlab{b}}){Bonnefoy}, {Marleau},
  {Galicher}, {Beust}, {Lagrange}, {Baudino}, {Chauvin}, {Borgniet}, {Meunier},
  {Rameau}, {Boccaletti}, {Cumming}, {Helling}, {Homeier}, {Allard}, \&
  {Delorme}}]{Bonnefoy2014bet}
{Bonnefoy}, M., {Marleau}, G.-D., {Galicher}, R., {et~al.} 2014{\natexlab{b}},
  \aap, 567, L9

\bibitem[{{Boss}(1998)}]{Boss1998}
{Boss}, A.~P. 1998, \apj, 503, 923

\bibitem[{{Bowler}(2016)}]{Bowler2016}
{Bowler}, B.~P. 2016, \pasp, 128, 102001

\bibitem[{{Bowler} {et~al.}(2020){Bowler}, {Blunt}, \& {Nielsen}}]{Bowler2020}
{Bowler}, B.~P., {Blunt}, S.~C., \& {Nielsen}, E.~L. 2020, \aj, 159, 63

\bibitem[{{Bowler} {et~al.}(2018){Bowler}, {Dupuy}, {Endl}, {Cochran},
  {MacQueen}, {Fulton}, {Petigura}, {Howard}, {Hirsch}, {Kratter}, {Crepp},
  {Biller}, {Johnson}, \& {Wittenmyer}}]{Bowler2018}
{Bowler}, B.~P., {Dupuy}, T.~J., {Endl}, M., {et~al.} 2018, \aj, 155, 159

\bibitem[{{Brandt} {et~al.}(2019){Brandt}, {Dupuy}, {Bowler}, {Bardalez
  Gagliuffi}, {Faherty}, {Mirek Brandt}, \& {Michalik}}]{Brandt2019}
{Brandt}, T.~D., {Dupuy}, T.~J., {Bowler}, B.~P., {et~al.} 2019, arXiv
  e-prints, arXiv:1910.01652

\bibitem[{{Brandt} {et~al.}(2014){Brandt}, {McElwain}, {Turner}, {Mede},
  {Spiegel}, {Kuzuhara}, {Schlieder}, {Wisniewski}, {Abe}, {Biller},
  {Brandner}, {Carson}, {Currie}, {Egner}, {Feldt}, {Golota}, {Goto}, {Grady},
  {Guyon}, {Hashimoto}, {Hayano}, {Hayashi}, {Hayashi}, {Henning}, {Hodapp},
  {Inutsuka}, {Ishii}, {Iye}, {Janson}, {Kand ori}, {Knapp}, {Kudo},
  {Kusakabe}, {Kwon}, {Matsuo}, {Miyama}, {Morino}, {Moro-Mart{\'\i}n},
  {Nishimura}, {Pyo}, {Serabyn}, {Suto}, {Suzuki}, {Takami}, {Takato},
  {Terada}, {Thalmann}, {Tomono}, {Watanabe}, {Yamada}, {Takami}, {Usuda}, \&
  {Tamura}}]{Brandt2014}
{Brandt}, T.~D., {McElwain}, M.~W., {Turner}, E.~L., {et~al.} 2014, \apj, 794,
  159

\bibitem[{{Burrows} {et~al.}(1997){Burrows}, {Marley}, {Hubbard}, {Lunine},
  {Guillot}, {Saumon}, {Freedman}, {Sudarsky}, \& {Sharp}}]{Burrows1997}
{Burrows}, A., {Marley}, M., {Hubbard}, W.~B., {et~al.} 1997, \apj, 491, 856

\bibitem[{{Cantalloube} {et~al.}(2015){Cantalloube}, {Mouillet}, {Mugnier},
  {Milli}, {Absil}, {Gomez Gonzalez}, {Chauvin}, {Beuzit}, \&
  {Cornia}}]{Cantalloube2015}
{Cantalloube}, F., {Mouillet}, D., {Mugnier}, L.~M., {et~al.} 2015, \aap, 582,
  A89

\bibitem[{{Chabrier} {et~al.}(2014){Chabrier}, {Johansen}, {Janson}, \&
  {Rafikov}}]{Chabrier2014}
{Chabrier}, G., {Johansen}, A., {Janson}, M., \& {Rafikov}, R. 2014, in
  Protostars and Planets VI, ed. H.~{Beuther}, R.~S. {Klessen}, C.~P.
  {Dullemond}, \& T.~{Henning}, 619

\bibitem[{{Chambers} {et~al.}(1996){Chambers}, {Wetherill}, \&
  {Boss}}]{Chambers1996}
{Chambers}, J.~E., {Wetherill}, G.~W., \& {Boss}, A.~P. 1996, \icarus, 119, 261

\bibitem[{{Chatterjee} {et~al.}(2008){Chatterjee}, {Ford}, {Matsumura}, \&
  {Rasio}}]{Chatterjee2008}
{Chatterjee}, S., {Ford}, E.~B., {Matsumura}, S., \& {Rasio}, F.~A. 2008, \apj,
  686, 580

\bibitem[{{Chauvin} {et~al.}(2017){Chauvin}, {Desidera}, {Lagrange}, {Vigan},
  {Gratton}, {Langlois}, {Bonnefoy}, {Beuzit}, {Feldt}, {Mouillet}, {Meyer},
  {Cheetham}, {Biller}, {Boccaletti}, {D'Orazi}, {Galicher}, {Hagelberg},
  {Maire}, {Mesa}, {Olofsson}, {Samland}, {Schmidt}, {Sissa}, {Bonavita},
  {Charnay}, {Cudel}, {Daemgen}, {Delorme}, {Janin-Potiron}, {Janson},
  {Keppler}, {Le Coroller}, {Ligi}, {Marleau}, {Messina}, {Molli{\`e}re},
  {Mordasini}, {M{\"u}ller}, {Peretti}, {Perrot}, {Rodet}, {Rouan}, {Zurlo},
  {Dominik}, {Henning}, {Menard}, {Schmid}, {Turatto}, {Udry}, {Vakili}, {Abe},
  {Antichi}, {Baruffolo}, {Baudoz}, {Baudrand}, {Blanchard}, {Bazzon}, {Buey},
  {Carbillet}, {Carle}, {Charton}, {Cascone}, {Claudi}, {Costille}, {Deboulbe},
  {De Caprio}, {Dohlen}, {Fantinel}, {Feautrier}, {Fusco}, {Gigan}, {Giro},
  {Gisler}, {Gluck}, {Hubin}, {Hugot}, {Jaquet}, {Kasper}, {Madec}, {Magnard},
  {Martinez}, {Maurel}, {Le Mignant}, {M{\"o}ller-Nilsson}, {Llored}, {Moulin},
  {Orign{\'e}}, {Pavlov}, {Perret}, {Petit}, {Pragt}, {Puget}, {Rabou},
  {Ramos}, {Rigal}, {Rochat}, {Roelfsema}, {Rousset}, {Roux}, {Salasnich},
  {Sauvage}, {Sevin}, {Soenke}, {Stadler}, {Suarez}, {Weber}, {Wildi},
  {Antoniucci}, {Augereau}, {Baudino}, {Brandner}, {Engler}, {Girard}, {Gry},
  {Kral}, {Kopytova}, {Lagadec}, {Milli}, {Moutou}, {Schlieder},
  {Szul{\'a}gyi}, {Thalmann}, \& {Wahhaj}}]{Chauvin2017}
{Chauvin}, G., {Desidera}, S., {Lagrange}, A.-M., {et~al.} 2017, \aap, 605, L9

\bibitem[{{Chauvin} {et~al.}(2018){Chauvin}, {Gratton}, {Bonnefoy}, {Lagrange},
  {de Boer}, {Vigan}, {Beust}, {Lazzoni}, {Boccaletti}, {Galicher}, {Desidera},
  {Delorme}, {Keppler}, {Lannier}, {Maire}, {Mesa}, {Meunier}, {Kral},
  {Henning}, {Menard}, {Moor}, {Avenhaus}, {Bazzon}, {Janson}, {Beuzit},
  {Bhowmik}, {Bonavita}, {Borgniet}, {Brandner}, {Cheetham}, {Cudel}, {Feldt},
  {Fontanive}, {Ginski}, {Hagelberg}, {Janin-Potiron}, {Lagadec}, {Langlois},
  {Le Coroller}, {Messina}, {Meyer}, {Mouillet}, {Peretti}, {Perrot}, {Rodet},
  {Samland}, {Sissa}, {Olofsson}, {Salter}, {Schmidt}, {Zurlo}, {Milli}, {van
  Boekel}, {Quanz}, {Feautrier}, {Le Mignant}, {Perret}, {Ramos}, \&
  {Rochat}}]{Chauvin2018}
{Chauvin}, G., {Gratton}, R., {Bonnefoy}, M., {et~al.} 2018, \aap, 617, A76

\bibitem[{{Chauvin} {et~al.}(2005{\natexlab{a}}){Chauvin}, {Lagrange},
  {Lacombe}, {Dumas}, {Mouillet}, {Zuckerman}, {Gendron}, {Song}, {Beuzit},
  {Lowrance}, \& {Fusco}}]{Chauvin2005b}
{Chauvin}, G., {Lagrange}, A.~M., {Lacombe}, F., {et~al.} 2005{\natexlab{a}},
  \aap, 430, 1027

\bibitem[{{Chauvin} {et~al.}(2005{\natexlab{b}}){Chauvin}, {Lagrange},
  {Zuckerman}, {Dumas}, {Mouillet}, {Song}, {Beuzit}, {Lowrance}, \&
  {Bessell}}]{Chauvin2005}
{Chauvin}, G., {Lagrange}, A.~M., {Zuckerman}, B., {et~al.} 2005{\natexlab{b}},
  \aap, 438, L29

\bibitem[{{Chauvin} {et~al.}(2015){Chauvin}, {Vigan}, {Bonnefoy}, {Desidera},
  {Bonavita}, {Mesa}, {Boccaletti}, {Buenzli}, {Carson}, {Delorme},
  {Hagelberg}, {Montagnier}, {Mordasini}, {Quanz}, {Segransan}, {Thalmann},
  {Beuzit}, {Biller}, {Covino}, {Feldt}, {Girard}, {Gratton}, {Henning},
  {Kasper}, {Lagrange}, {Messina}, {Meyer}, {Mouillet}, {Moutou}, {Reggiani},
  {Schlieder}, \& {Zurlo}}]{Chauvin2015}
{Chauvin}, G., {Vigan}, A., {Bonnefoy}, M., {et~al.} 2015, \aap, 573, A127

\bibitem[{{Cheetham} {et~al.}(2018){Cheetham}, {Bonnefoy}, {Desidera},
  {Langlois}, {Vigan}, {Schmidt}, {Olofsson}, {Chauvin}, {Klahr}, {Gratton},
  {D'Orazi}, {Henning}, {Janson}, {Biller}, {Peretti}, {Hagelberg},
  {S{\'e}gransan}, {Udry}, {Mesa}, {Sissa}, {Kral}, {Schlieder}, {Maire},
  {Mordasini}, {Menard}, {Zurlo}, {Beuzit}, {Feldt}, {Mouillet}, {Meyer},
  {Lagrange}, {Boccaletti}, {Keppler}, {Kopytova}, {Ligi}, {Rouan}, {Le
  Coroller}, {Dominik}, {Lagadec}, {Turatto}, {Abe}, {Antichi}, {Baruffolo},
  {Baudoz}, {Blanchard}, {Buey}, {Carbillet}, {Carle}, {Cascone}, {Claudi},
  {Costille}, {Delboulb{\'e}}, {De Caprio}, {Dohlen}, {Fantinel}, {Feautrier},
  {Fusco}, {Giro}, {Gluck}, {Hubin}, {Hugot}, {Jaquet}, {Kasper}, {Llored},
  {Madec}, {Magnard}, {Martinez}, {Maurel}, {Le Mignant}, {M{\"o}ller-Nilsson},
  {Moulin}, {Orign{\'e}}, {Pavlov}, {Perret}, {Petit}, {Pragt}, {Puget},
  {Rabou}, {Ramos}, {Rigal}, {Rochat}, {Roelfsema}, {Rousset}, {Roux},
  {Salasnich}, {Sauvage}, {Sevin}, {Soenke}, {Stadler}, {Suarez}, {Weber}, \&
  {Wildi}}]{Cheetham2018}
{Cheetham}, A., {Bonnefoy}, M., {Desidera}, S., {et~al.} 2018, \aap, 615, A160

\bibitem[{{Cheetham} {et~al.}(2019){Cheetham}, {Samland}, {Brems}, {Launhardt},
  {Chauvin}, {S{\'e}gransan}, {Henning}, {Quirrenbach}, {Avenhaus}, {Cugno},
  {Girard}, {Godoy}, {Kennedy}, {Maire}, {Metchev}, {M{\"u}ller}, {Musso
  Barcucci}, {Olofsson}, {Pepe}, {Quanz}, {Queloz}, {Reffert}, {Rickman}, {van
  Boekel}, {Boccaletti}, {Bonnefoy}, {Cantalloube}, {Charnay}, {Delorme},
  {Janson}, {Keppler}, {Lagrange}, {Langlois}, {Lazzoni}, {Menard}, {Mesa},
  {Meyer}, {Schmidt}, {Sissa}, {Udry}, \& {Zurlo}}]{Cheetham2019}
{Cheetham}, A.~C., {Samland}, M., {Brems}, S.~S., {et~al.} 2019, \aap, 622, A80

\bibitem[{{Christiaens} {et~al.}(2019){Christiaens}, {Cantalloube}, {Casassus},
  {Price}, {Absil}, {Pinte}, {Girard}, \& {Montesinos}}]{Christiaens2019b}
{Christiaens}, V., {Cantalloube}, F., {Casassus}, S., {et~al.} 2019, \apjl,
  877, L33

\bibitem[{{Clarke}(2009)}]{Clarke2009}
{Clarke}, C.~J. 2009, \mnras, 396, 1066

\bibitem[{{Clarke} {et~al.}(2001){Clarke}, {Gendrin}, \&
  {Sotomayor}}]{Clarke2001}
{Clarke}, C.~J., {Gendrin}, A., \& {Sotomayor}, M. 2001, \mnras, 328, 485

\bibitem[{{Crepp} {et~al.}(2012){Crepp}, {Johnson}, {Fischer}, {Howard},
  {Marcy}, {Wright}, {Isaacson}, {Boyajian}, {von Braun}, {Hillenbrand},
  {Hinkley}, {Carpenter}, \& {Brewer}}]{Crepp2012}
{Crepp}, J.~R., {Johnson}, J.~A., {Fischer}, D.~A., {et~al.} 2012, \apj, 751,
  97

\bibitem[{{Cumming} {et~al.}(2008){Cumming}, {Butler}, {Marcy}, {Vogt},
  {Wright}, \& {Fischer}}]{Cumming2008}
{Cumming}, A., {Butler}, R.~P., {Marcy}, G.~W., {et~al.} 2008, \pasp, 120, 531

\bibitem[{{Cumming} {et~al.}(2018){Cumming}, {Helled}, \&
  {Venturini}}]{Cumming2018}
{Cumming}, A., {Helled}, R., \& {Venturini}, J. 2018, \mnras, 477, 4817

\bibitem[{{Dawson} \& {Murray-Clay}(2013)}]{Dawson2013}
{Dawson}, R.~I. \& {Murray-Clay}, R.~A. 2013, \apjl, 767, L24

\bibitem[{{De Rosa} {et~al.}(2014){De Rosa}, {Patience}, {Wilson}, {Schneider},
  {Wiktorowicz}, {Vigan}, {Marois}, {Song}, {Macintosh}, {Graham}, {Doyon},
  {Bessell}, {Thomas}, \& {Lai}}]{DeRosa2014}
{De Rosa}, R.~J., {Patience}, J., {Wilson}, P.~A., {et~al.} 2014, \mnras, 437,
  1216

\bibitem[{{De Rosa} {et~al.}(2016){De Rosa}, {Rameau}, {Patience}, {Graham},
  {Doyon}, {Lafreni{\`e}re}, {Macintosh}, {Pueyo}, {Rajan}, {Wang},
  {Ward-Duong}, {Hung}, {Maire}, {Nielsen}, {Ammons}, {Bulger}, {Cardwell},
  {Chilcote}, {Galvez}, {Gerard}, {Goodsell}, {Hartung}, {Hibon}, {Ingraham},
  {Johnson-Groh}, {Kalas}, {Konopacky}, {Marchis}, {Marois}, {Metchev},
  {Morzinski}, {Oppenheimer}, {Perrin}, {Rantakyr{\"o}}, {Savransky}, \&
  {Thomas}}]{DeRosa2016}
{De Rosa}, R.~J., {Rameau}, J., {Patience}, J., {et~al.} 2016, \apj, 824, 121

\bibitem[{{Delorme} {et~al.}(2017{\natexlab{a}}){Delorme}, {Meunier}, {Albert},
  {Lagadec}, {Le Coroller}, {Galicher}, {Mouillet}, {Boccaletti}, {Mesa},
  {Meunier}, {Beuzit}, {Lagrange}, {Chauvin}, {Sapone}, {Langlois}, {Maire},
  {Montarg{\`e}s}, {Gratton}, {Vigan}, \& {Surace}}]{Delorme2017b}
{Delorme}, P., {Meunier}, N., {Albert}, D., {et~al.} 2017{\natexlab{a}}, in
  SF2A-2017: Proceedings of the Annual meeting of the French Society of
  Astronomy and Astrophysics, Di

\bibitem[{{Delorme} {et~al.}(2017{\natexlab{b}}){Delorme}, {Schmidt},
  {Bonnefoy}, {Desidera}, {Ginski}, {Charnay}, {Lazzoni}, {Christiaens},
  {Messina}, {D'Orazi}, {Milli}, {Schlieder}, {Gratton}, {Rodet}, {Lagrange},
  {Absil}, {Vigan}, {Galicher}, {Hagelberg}, {Bonavita}, {Lavie}, {Zurlo},
  {Olofsson}, {Boccaletti}, {Cantalloube}, {Mouillet}, {Chauvin}, {Hambsch},
  {Langlois}, {Udry}, {Henning}, {Beuzit}, {Mordasini}, {Lucas}, {Marocco},
  {Biller}, {Carson}, {Cheetham}, {Covino}, {De Caprio}, {Delboulbe}, {Feldt},
  {Girard}, {Hubin}, {Maire}, {Pavlov}, {Petit}, {Rouan}, {Roelfsema}, \&
  {Wildi}}]{Delorme2017a}
{Delorme}, P., {Schmidt}, T., {Bonnefoy}, M., {et~al.} 2017{\natexlab{b}},
  \aap, 608, A79

\bibitem[{{Desidera} {et~al.}({submitted}){Desidera}, {Chauvin}, {Bonavita}, \&
  {SHINE consortium}}]{SHINEPaperI}
{Desidera}, S., {Chauvin}, G., {Bonavita}, M., \& {SHINE consortium}.
  {submitted}, \aap

\bibitem[{{Dupuy} {et~al.}(2019){Dupuy}, {Brandt}, {Kratter}, \&
  {Bowler}}]{Dupuy2019}
{Dupuy}, T.~J., {Brandt}, T.~D., {Kratter}, K.~M., \& {Bowler}, B.~P. 2019,
  \apjl, 871, L4

\bibitem[{{Dupuy} \& {Liu}(2017)}]{Dupuy2017}
{Dupuy}, T.~J. \& {Liu}, M.~C. 2017, \apjs, 231, 15

\bibitem[{{Emsenhuber} {et~al.}({submitted~a}){Emsenhuber}, {Mordasini},
  {Burn}, {Alibert}, {Benz}, \& {Asphaug}}]{Emsenhuber2020A}
{Emsenhuber}, A., {Mordasini}, C., {Burn}, R., {et~al.} {submitted~a}, \aap

\bibitem[{{Emsenhuber} {et~al.}({submitted~b}){Emsenhuber}, {Mordasini},
  {Burn}, {Alibert}, {Benz}, \& {Asphaug}}]{Emsenhuber2020B}
{Emsenhuber}, A., {Mordasini}, C., {Burn}, R., {et~al.} {submitted~b}, \aap

\bibitem[{{Fernandes} {et~al.}(2019){Fernandes}, {Mulders}, {Pascucci},
  {Mordasini}, \& {Emsenhuber}}]{Fernandes2019}
{Fernandes}, R.~B., {Mulders}, G.~D., {Pascucci}, I., {Mordasini}, C., \&
  {Emsenhuber}, A. 2019, \apj, 874, 81

\bibitem[{{Flasseur} {et~al.}(2018){Flasseur}, {Denis}, {Thi{\'e}baut}, \&
  {Langlois}}]{Flasseur2018}
{Flasseur}, O., {Denis}, L., {Thi{\'e}baut}, {\'E}., \& {Langlois}, M. 2018,
  \aap, 618, A138

\bibitem[{{Fontanive} {et~al.}(2018){Fontanive}, {Biller}, {Bonavita}, \&
  {Allers}}]{Fontanive2018}
{Fontanive}, C., {Biller}, B., {Bonavita}, M., \& {Allers}, K. 2018, \mnras,
  479, 2702

\bibitem[{{Fontanive} {et~al.}(2019){Fontanive}, {Rice}, {Bonavita}, {Lopez},
  {Mu{\v z}i{\'c}}, {}, \& {Biller}}]{Fontanive2019}
{Fontanive}, C., {Rice}, K., {Bonavita}, M., {et~al.} 2019, \mnras, 485, 4967

\bibitem[{{Foreman-Mackey} {et~al.}(2013){Foreman-Mackey}, {Hogg}, {Lang}, \&
  {Goodman}}]{Foreman-Mackey2013}
{Foreman-Mackey}, D., {Hogg}, D.~W., {Lang}, D., \& {Goodman}, J. 2013, \pasp,
  125, 306

\bibitem[{{Forgan} {et~al.}(2015){Forgan}, {Parker}, \& {Rice}}]{Forgan2015}
{Forgan}, D., {Parker}, R.~J., \& {Rice}, K. 2015, \mnras, 447, 836

\bibitem[{{Forgan} \& {Rice}(2012)}]{Forgan2012}
{Forgan}, D. \& {Rice}, K. 2012, \mnras, 420, 299

\bibitem[{{Forgan} \& {Rice}(2013)}]{Forgan2013}
{Forgan}, D. \& {Rice}, K. 2013, \mnras, 432, 3168

\bibitem[{{Forgan} {et~al.}(2018){Forgan}, {Hall}, {Meru}, \&
  {Rice}}]{Forgan2018}
{Forgan}, D.~H., {Hall}, C., {Meru}, F., \& {Rice}, W.~K.~M. 2018, \mnras, 474,
  5036

\bibitem[{{Fortney} {et~al.}(2011){Fortney}, {Ikoma}, {Nettelmann}, {Guillot},
  \& {Marley}}]{Fortney2011}
{Fortney}, J.~J., {Ikoma}, M., {Nettelmann}, N., {Guillot}, T., \& {Marley},
  M.~S. 2011, \apj, 729, 32

\bibitem[{{Fortney} {et~al.}(2008){Fortney}, {Marley}, {Saumon}, \&
  {Lodders}}]{Fortney2008}
{Fortney}, J.~J., {Marley}, M.~S., {Saumon}, D., \& {Lodders}, K. 2008, \apj,
  683, 1104

\bibitem[{{Gaia Collaboration} {et~al.}(2018){Gaia Collaboration}, {Brown},
  {Vallenari}, {Prusti}, {de Bruijne}, {Babusiaux}, {Bailer-Jones}, {Biermann},
  {Evans}, {Eyer}, {Jansen}, {Jordi}, {Klioner}, {Lammers}, {Lindegren},
  {Luri}, {Mignard}, {Panem}, {Pourbaix}, {Randich}, {Sartoretti}, {Siddiqui},
  {Soubiran}, {van Leeuwen}, {Walton}, {Arenou}, {Bastian}, {Cropper},
  {Drimmel}, {Katz}, {Lattanzi}, {Bakker}, {Cacciari}, {Casta{\~n}eda},
  {Chaoul}, {Cheek}, {De Angeli}, {Fabricius}, {Guerra}, {Holl}, {Masana},
  {Messineo}, {Mowlavi}, {Nienartowicz}, {Panuzzo}, {Portell}, {Riello},
  {Seabroke}, {Tanga}, {Th{\'e}venin}, {Gracia-Abril}, {Comoretto},
  {Garcia-Reinaldos}, {Teyssier}, {Altmann}, {Andrae}, {Audard},
  {Bellas-Velidis}, {Benson}, {Berthier}, {Blomme}, {Burgess}, {Busso},
  {Carry}, {Cellino}, {Clementini}, {Clotet}, {Creevey}, {Davidson}, {De
  Ridder}, {Delchambre}, {Dell'Oro}, {Ducourant},
  {Fern{\'a}ndez-Hern{\'a}ndez}, {Fouesneau}, {Fr{\'e}mat}, {Galluccio},
  {Garc{\'\i}a-Torres}, {Gonz{\'a}lez-N{\'u}{\~n}ez}, {Gonz{\'a}lez-Vidal},
  {Gosset}, {Guy}, {Halbwachs}, {Hambly}, {Harrison}, {Hern{\'a}ndez},
  {Hestroffer}, {Hodgkin}, {Hutton}, {Jasniewicz}, {Jean-Antoine-Piccolo},
  {Jordan}, {Korn}, {Krone-Martins}, {Lanzafame}, {Lebzelter}, {L{\"o}ffler},
  {Manteiga}, {Marrese}, {Mart{\'\i}n-Fleitas}, {Moitinho}, {Mora}, {Muinonen},
  {Osinde}, {Pancino}, {Pauwels}, {Petit}, {Recio-Blanco}, {Richards},
  {Rimoldini}, {Robin}, {Sarro}, {Siopis}, {Smith}, {Sozzetti}, {S{\"u}veges},
  {Torra}, {van Reeven}, {Abbas}, {Abreu Aramburu}, {Accart}, {Aerts},
  {Altavilla}, {{\'A}lvarez}, {Alvarez}, {Alves}, {Anderson}, {Andrei},
  {Anglada Varela}, {Antiche}, {Antoja}, {Arcay}, {Astraatmadja}, {Bach},
  {Baker}, {Balaguer-N{\'u}{\~n}ez}, {Balm}, {Barache}, {Barata}, {Barbato},
  {Barblan}, {Barklem}, {Barrado}, {Barros}, {Barstow}, {Bartholom{\'e}
  Mu{\~n}oz}, {Bassilana}, {Becciani}, {Bellazzini}, {Berihuete}, {Bertone},
  {Bianchi}, {Bienaym{\'e}}, {Blanco-Cuaresma}, {Boch}, {Boeche}, {Bombrun},
  {Borrachero}, {Bossini}, {Bouquillon}, {Bourda}, {Bragaglia}, {Bramante},
  {Breddels}, {Bressan}, {Brouillet}, {Br{\"u}semeister}, {Brugaletta},
  {Bucciarelli}, {Burlacu}, {Busonero}, {Butkevich}, {Buzzi}, {Caffau},
  {Cancelliere}, {Cannizzaro}, {Cantat-Gaudin}, {Carballo}, {Carlucci},
  {Carrasco}, {Casamiquela}, {Castellani}, {Castro-Ginard}, {Charlot},
  {Chemin}, {Chiavassa}, {Cocozza}, {Costigan}, {Cowell}, {Crifo}, {Crosta},
  {Crowley}, {Cuypers}, {Dafonte}, {Damerdji}, {Dapergolas}, {David}, {David},
  {de Laverny}, {De Luise}, {De March}, {de Martino}, {de Souza}, {de Torres},
  {Debosscher}, {del Pozo}, {Delbo}, {Delgado}, {Delgado}, {Di Matteo},
  {Diakite}, {Diener}, {Distefano}, {Dolding}, {Drazinos}, {Dur{\'a}n},
  {Edvardsson}, {Enke}, {Eriksson}, {Esquej}, {Eynard Bontemps}, {Fabre},
  {Fabrizio}, {Faigler}, {Falc{\~a}o}, {Farr{\`a}s Casas}, {Federici},
  {Fedorets}, {Fernique}, {Figueras}, {Filippi}, {Findeisen}, {Fonti},
  {Fraile}, {Fraser}, {Fr{\'e}zouls}, {Gai}, {Galleti}, {Garabato},
  {Garc{\'\i}a-Sedano}, {Garofalo}, {Garralda}, {Gavel}, {Gavras}, {Gerssen},
  {Geyer}, {Giacobbe}, {Gilmore}, {Girona}, {Giuffrida}, {Glass}, {Gomes},
  {Granvik}, {Gueguen}, {Guerrier}, {Guiraud}, {Guti{\'e}rrez-S{\'a}nchez},
  {Haigron}, {Hatzidimitriou}, {Hauser}, {Haywood}, {Heiter}, {Helmi}, {Heu},
  {Hilger}, {Hobbs}, {Hofmann}, {Holland}, {Huckle}, {Hypki}, {Icardi},
  {Jan{\ss}en}, {Jevardat de Fombelle}, {Jonker}, {Juh{\'a}sz}, {Julbe},
  {Karampelas}, {Kewley}, {Klar}, {Kochoska}, {Kohley}, {Kolenberg},
  {Kontizas}, {Kontizas}, {Koposov}, {Kordopatis}, {Kostrzewa-Rutkowska},
  {Koubsky}, {Lambert}, {Lanza}, {Lasne}, {Lavigne}, {Le Fustec}, {Le
  Poncin-Lafitte}, {Lebreton}, {Leccia}, {Leclerc}, {Lecoeur-Taibi},
  {Lenhardt}, {Leroux}, {Liao}, {Licata}, {Lindstr{\o}m}, {Lister}, {Livanou},
  {Lobel}, {L{\'o}pez}, {Managau}, {Mann}, {Mantelet}, {Marchal}, {Marchant},
  {Marconi}, {Marinoni}, {Marschalk{\'o}}, {Marshall}, {Martino}, {Marton},
  {Mary}, {Massari}, {Matijevi{\v{c}}}, {Mazeh}, {McMillan}, {Messina},
  {Michalik}, {Millar}, {Molina}, {Molinaro}, {Moln{\'a}r}, {Montegriffo},
  {Mor}, {Morbidelli}, {Morel}, {Morris}, {Mulone}, {Muraveva}, {Musella},
  {Nelemans}, {Nicastro}, {Noval}, {O'Mullane}, {Ord{\'e}novic},
  {Ord{\'o}{\~n}ez-Blanco}, {Osborne}, {Pagani}, {Pagano}, {Pailler},
  {Palacin}, {Palaversa}, {Panahi}, {Pawlak}, {Piersimoni}, {Pineau}, {Plachy},
  {Plum}, {Poggio}, {Poujoulet}, {Pr{\v{s}}a}, {Pulone}, {Racero}, {Ragaini},
  {Rambaux}, {Ramos-Lerate}, {Regibo}, {Reyl{\'e}}, {Riclet}, {Ripepi}, {Riva},
  {Rivard}, {Rixon}, {Roegiers}, {Roelens}, {Romero-G{\'o}mez}, {Rowell},
  {Royer}, {Ruiz-Dern}, {Sadowski}, {Sagrist{\`a} Sell{\'e}s}, {Sahlmann},
  {Salgado}, {Salguero}, {Sanna}, {Santana-Ros}, {Sarasso}, {Savietto},
  {Schultheis}, {Sciacca}, {Segol}, {Segovia}, {S{\'e}gransan}, {Shih},
  {Siltala}, {Silva}, {Smart}, {Smith}, {Solano}, {Solitro}, {Sordo}, {Soria
  Nieto}, {Souchay}, {Spagna}, {Spoto}, {Stampa}, {Steele},
  {Steidelm{\"u}ller}, {Stephenson}, {Stoev}, {Suess}, {Surdej}, {Szabados},
  {Szegedi-Elek}, {Tapiador}, {Taris}, {Tauran}, {Taylor}, {Teixeira},
  {Terrett}, {Teyssand ier}, {Thuillot}, {Titarenko}, {Torra Clotet}, {Turon},
  {Ulla}, {Utrilla}, {Uzzi}, {Vaillant}, {Valentini}, {Valette}, {van Elteren},
  {Van Hemelryck}, {van Leeuwen}, {Vaschetto}, {Vecchiato}, {Veljanoski},
  {Viala}, {Vicente}, {Vogt}, {von Essen}, {Voss}, {Votruba}, {Voutsinas},
  {Walmsley}, {Weiler}, {Wertz}, {Wevers}, {Wyrzykowski}, {Yoldas},
  {{\v{Z}}erjal}, {Ziaeepour}, {Zorec}, {Zschocke}, {Zucker}, {Zurbach}, \&
  {Zwitter}}]{Gaia2018}
{Gaia Collaboration}, {Brown}, A.~G.~A., {Vallenari}, A., {et~al.} 2018, \aap,
  616, A1

\bibitem[{{Gaia Collaboration} {et~al.}(2016){Gaia Collaboration}, {Prusti},
  {de Bruijne}, {Brown}, {Vallenari}, {Babusiaux}, {Bailer-Jones}, {Bastian},
  {Biermann}, {Evans}, \& et~al.}]{Gaia2016}
{Gaia Collaboration}, {Prusti}, T., {de Bruijne}, J.~H.~J., {et~al.} 2016,
  \aap, 595, A1

\bibitem[{{Galicher} {et~al.}(2018){Galicher}, {Boccaletti}, {Mesa}, {Delorme},
  {Gratton}, {Langlois}, {Lagrange}, {Maire}, {Le Coroller}, {Chauvin},
  {Biller}, {Cantalloube}, {Janson}, {Lagadec}, {Meunier}, {Vigan},
  {Hagelberg}, {Bonnefoy}, {Zurlo}, {Rocha}, {Maurel}, {Jaquet}, {Buey}, \&
  {Weber}}]{Galicher2018}
{Galicher}, R., {Boccaletti}, A., {Mesa}, D., {et~al.} 2018, \aap, 615, A92

\bibitem[{{Galicher} {et~al.}(2016){Galicher}, {Marois}, {Macintosh},
  {Zuckerman}, {Barman}, {Konopacky}, {Song}, {Patience}, {Lafreni{\`e}re},
  {Doyon}, \& {Nielsen}}]{Galicher2016}
{Galicher}, R., {Marois}, C., {Macintosh}, B., {et~al.} 2016, \aap, 594, A63

\bibitem[{{Ginski} {et~al.}(2014){Ginski}, {Mugrauer}, {Neuh{\"a}user}, \&
  {Schmidt}}]{Ginski2014}
{Ginski}, C., {Mugrauer}, M., {Neuh{\"a}user}, R., \& {Schmidt}, T.~O.~B. 2014,
  \mnras, 438, 1102

\bibitem[{{Goodman} \& {Weare}(2010)}]{Goodman2010}
{Goodman}, J. \& {Weare}, J. 2010, Communications in Applied Mathematics and
  Computational Science, 5, 65

\bibitem[{{Grandjean} {et~al.}(2019){Grandjean}, {Lagrange}, {Beust}, {Rodet},
  {Milli}, {Rubini}, {Babusiaux}, {Meunier}, {Delorme}, {Aigrain}, {Zicher},
  {Bonnefoy}, {Biller}, {Baudino}, {Bonavita}, {Boccaletti}, {Cheetham},
  {Girard}, {Hagelberg}, {Janson}, {Lannier}, {Lazzoni}, {Ligi}, {Maire},
  {Mesa}, {Perrot}, {Rouan}, \& {Zurlo}}]{Grandjean2019}
{Grandjean}, A., {Lagrange}, A.~M., {Beust}, H., {et~al.} 2019, \aap, 627, L9

\bibitem[{{Gressel} {et~al.}(2013){Gressel}, {Nelson}, {Turner}, \&
  {Ziegler}}]{Gressel2013}
{Gressel}, O., {Nelson}, R.~P., {Turner}, N.~J., \& {Ziegler}, U. 2013, \apj,
  779, 59

\bibitem[{{Haffert} {et~al.}(2019){Haffert}, {Bohn}, {de Boer}, {Snellen},
  {Brinchmann}, {Girard}, {Keller}, \& {Bacon}}]{Haffert2019}
{Haffert}, S.~Y., {Bohn}, A.~J., {de Boer}, J., {et~al.} 2019, Nature
  Astronomy, 3, 749

\bibitem[{{Hashimoto} {et~al.}(2020){Hashimoto}, {Aoyama}, {Konishi}, {Uyama},
  {Takasao}, {Ikoma}, \& {Tanigawa}}]{Hashimoto2020}
{Hashimoto}, J., {Aoyama}, Y., {Konishi}, M., {et~al.} 2020, \aj, 159, 222

\bibitem[{{Heinze} {et~al.}(2010){Heinze}, {Hinz}, {Sivanandam}, {Kenworthy},
  {Meyer}, \& {Miller}}]{Heinze2010}
{Heinze}, A.~N., {Hinz}, P.~M., {Sivanandam}, S., {et~al.} 2010, \apj, 714,
  1551

\bibitem[{{Hogg} {et~al.}(2010){Hogg}, {Myers}, \& {Bovy}}]{Hogg2010}
{Hogg}, D.~W., {Myers}, A.~D., \& {Bovy}, J. 2010, \apj, 725, 2166

\bibitem[{{Ida} \& {Makino}(1993)}]{IdaMakino1993}
{Ida}, S. \& {Makino}, J. 1993, \icarus, 106, 210

\bibitem[{{Janson} {et~al.}(2011){Janson}, {Bonavita}, {Klahr},
  {Lafreni{\`e}re}, {Jayawardhana}, \& {Zinnecker}}]{Janson2011}
{Janson}, M., {Bonavita}, M., {Klahr}, H., {et~al.} 2011, \apj, 736, 89

\bibitem[{{Kasper} {et~al.}(2007){Kasper}, {Apai}, {Janson}, \&
  {Brandner}}]{Kasper2007}
{Kasper}, M., {Apai}, D., {Janson}, M., \& {Brandner}, W. 2007, \aap, 472, 321

\bibitem[{{Kennedy} \& {Kenyon}(2008)}]{kennedy2008}
{Kennedy}, G.~M. \& {Kenyon}, S.~J. 2008, \apj, 673, 502

\bibitem[{{Keppler} {et~al.}(2018){Keppler}, {Benisty}, {M{\"u}ller},
  {Henning}, {van Boekel}, {Cantalloube}, {Ginski}, {van Holstein}, {Maire},
  {Pohl}, {Samland }, {Avenhaus}, {Baudino}, {Boccaletti}, {de Boer},
  {Bonnefoy}, {Chauvin}, {Desidera}, {Langlois}, {Lazzoni}, {Marleau},
  {Mordasini}, {Pawellek}, {Stolker}, {Vigan}, {Zurlo}, {Birnstiel},
  {Brandner}, {Feldt}, {Flock}, {Girard}, {Gratton}, {Hagelberg}, {Isella},
  {Janson}, {Juhasz}, {Kemmer}, {Kral}, {Lagrange}, {Launhardt}, {Matter},
  {M{\'e}nard}, {Milli}, {Molli{\`e}re}, {Olofsson}, {P{\'e}rez}, {Pinilla},
  {Pinte}, {Quanz}, {Schmidt}, {Udry}, {Wahhaj}, {Williams}, {Buenzli},
  {Cudel}, {Dominik}, {Galicher}, {Kasper}, {Lannier}, {Mesa}, {Mouillet},
  {Peretti}, {Perrot}, {Salter}, {Sissa}, {Wildi}, {Abe}, {Antichi},
  {Augereau}, {Baruffolo}, {Baudoz}, {Bazzon}, {Beuzit}, {Blanchard}, {Brems},
  {Buey}, {De Caprio}, {Carbillet}, {Carle}, {Cascone}, {Cheetham}, {Claudi},
  {Costille}, {Delboulb{\'e}}, {Dohlen}, {Fantinel}, {Feautrier}, {Fusco},
  {Giro}, {Gluck}, {Gry}, {Hubin}, {Hugot}, {Jaquet}, {Le Mignant}, {Llored},
  {Madec}, {Magnard}, {Martinez}, {Maurel}, {Meyer}, {M{\"o}ller-Nilsson},
  {Moulin}, {Mugnier}, {Orign{\'e}}, {Pavlov}, {Perret}, {Petit}, {Pragt},
  {Puget}, {Rabou}, {Ramos}, {Rigal}, {Rochat}, {Roelfsema}, {Rousset}, {Roux},
  {Salasnich}, {Sauvage}, {Sevin}, {Soenke}, {Stadler}, {Suarez}, {Turatto}, \&
  {Weber}}]{Keppler2018}
{Keppler}, M., {Benisty}, M., {M{\"u}ller}, A., {et~al.} 2018, \aap, 617, A44

\bibitem[{{Kley} \& {Nelson}(2012)}]{Kley2012}
{Kley}, W. \& {Nelson}, R.~P. 2012, \araa, 50, 211

\bibitem[{{Konopacky} {et~al.}(2010){Konopacky}, {Ghez}, {Barman}, {Rice},
  {Bailey}, {White}, {McLean}, \& {Duch{\^e}ne}}]{Konopacky2010}
{Konopacky}, Q.~M., {Ghez}, A.~M., {Barman}, T.~S., {et~al.} 2010, \apj, 711,
  1087

\bibitem[{{Konopacky} {et~al.}(2016){Konopacky}, {Rameau}, {Duch{\^e}ne},
  {Filippazzo}, {Giorla Godfrey}, {Marois}, {Nielsen}, {Pueyo}, {Rafikov},
  {Rice}, {Wang}, {Ammons}, {Bailey}, {Barman}, {Bulger}, {Bruzzone},
  {Chilcote}, {Cotten}, {Dawson}, {De Rosa}, {Doyon}, {Esposito}, {Fitzgerald},
  {Follette}, {Goodsell}, {Graham}, {Greenbaum}, {Hibon}, {Hung}, {Ingraham},
  {Kalas}, {Lafreni{\`e}re}, {Larkin}, {Macintosh}, {Maire}, {Marchis},
  {Marley}, {Matthews}, {Metchev}, {Millar-Blanchaer}, {Oppenheimer}, {Palmer},
  {Patience}, {Perrin}, {Poyneer}, {Rajan}, {Rantakyr{\"o}}, {Savransky},
  {Schneider}, {Sivaramakrishnan}, {Song}, {Soummer}, {Thomas}, {Wallace},
  {Ward-Duong}, {Wiktorowicz}, \& {Wolff}}]{Konopacky2016}
{Konopacky}, Q.~M., {Rameau}, J., {Duch{\^e}ne}, G., {et~al.} 2016, \apjl, 829,
  L4

\bibitem[{{Kratter} {et~al.}(2010){Kratter}, {Murray-Clay}, \&
  {Youdin}}]{Kratter2010}
{Kratter}, K.~M., {Murray-Clay}, R.~A., \& {Youdin}, A.~N. 2010, \apj, 710,
  1375

\bibitem[{{Lafreni{\`e}re} {et~al.}(2007){Lafreni{\`e}re}, {Doyon}, {Marois},
  {Nadeau}, {Oppenheimer}, {Roche}, {Rigaut}, {Graham}, {Jayawardhana},
  {Johnstone}, {Kalas}, {Macintosh}, \& {Racine}}]{Lafreniere2007}
{Lafreni{\`e}re}, D., {Doyon}, R., {Marois}, C., {et~al.} 2007, \apj, 670, 1367

\bibitem[{{Lafreni{\`e}re} {et~al.}(2011){Lafreni{\`e}re}, {Jayawardhana},
  {Janson}, {Helling}, {Witte}, \& {Hauschildt}}]{Lafreniere2011}
{Lafreni{\`e}re}, D., {Jayawardhana}, R., {Janson}, M., {et~al.} 2011, \apj,
  730, 42

\bibitem[{{Lagrange} {et~al.}(2019){Lagrange}, {Boccaletti}, {Langlois},
  {Chauvin}, {Gratton}, {Beust}, {Desidera}, {Milli}, {Bonnefoy}, {Cheetham},
  {Feldt}, {Meyer}, {Vigan}, {Biller}, {Bonavita}, {Baudino}, {Cantalloube},
  {Cudel}, {Daemgen}, {Delorme}, {D'Orazi}, {Girard}, {Fontanive}, {Hagelberg},
  {Janson}, {Keppler}, {Koypitova}, {Galicher}, {Lannier}, {Le Coroller},
  {Ligi}, {Maire}, {Mesa}, {Messina}, {M{\"u}eller}, {Peretti}, {Perrot},
  {Rouan}, {Salter}, {Samland}, {Schmidt}, {Sissa}, {Zurlo}, {Beuzit},
  {Mouillet}, {Dominik}, {Henning}, {Lagadec}, {M{\'e}nard}, {Schmid},
  {Turatto}, {Udry}, {Bohn}, {Charnay}, {Gomez Gonzales}, {Gry}, {Kenworthy},
  {Kral}, {Mordasini}, {Moutou}, {van der Plas}, {Schlieder}, {Abe}, {Antichi},
  {Baruffolo}, {Baudoz}, {Baudrand}, {Blanchard}, {Bazzon}, {Buey},
  {Carbillet}, {Carle}, {Charton}, {Cascone}, {Claudi}, {Costille}, {Deboulbe},
  {De Caprio}, {Dohlen}, {Fantinel}, {Feautrier}, {Fusco}, {Gigan}, {Giro},
  {Gisler}, {Gluck}, {Hubin}, {Hugot}, {Jaquet}, {Kasper}, {Madec}, {Magnard},
  {Martinez}, {Maurel}, {Le Mignant}, {M{\"o}ller-Nilsson}, {Llored}, {Moulin},
  {Orign{\'e}}, {Pavlov}, {Perret}, {Petit}, {Pragt}, {Szulagyi}, \&
  {Wildi}}]{Lagrange2019}
{Lagrange}, A.~M., {Boccaletti}, A., {Langlois}, M., {et~al.} 2019, \aap, 621,
  L8

\bibitem[{{Lambrechts} \& {Johansen}(2012)}]{Lambrechts2012}
{Lambrechts}, M. \& {Johansen}, A. 2012, \aap, 544, A32

\bibitem[{{Langlois} {et~al.}({submitted}){Langlois}, {Gratton}, {Lagrange},
  {Delorme}, {Boccaletti}, {Bonnefoy}, {Maire}, {Mesa}, {Chauvin}, {Desidera},
  {Vigan}, \& {SHINE consortium}}]{SHINEPaperII}
{Langlois}, M., {Gratton}, R., {Lagrange}, A.-M., {et~al.} {submitted}, \aap

\bibitem[{{Lannier} {et~al.}(2016){Lannier}, {Delorme}, {Lagrange}, {Borgniet},
  {Rameau}, {Schlieder}, {Gagn{\'e}}, {Bonavita}, {Malo}, {Chauvin},
  {Bonnefoy}, \& {Girard}}]{Lannier2016}
{Lannier}, J., {Delorme}, P., {Lagrange}, A.~M., {et~al.} 2016, \aap, 596, A83

\bibitem[{{Levison} {et~al.}(2015){Levison}, {Kretke}, \&
  {Duncan}}]{Levison2015}
{Levison}, H.~F., {Kretke}, K.~A., \& {Duncan}, M.~J. 2015, \nat, 524, 322

\bibitem[{{Lewis}(1974)}]{Lewis1974}
{Lewis}, J.~S. 1974, Science, 186, 440

\bibitem[{{Linder} {et~al.}(2019){Linder}, {Mordasini}, {Molli{\`e}re},
  {Marleau}, {Malik}, {Quanz}, \& {Meyer}}]{Linder2019}
{Linder}, E.~F., {Mordasini}, C., {Molli{\`e}re}, P., {et~al.} 2019, \aap, 623,
  A85

\bibitem[{{Lodders}(2003)}]{Lodders2003}
{Lodders}, K. 2003, \apj, 591, 1220

\bibitem[{{Lowrance} {et~al.}(2000){Lowrance}, {Schneider}, {Kirkpatrick},
  {Becklin}, {Weinberger}, {Zuckerman}, {Plait}, {Malmuth}, {Heap}, {Schultz},
  {Smith}, {Terrile}, \& {Hines}}]{Lowrance2000}
{Lowrance}, P.~J., {Schneider}, G., {Kirkpatrick}, J.~D., {et~al.} 2000, \apj,
  541, 390

\bibitem[{{Lynden-Bell} \& {Pringle}(1974)}]{LyndenBellPringle1974}
{Lynden-Bell}, D. \& {Pringle}, J.~E. 1974, \mnras, 168, 603

\bibitem[{{Macintosh} {et~al.}(2015){Macintosh}, {Graham}, {Barman}, {De Rosa},
  {Konopacky}, {Marley}, {Marois}, {Nielsen}, {Pueyo}, {Rajan}, {Rameau},
  {Saumon}, {Wang}, {Patience}, {Ammons}, {Arriaga}, {Artigau}, {Beckwith},
  {Brewster}, {Bruzzone}, {Bulger}, {Burningham}, {Burrows}, {Chen}, {Chiang},
  {Chilcote}, {Dawson}, {Dong}, {Doyon}, {Draper}, {Duch{\^e}ne}, {Esposito},
  {Fabrycky}, {Fitzgerald}, {Follette}, {Fortney}, {Gerard}, {Goodsell},
  {Greenbaum}, {Hibon}, {Hinkley}, {Cotten}, {Hung}, {Ingraham},
  {Johnson-Groh}, {Kalas}, {Lafreniere}, {Larkin}, {Lee}, {Line}, {Long},
  {Maire}, {Marchis}, {Matthews}, {Max}, {Metchev}, {Millar-Blanchaer},
  {Mittal}, {Morley}, {Morzinski}, {Murray-Clay}, {Oppenheimer}, {Palmer},
  {Patel}, {Perrin}, {Poyneer}, {Rafikov}, {Rantakyr{\"o}}, {Rice}, {Rojo},
  {Rudy}, {Ruffio}, {Ruiz}, {Sadakuni}, {Saddlemyer}, {Salama}, {Savransky},
  {Schneider}, {Sivaramakrishnan}, {Song}, {Soummer}, {Thomas}, {Vasisht},
  {Wallace}, {Ward-Duong}, {Wiktorowicz}, {Wolff}, \&
  {Zuckerman}}]{Macintosh2015}
{Macintosh}, B., {Graham}, J.~R., {Barman}, T., {et~al.} 2015, Science, 350, 64

\bibitem[{{Maire} {et~al.}(2016){Maire}, {Bonnefoy}, {Ginski}, {Vigan},
  {Messina}, {Mesa}, {Galicher}, {Gratton}, {Desidera}, {Kopytova}, {Millward},
  {Thalmann}, {Claudi}, {Ehrenreich}, {Zurlo}, {Chauvin}, {Antichi},
  {Baruffolo}, {Bazzon}, {Beuzit}, {Blanchard}, {Boccaletti}, {de Boer},
  {Carle}, {Cascone}, {Costille}, {De Caprio}, {Delboulb{\'e}}, {Dohlen},
  {Dominik}, {Feldt}, {Fusco}, {Girard}, {Giro}, {Gisler}, {Gluck}, {Gry},
  {Henning}, {Hubin}, {Hugot}, {Jaquet}, {Kasper}, {Lagrange}, {Langlois}, {Le
  Mignant}, {Llored}, {Madec}, {Martinez}, {Mawet}, {Milli},
  {M{\"o}ller-Nilsson}, {Mouillet}, {Moulin}, {Moutou}, {Orign{\'e}}, {Pavlov},
  {Petit}, {Pragt}, {Puget}, {Ramos}, {Rochat}, {Roelfsema}, {Salasnich},
  {Sauvage}, {Schmid}, {Turatto}, {Udry}, {Vakili}, {Wahhaj}, {Weber}, \&
  {Wildi}}]{Maire2016}
{Maire}, A.~L., {Bonnefoy}, M., {Ginski}, C., {et~al.} 2016, \aap, 587, A56

\bibitem[{{Maire} {et~al.}(2019){Maire}, {Rodet}, {Cantalloube}, {Galicher},
  {Brandner}, {Messina}, {Lazzoni}, {Mesa}, {Melnick}, {Carson}, {Samland},
  {Biller}, {Boccaletti}, {Wahhaj}, {Beust}, {Bonnefoy}, {Chauvin}, {Desidera},
  {Langlois}, {Henning}, {Janson}, {Olofsson}, {Rouan}, {M{\'e}nard},
  {Lagrange}, {Gratton}, {Vigan}, {Meyer}, {Cheetham}, {Beuzit}, {Dohlen},
  {Avenhaus}, {Bonavita}, {Claudi}, {Cudel}, {Daemgen}, {D'Orazi}, {Fontanive},
  {Hagelberg}, {Le Coroller}, {Perrot}, {Rickman}, {Schmidt}, {Sissa}, {Udry},
  {Zurlo}, {Abe}, {Orign{\'e}}, {Rigal}, {Rousset}, {Roux}, \&
  {Weber}}]{Maire2019}
{Maire}, A.~L., {Rodet}, L., {Cantalloube}, F., {et~al.} 2019, \aap, 624, A118

\bibitem[{{Marleau} {et~al.}(2019{\natexlab{a}}){Marleau}, {Coleman}, {Leleu},
  \& {Mordasini}}]{Marleau2019}
{Marleau}, G.-D., {Coleman}, G. A.~L., {Leleu}, A., \& {Mordasini}, C.
  2019{\natexlab{a}}, \aap, 624, A20

\bibitem[{{Marleau} \& {Cumming}(2014)}]{Marleau2014}
{Marleau}, G.-D. \& {Cumming}, A. 2014, \mnras, 437, 1378

\bibitem[{{Marleau} {et~al.}(2017){Marleau}, {Klahr}, {Kuiper}, \&
  {Mordasini}}]{Marleau2017}
{Marleau}, G.-D., {Klahr}, H., {Kuiper}, R., \& {Mordasini}, C. 2017, \apj,
  836, 221

\bibitem[{{Marleau} {et~al.}(2019{\natexlab{b}}){Marleau}, {Mordasini}, \&
  {Kuiper}}]{Marleau2019shock}
{Marleau}, G.-D., {Mordasini}, C., \& {Kuiper}, R. 2019{\natexlab{b}}, \apj,
  881, 144

\bibitem[{{Marley} {et~al.}(2007){Marley}, {Fortney}, {Hubickyj},
  {Bodenheimer}, \& {Lissauer}}]{Marley2007}
{Marley}, M.~S., {Fortney}, J.~J., {Hubickyj}, O., {Bodenheimer}, P., \&
  {Lissauer}, J.~J. 2007, \apj, 655, 541

\bibitem[{{Matsuyama} {et~al.}(2003){Matsuyama}, {Johnstone}, \&
  {Hartmann}}]{Matsuyama2003}
{Matsuyama}, I., {Johnstone}, D., \& {Hartmann}, L. 2003, \apj, 582, 893

\bibitem[{{Mawet} {et~al.}(2014){Mawet}, {Milli}, {Wahhaj}, {Pelat}, {Absil},
  {Delacroix}, {Boccaletti}, {Kasper}, {Kenworthy}, {Marois}, {Mennesson}, \&
  {Pueyo}}]{Mawet2014}
{Mawet}, D., {Milli}, J., {Wahhaj}, Z., {et~al.} 2014, \apj, 792, 97

\bibitem[{{Meru} \& {Bate}(2011)}]{Meru2011}
{Meru}, F. \& {Bate}, M.~R. 2011, \mnras, 411, L1

\bibitem[{{Metchev} \& {Hillenbrand}(2009)}]{Metchev2009}
{Metchev}, S.~A. \& {Hillenbrand}, L.~A. 2009, \apjs, 181, 62

\bibitem[{{Meyer} {et~al.}({in prep.}){Meyer}, {Amara}, {Susemiehl}, \&
  {Peterson}}]{Meyer2020}
{Meyer}, M., {Amara}, A., {Susemiehl}, N., \& {Peterson}, A. {in prep.}

\bibitem[{{Meyer} {et~al.}(2018){Meyer}, {Amara}, {Reggiani}, \&
  {Quanz}}]{Meyer2018}
{Meyer}, M.~R., {Amara}, A., {Reggiani}, M., \& {Quanz}, S.~P. 2018, \aap, 612,
  L3

\bibitem[{{Milli} {et~al.}(2017){Milli}, {Hibon}, {Christiaens}, {Choquet},
  {Bonnefoy}, {Kennedy}, {Wyatt}, {Absil}, {G{\'o}mez Gonz{\'a}lez}, {del
  Burgo}, {Matr{\`a}}, {Augereau}, {Boccaletti}, {Delacroix}, {Ertel}, {Dent},
  {Forsberg}, {Fusco}, {Girard}, {Habraken}, {Huby}, {Karlsson}, {Lagrange},
  {Mawet}, {Mouillet}, {Perrin}, {Pinte}, {Pueyo}, {Reyes}, {Soummer},
  {Surdej}, {Tarricq}, \& {Wahhaj}}]{Milli2017}
{Milli}, J., {Hibon}, P., {Christiaens}, V., {et~al.} 2017, \aap, 597, L2

\bibitem[{{Mizuno}(1980)}]{Mizuno1980}
{Mizuno}, H. 1980, Progress of Theoretical Physics, 64, 544

\bibitem[{{Molli{\`e}re} \& {Mordasini}(2012)}]{Molliere2012}
{Molli{\`e}re}, P. \& {Mordasini}, C. 2012, \aap, 547, A105

\bibitem[{{Mordasini}(2013)}]{Mordasini2013}
{Mordasini}, C. 2013, \aap, 558, A113

\bibitem[{{Mordasini}(2018)}]{Mordasini2018}
{Mordasini}, C. 2018, {Planetary Population Synthesis}, 143

\bibitem[{{Mordasini} {et~al.}(2009){Mordasini}, {Alibert}, \&
  {Benz}}]{Mordasini2009}
{Mordasini}, C., {Alibert}, Y., \& {Benz}, W. 2009, \aap, 501, 1139

\bibitem[{{Mordasini} {et~al.}(2012){Mordasini}, {Alibert}, {Klahr}, \&
  {Henning}}]{Mordasini2012}
{Mordasini}, C., {Alibert}, Y., {Klahr}, H., \& {Henning}, T. 2012, \aap, 547,
  A111

\bibitem[{{Mordasini} {et~al.}(2017){Mordasini}, {Marleau}, \&
  {Molli{\`e}re}}]{Mordasini2017}
{Mordasini}, C., {Marleau}, G.~D., \& {Molli{\`e}re}, P. 2017, \aap, 608, A72

\bibitem[{{Mordasini} {et~al.}(2016){Mordasini}, {van Boekel}, {Molli{\`e}re},
  {Henning}, \& {Benneke}}]{Mordasini2016}
{Mordasini}, C., {van Boekel}, R., {Molli{\`e}re}, P., {Henning}, T., \&
  {Benneke}, B. 2016, \apj, 832, 41

\bibitem[{{M{\"u}ller} {et~al.}(2018){M{\"u}ller}, {Keppler}, {Henning},
  {Samland}, {Chauvin}, {Beust}, {Maire}, {Molaverdikhani}, {van Boekel},
  {Benisty}, {Boccaletti}, {Bonnefoy}, {Cantalloube}, {Charnay}, {Baudino},
  {Gennaro}, {Long}, {Cheetham}, {Desidera}, {Feldt}, {Fusco}, {Girard},
  {Gratton}, {Hagelberg}, {Janson}, {Lagrange}, {Langlois}, {Lazzoni}, {Ligi},
  {M{\'e}nard}, {Mesa}, {Meyer}, {Molli{\`e}re}, {Mordasini}, {Moulin},
  {Pavlov}, {Pawellek}, {Quanz}, {Ramos}, {Rouan}, {Sissa}, {Stadler}, {Vigan},
  {Wahhaj}, {Weber}, \& {Zurlo}}]{Mueller2018}
{M{\"u}ller}, A., {Keppler}, M., {Henning}, T., {et~al.} 2018, \aap, 617, L2

\bibitem[{{Nakamoto} \& {Nakagawa}(1994)}]{NakamotoNakagawa1994}
{Nakamoto}, T. \& {Nakagawa}, Y. 1994, \apj, 421, 640

\bibitem[{{Nayakshin}(2010)}]{Nayakshin2010}
{Nayakshin}, S. 2010, \mnras, 408, 2381

\bibitem[{{Ndugu} {et~al.}(2018){Ndugu}, {Bitsch}, \& {Jurua}}]{Ndugu2018}
{Ndugu}, N., {Bitsch}, B., \& {Jurua}, E. 2018, \mnras, 474, 886

\bibitem[{{Neuh{\"a}user} {et~al.}(2011){Neuh{\"a}user}, {Ginski}, {Schmidt},
  \& {Mugrauer}}]{Neuhauser2011}
{Neuh{\"a}user}, R., {Ginski}, C., {Schmidt}, T.~O.~B., \& {Mugrauer}, M. 2011,
  \mnras, 416, 1430

\bibitem[{{Nielsen} \& {Close}(2010)}]{Nielsen2010}
{Nielsen}, E.~L. \& {Close}, L.~M. 2010, \apj, 717, 878

\bibitem[{{Nielsen} {et~al.}(2019){Nielsen}, {De Rosa}, {Macintosh}, {Wang},
  {Ruffio}, {Chiang}, {Marley}, {Saumon}, {Savransky}, {Ammons}, {Bailey},
  {Barman}, {Blain}, {Bulger}, {Burrows}, {Chilcote}, {Cotten}, {Czekala},
  {Doyon}, {Duch{\^e}ne}, {Esposito}, {Fabrycky}, {Fitzgerald}, {Follette},
  {Fortney}, {Gerard}, {Goodsell}, {Graham}, {Greenbaum}, {Hibon}, {Hinkley},
  {Hirsch}, {Hom}, {Hung}, {Dawson}, {Ingraham}, {Kalas}, {Konopacky},
  {Larkin}, {Lee}, {Lin}, {Maire}, {Marchis}, {Marois}, {Metchev},
  {Millar-Blanchaer}, {Morzinski}, {Oppenheimer}, {Palmer}, {Patience},
  {Perrin}, {Poyneer}, {Pueyo}, {Rafikov}, {Rajan}, {Rameau}, {Rantakyr{\"o}},
  {Ren}, {Schneider}, {Sivaramakrishnan}, {Song}, {Soummer}, {Tallis},
  {Thomas}, {Ward-Duong}, \& {Wolff}}]{Nielsen2019}
{Nielsen}, E.~L., {De Rosa}, R.~J., {Macintosh}, B., {et~al.} 2019, \aj, 158,
  13

\bibitem[{{{\"O}berg} {et~al.}(2011){{\"O}berg}, {Murray-Clay}, \&
  {Bergin}}]{Oberg2011}
{{\"O}berg}, K.~I., {Murray-Clay}, R., \& {Bergin}, E.~A. 2011, \apjl, 743, L16

\bibitem[{{Offner} {et~al.}(2010){Offner}, {Kratter}, {Matzner}, {Krumholz}, \&
  {Klein}}]{Offner2010}
{Offner}, S. S.~R., {Kratter}, K.~M., {Matzner}, C.~D., {Krumholz}, M.~R., \&
  {Klein}, R.~I. 2010, \apj, 725, 1485

\bibitem[{{Ohtsuki} {et~al.}(2002){Ohtsuki}, {Stewart}, \& {Ida}}]{Ohtsuki2002}
{Ohtsuki}, K., {Stewart}, G.~R., \& {Ida}, S. 2002, \icarus, 155, 436

\bibitem[{{Owen} {et~al.}(2011){Owen}, {Ercolano}, \& {Clarke}}]{Owen2011}
{Owen}, J.~E., {Ercolano}, B., \& {Clarke}, C.~J. 2011, \mnras, 412, 13

\bibitem[{{Paardekooper}(2012)}]{Paardekooper2012}
{Paardekooper}, S.-J. 2012, \mnras, 421, 3286

\bibitem[{{Pecaut} \& {Mamajek}(2016)}]{Pecaut2016}
{Pecaut}, M.~J. \& {Mamajek}, E.~E. 2016, \mnras, 461, 794

\bibitem[{{Peretti} {et~al.}(2019){Peretti}, {S{\'e}gransan}, {Lavie},
  {Desidera}, {Maire}, {D'Orazi}, {Vigan}, {Baudino}, {Cheetham}, {Janson},
  {Chauvin}, {Hagelberg}, {Menard}, {Heng}, {Udry}, {Boccaletti}, {Daemgen},
  {Le Coroller}, {Mesa}, {Rouan}, {Samland}, {Schmidt}, {Zurlo}, {Bonnefoy},
  {Feldt}, {Gratton}, {Lagrange}, {Langlois}, {Meyer}, {Carbillet}, {Carle},
  {De Caprio}, {Gluck}, {Hugot}, {Magnard}, {Moulin}, {Pavlov}, {Pragt},
  {Rabou}, {Ramos}, {Rousset}, {Sevin}, {Soenke}, {Stadler}, {Weber}, \&
  {Wildi}}]{Peretti2019}
{Peretti}, S., {S{\'e}gransan}, D., {Lavie}, B., {et~al.} 2019, \aap, 631, A107

\bibitem[{{Piso} {et~al.}(2015{\natexlab{a}}){Piso}, {{\"O}berg}, {Birnstiel},
  \& {Murray-Clay}}]{Piso2015b}
{Piso}, A.-M.~A., {{\"O}berg}, K.~I., {Birnstiel}, T., \& {Murray-Clay}, R.~A.
  2015{\natexlab{a}}, \apj, 815, 109

\bibitem[{{Piso} {et~al.}(2015{\natexlab{b}}){Piso}, {Youdin}, \&
  {Murray-Clay}}]{Piso2015a}
{Piso}, A.-M.~A., {Youdin}, A.~N., \& {Murray-Clay}, R.~A. 2015{\natexlab{b}},
  \apj, 800, 82

\bibitem[{{Pollack} {et~al.}(1996){Pollack}, {Hubickyj}, {Bodenheimer},
  {Lissauer}, {Podolak}, \& {Greenzweig}}]{Pollack1996}
{Pollack}, J.~B., {Hubickyj}, O., {Bodenheimer}, P., {et~al.} 1996, \icarus,
  124, 62

\bibitem[{{Rafikov}(2005)}]{Rafikov2005}
{Rafikov}, R.~R. 2005, \apjl, 621, L69

\bibitem[{{Raghavan} {et~al.}(2010){Raghavan}, {McAlister}, {Henry}, {Latham},
  {Marcy}, {Mason}, {Gies}, {White}, \& {ten Brummelaar}}]{Raghavan2010}
{Raghavan}, D., {McAlister}, H.~A., {Henry}, T.~J., {et~al.} 2010, \apjs, 190,
  1

\bibitem[{{Rajan} {et~al.}(2017){Rajan}, {Rameau}, {De Rosa}, {Marley},
  {Graham}, {Macintosh}, {Marois}, {Morley}, {Patience}, {Pueyo}, {Saumon},
  {Ward-Duong}, {Ammons}, {Arriaga}, {Bailey}, {Barman}, {Bulger}, {Burrows},
  {Chilcote}, {Cotten}, {Czekala}, {Doyon}, {Duch{\^e}ne}, {Esposito},
  {Fitzgerald}, {Follette}, {Fortney}, {Goodsell}, {Greenbaum}, {Hibon},
  {Hung}, {Ingraham}, {Johnson-Groh}, {Kalas}, {Konopacky}, {Lafreni{\`e}re},
  {Larkin}, {Maire}, {Marchis}, {Metchev}, {Millar-Blanchaer}, {Morzinski},
  {Nielsen}, {Oppenheimer}, {Palmer}, {Patel}, {Perrin}, {Poyneer},
  {Rantakyr{\"o}}, {Ruffio}, {Savransky}, {Schneider}, {Sivaramakrishnan},
  {Song}, {Soummer}, {Thomas}, {Vasisht}, {Wallace}, {Wang}, {Wiktorowicz}, \&
  {Wolff}}]{Rajan2017}
{Rajan}, A., {Rameau}, J., {De Rosa}, R.~J., {et~al.} 2017, \aj, 154, 10

\bibitem[{{Rameau} {et~al.}(2013){Rameau}, {Chauvin}, {Lagrange}, {Klahr},
  {Bonnefoy}, {Mordasini}, {Bonavita}, {Desidera}, {Dumas}, \&
  {Girard}}]{Rameau2013}
{Rameau}, J., {Chauvin}, G., {Lagrange}, A.~M., {et~al.} 2013, \aap, 553, A60

\bibitem[{{Reggiani} \& {Meyer}(2013)}]{Reggiani2013}
{Reggiani}, M. \& {Meyer}, M.~R. 2013, \aap, 553, A124

\bibitem[{{Reggiani} {et~al.}(2016){Reggiani}, {Meyer}, {Chauvin}, {Vigan},
  {Quanz}, {Biller}, {Bonavita}, {Desidera}, {Delorme}, {Hagelberg}, {Maire},
  {Boccaletti}, {Beuzit}, {Buenzli}, {Carson}, {Covino}, {Feldt}, {Girard},
  {Gratton}, {Henning}, {Kasper}, {Lagrange}, {Mesa}, {Messina}, {Montagnier},
  {Mordasini}, {Mouillet}, {Schlieder}, {Segransan}, {Thalmann}, \&
  {Zurlo}}]{Reggiani2016}
{Reggiani}, M., {Meyer}, M.~R., {Chauvin}, G., {et~al.} 2016, \aap, 586, A147

\bibitem[{{Reggiani} \& {Meyer}(2011)}]{Reggiani2011}
{Reggiani}, M.~M. \& {Meyer}, M.~R. 2011, \apj, 738, 60

\bibitem[{{Rice} \& {Armitage}(2009)}]{Rice2009}
{Rice}, W.~K.~M. \& {Armitage}, P.~J. 2009, \mnras, 396, 2228

\bibitem[{{Rice} {et~al.}(2012){Rice}, {Forgan}, \& {Armitage}}]{Rice2012}
{Rice}, W.~K.~M., {Forgan}, D.~H., \& {Armitage}, P.~J. 2012, \mnras, 420, 1640

\bibitem[{{Rice} {et~al.}(2014){Rice}, {Paardekooper}, {Forgan}, \&
  {Armitage}}]{Rice2014}
{Rice}, W.~K.~M., {Paardekooper}, S.~J., {Forgan}, D.~H., \& {Armitage}, P.~J.
  2014, \mnras, 438, 1593

\bibitem[{{Ruffio} {et~al.}(2017){Ruffio}, {Macintosh}, {Wang}, {Pueyo},
  {Nielsen}, {De Rosa}, {Czekala}, {Marley}, {Arriaga}, {Bailey}, {Barman},
  {Bulger}, {Chilcote}, {Cotten}, {Doyon}, {Duch{\^e}ne}, {Fitzgerald},
  {Follette}, {Gerard}, {Goodsell}, {Graham}, {Greenbaum}, {Hibon}, {Hung},
  {Ingraham}, {Kalas}, {Konopacky}, {Larkin}, {Maire}, {Marchis}, {Marois},
  {Metchev}, {Millar-Blanchaer}, {Morzinski}, {Oppenheimer}, {Palmer},
  {Patience}, {Perrin}, {Poyneer}, {Rajan}, {Rameau}, {Rantakyr{\"o}},
  {Savransky}, {Schneider}, {Sivaramakrishnan}, {Song}, {Soummer}, {Thomas},
  {Wallace}, {Ward-Duong}, {Wiktorowicz}, \& {Wolff}}]{Ruffio2017}
{Ruffio}, J.-B., {Macintosh}, B., {Wang}, J.~J., {et~al.} 2017, \apj, 842, 14

\bibitem[{{Samland} {et~al.}(2017){Samland}, {Molli{\`e}re}, {Bonnefoy},
  {Maire}, {Cantalloube}, {Cheetham}, {Mesa}, {Gratton}, {Biller}, {Wahhaj},
  {Bouwman}, {Brandner}, {Melnick}, {Carson}, {Janson}, {Henning}, {Homeier},
  {Mordasini}, {Langlois}, {Quanz}, {van Boekel}, {Zurlo}, {Schlieder},
  {Avenhaus}, {Beuzit}, {Boccaletti}, {Bonavita}, {Chauvin}, {Claudi}, {Cudel},
  {Desidera}, {Feldt}, {Fusco}, {Galicher}, {Kopytova}, {Lagrange}, {Le
  Coroller}, {Martinez}, {Moeller-Nilsson}, {Mouillet}, {Mugnier}, {Perrot},
  {Sevin}, {Sissa}, {Vigan}, \& {Weber}}]{Samland2017}
{Samland}, M., {Molli{\`e}re}, P., {Bonnefoy}, M., {et~al.} 2017, \aap, 603,
  A57

\bibitem[{{Sauvage} {et~al.}(2016){Sauvage}, {Fusco}, {Petit}, {Costille},
  {Mouillet}, {Beuzit}, {Dohlen}, {Kasper}, {Suarez}, {Soenke}, {Baruffolo},
  {Salasnich}, {Rochat}, {Fedrigo}, {Baudoz}, {Hugot}, {Sevin}, {Perret},
  {Wildi}, {Downing}, {Feautrier}, {Puget}, {Vigan}, {O'Neal}, {Girard},
  {Mawet}, {Schmid}, \& {Roelfsema}}]{Sauvage2016}
{Sauvage}, J.-F., {Fusco}, T., {Petit}, C., {et~al.} 2016, Journal of
  Astronomical Telescopes, Instruments, and Systems, 2, 025003

\bibitem[{{Schulik} {et~al.}(2019){Schulik}, {Johansen}, {Bitsch}, \&
  {Lega}}]{Schulik2019}
{Schulik}, M., {Johansen}, A., {Bitsch}, B., \& {Lega}, E. 2019, \aap, 632,
  A118

\bibitem[{{Shakura} \& {Sunyaev}(1973)}]{ShakuraSunyaev1973}
{Shakura}, N.~I. \& {Sunyaev}, R.~A. 1973, \aap, 500, 33

\bibitem[{{Spiegel} \& {Burrows}(2012)}]{Spiegel2012}
{Spiegel}, D.~S. \& {Burrows}, A. 2012, \apj, 745, 174

\bibitem[{{Stamatellos} {et~al.}(2011){Stamatellos}, {Maury}, {Whitworth}, \&
  {Andr{\'e}}}]{Stamatellos2011}
{Stamatellos}, D., {Maury}, A., {Whitworth}, A., \& {Andr{\'e}}, P. 2011,
  \mnras, 413, 1787

\bibitem[{{Stamatellos} \& {Whitworth}(2008)}]{Stamatellos2008}
{Stamatellos}, D. \& {Whitworth}, A.~P. 2008, \aap, 480, 879

\bibitem[{{Stone} {et~al.}(2018){Stone}, {Skemer}, {Hinz}, {Bonavita},
  {Kratter}, {Maire}, {Defrere}, {Bailey}, {Spalding}, {Leisenring},
  {Desidera}, {Bonnefoy}, {Biller}, {Woodward}, {Henning}, {Skrutskie},
  {Eisner}, {Crepp}, {Patience}, {Weigelt}, {De Rosa}, {Schlieder}, {Brandner},
  {Apai}, {Su}, {Ertel}, {Ward-Duong}, {Morzinski}, {Schertl}, {Hofmann},
  {Close}, {Brems}, {Fortney}, {Oza}, {Buenzli}, \& {Bass}}]{Stone2018}
{Stone}, J.~M., {Skemer}, A.~J., {Hinz}, P.~M., {et~al.} 2018, \aj, 156, 286

\bibitem[{{Szul{\'a}gyi}(2017)}]{Szulagyi2017b}
{Szul{\'a}gyi}, J. 2017, \apj, 842, 103

\bibitem[{{Thanathibodee} {et~al.}(2019){Thanathibodee}, {Calvet}, {Bae},
  {Muzerolle}, \& {Hern{\'a}ndez}}]{Thanathibodee2019}
{Thanathibodee}, T., {Calvet}, N., {Bae}, J., {Muzerolle}, J., \&
  {Hern{\'a}ndez}, R.~F. 2019, \apj, 885, 94

\bibitem[{{Thommes} {et~al.}(2003){Thommes}, {Duncan}, \&
  {Levison}}]{Thommes2003}
{Thommes}, E.~W., {Duncan}, M.~J., \& {Levison}, H.~F. 2003, \icarus, 161, 431

\bibitem[{{Tychoniec} {et~al.}(2018){Tychoniec}, {Tobin}, {Karska}, {Chandler},
  {Dunham}, {Harris}, {Kratter}, {Li}, {Looney}, \& {Melis}}]{Tychoniec2018}
{Tychoniec}, {\L}., {Tobin}, J.~J., {Karska}, A., {et~al.} 2018, \apjs, 238, 19

\bibitem[{{Venturini} {et~al.}(2015){Venturini}, {Alibert}, {Benz}, \&
  {Ikoma}}]{Venturini2015}
{Venturini}, J., {Alibert}, Y., {Benz}, W., \& {Ikoma}, M. 2015, \aap, 576,
  A114

\bibitem[{{Venuti} {et~al.}(2017){Venuti}, {Bouvier}, {Cody}, {Stauffer},
  {Micela}, {Rebull}, {Alencar}, {Sousa}, {Hillenbrand}, \&
  {Flaccomio}}]{Venuti2017}
{Venuti}, L., {Bouvier}, J., {Cody}, A.~M., {et~al.} 2017, \aap, 599, A23

\bibitem[{{Veras} {et~al.}(2009){Veras}, {Crepp}, \& {Ford}}]{Veras2009}
{Veras}, D., {Crepp}, J.~R., \& {Ford}, E.~B. 2009, \apj, 696, 1600

\bibitem[{{Vigan} {et~al.}(2017){Vigan}, {Bonavita}, {Biller}, {Forgan},
  {Rice}, {Chauvin}, {Desidera}, {Meunier}, {Delorme}, {Schlieder}, {Bonnefoy},
  {Carson}, {Covino}, {Hagelberg}, {Henning}, {Janson}, {Lagrange}, {Quanz},
  {Zurlo}, {Beuzit}, {Boccaletti}, {Buenzli}, {Feldt}, {Girard}, {Gratton},
  {Kasper}, {Le Coroller}, {Mesa}, {Messina}, {Meyer}, {Montagnier},
  {Mordasini}, {Mouillet}, {Moutou}, {Reggiani}, {Segransan}, \&
  {Thalmann}}]{Vigan2017}
{Vigan}, A., {Bonavita}, M., {Biller}, B., {et~al.} 2017, \aap, 603, A3

\bibitem[{{Vigan} {et~al.}(2012){Vigan}, {Patience}, {Marois}, {Bonavita}, {De
  Rosa}, {Macintosh}, {Song}, {Doyon}, {Zuckerman}, {Lafreni{\`e}re}, \&
  {Barman}}]{Vigan2012}
{Vigan}, A., {Patience}, J., {Marois}, C., {et~al.} 2012, \aap, 544, A9

\bibitem[{{Vorobyov}(2013)}]{Vorobyov2013}
{Vorobyov}, E.~I. 2013, \aap, 552, A129

\bibitem[{{Wagner} {et~al.}(2019){Wagner}, {Apai}, \& {Kratter}}]{Wagner2019}
{Wagner}, K., {Apai}, D., \& {Kratter}, K.~M. 2019, \apj, 877, 46

\bibitem[{{Wahhaj} {et~al.}(2011){Wahhaj}, {Liu}, {Biller}, {Clarke},
  {Nielsen}, {Close}, {Hayward}, {Mamajek}, {Cushing}, {Dupuy}, {Tecza},
  {Thatte}, {Chun}, {Ftaclas}, {Hartung}, {Reid}, {Shkolnik}, {Alencar},
  {Artymowicz}, {Boss}, {de Gouveia Dal Pino}, {Gregorio-Hetem}, {Ida},
  {Kuchner}, {Lin}, \& {Toomey}}]{Wahhaj2011}
{Wahhaj}, Z., {Liu}, M.~C., {Biller}, B.~A., {et~al.} 2011, \apj, 729, 139

\bibitem[{{Wahhaj} {et~al.}(2013){Wahhaj}, {Liu}, {Nielsen}, {Biller},
  {Hayward}, {Close}, {Males}, {Skemer}, {Ftaclas}, {Chun}, {Thatte}, {Tecza},
  {Shkolnik}, {Kuchner}, {Reid}, {de Gouveia Dal Pino}, {Alencar},
  {Gregorio-Hetem}, {Boss}, {Lin}, \& {Toomey}}]{Wahhaj2013}
{Wahhaj}, Z., {Liu}, M.~C., {Nielsen}, E.~L., {et~al.} 2013, \apj, 773, 179

\bibitem[{Wallis(2014)}]{Wallis2014}
Wallis, K.~F. 2014, Statist. Sci., 29, 106

\bibitem[{{Wang} {et~al.}(2018){Wang}, {Graham}, {Dawson}, {Fabrycky}, {De
  Rosa}, {Pueyo}, {Konopacky}, {Macintosh}, {Marois}, {Chiang}, {Ammons},
  {Arriaga}, {Bailey}, {Barman}, {Bulger}, {Chilcote}, {Cotten}, {Doyon},
  {Duch{\^e}ne}, {Esposito}, {Fitzgerald}, {Follette}, {Gerard}, {Goodsell},
  {Greenbaum}, {Hibon}, {Hung}, {Ingraham}, {Kalas}, {Larkin}, {Maire},
  {Marchis}, {Marley}, {Metchev}, {Millar-Blanchaer}, {Nielsen}, {Oppenheimer},
  {Palmer}, {Patience}, {Perrin}, {Poyneer}, {Rajan}, {Rameau},
  {Rantakyr{\"o}}, {Ruffio}, {Savransky}, {Schneider}, {Sivaramakrishnan},
  {Song}, {Soummer}, {Thomas}, {Wallace}, {Ward-Duong}, {Wiktorowicz}, \&
  {Wolff}}]{Wang2018}
{Wang}, J.~J., {Graham}, J.~R., {Dawson}, R., {et~al.} 2018, \aj, 156, 192

\bibitem[{{Winters} {et~al.}(2019){Winters}, {Henry}, {Jao}, {Subasavage},
  {Chatelain}, {Slatten}, {Riedel}, {Silverstein}, \& {Payne}}]{Winters2019}
{Winters}, J.~G., {Henry}, T.~J., {Jao}, W.-C., {et~al.} 2019, \aj, 157, 216

\bibitem[{{Young} \& {Clarke}(2016)}]{Young2016}
{Young}, M.~D. \& {Clarke}, C.~J. 2016, \mnras, 455, 1438

\end{thebibliography}

\appendix
\onecolumn

\section{Undefined candidates}
\label{sec:candidates_histo}

\begin{figure}
    \centering 
    \includegraphics[width=0.5\textwidth]{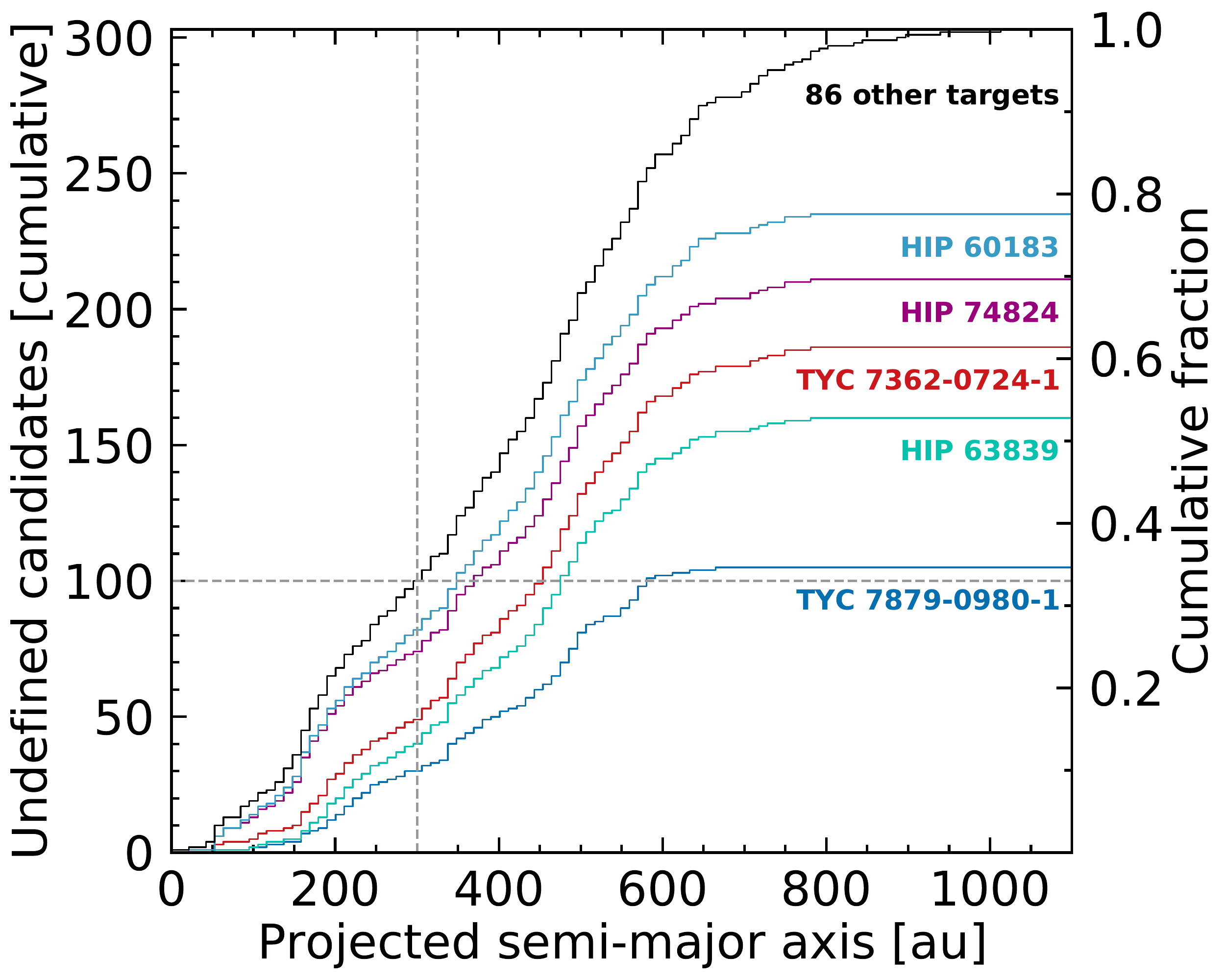}
    \caption{Cumulative histogram of the number of undefined candidates as a function of projected semimajor axis. Only the five targets that contribute the highest number of undefined candidates are labeled in the plot for clarity. With a cutoff of 300\,au for our analysis, only 96 undefined candidates remain (32\% of the total number).}
    \label{fig:candidates_histo}
\end{figure}

Figure~\ref{fig:candidates_histo} provides a cumulative histogram of the number of undefined candidates as a function of projected semimajors axes. Globally, TYC\,7879-0980-1 and HIP\,63839 are responsible for more than 50\% of the total number of undefined candidates: they are both close to the galactic plane and  have only one single-epoch observation, which explains why they contribute such a large fraction of the total number of undefined candidates. In total, five targets are responsible for $\sim$80\% of undefined candidates. With a cutoff at 300\,au, a total of 96 undefined candidates remain, which represents 32\% of the total number. The way they are handled in the analysis is detailed in Sect.~\ref{sec:planetary_candidates}.

\clearpage
\section{SHINE depth of search}
\label{sec:depth_of_search}

\begin{figure*}
    \centering 
    \includegraphics[width=1\textwidth]{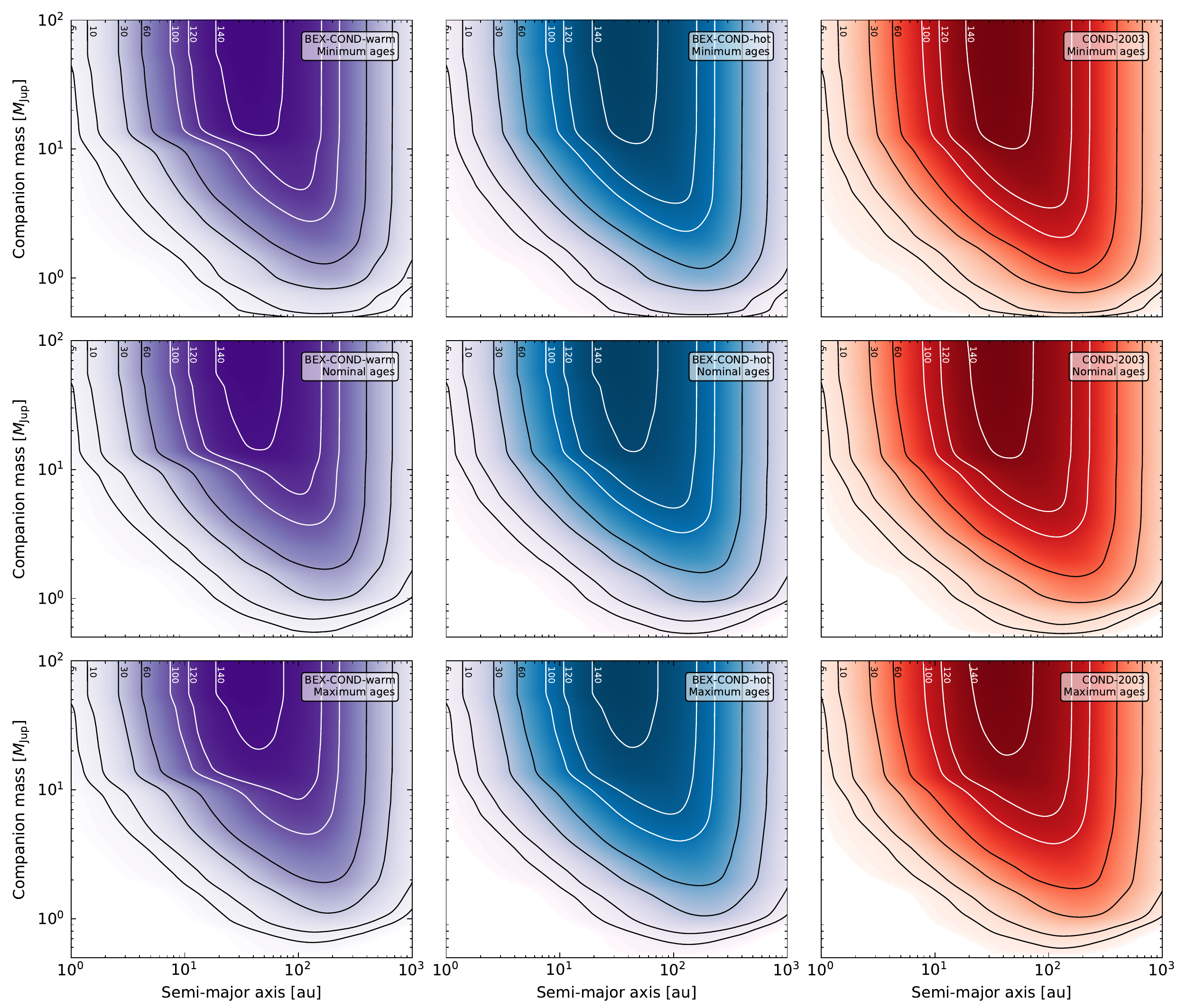}
    \caption{Depth of search of the SHINE survey for the 150 stars in the sample computed using detection limits converted into mass using three different sets of evolutionary models (left: BEX-COND-warm; center: BEX-COND-hot; right: COND-2003) and for different stellar ages (top: minimum; middle: nominal; bottom: maximum). Each plot gives the numbers of stars around which the survey is sensitive for substellar companions as a function of mass and semimajor axis.}
    \label{fig:shine_sensitivity_grid}
\end{figure*}

Figure~\ref{fig:shine_sensitivity_grid} shows the depth of search of the SHINE survey for the 150 stars in the sample based on different assumptions for the stellar ages and evolutionary models. The depth of search gives the number of stars in the sample around which the survey is sensitive for substellar companions as a function of mass and semimajor axis. We computed it using the nominal, minimum, and maximum ages and for the BEX-COND-warm and BEX-COND-hot (see Equation~\ref{eq:LpfBEX}), and COND-2003 evolutionary models.

\clearpage
\section{Semimajor axis cutoff}
\label{sec:sma_cutoff_test}

\begin{figure*}
    \centering 
    \includegraphics[width=0.49\textwidth]{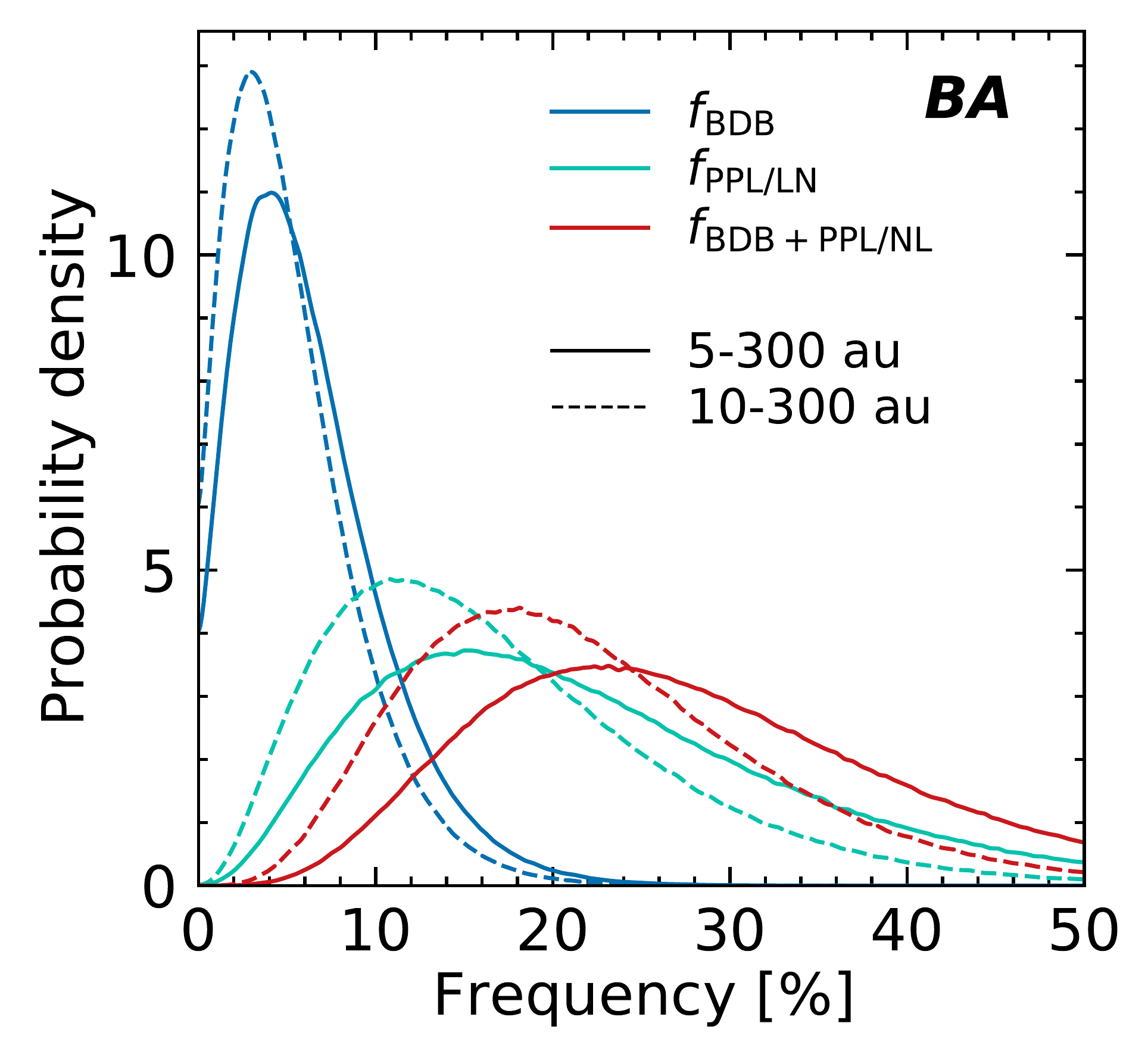}
    \includegraphics[width=0.49\textwidth]{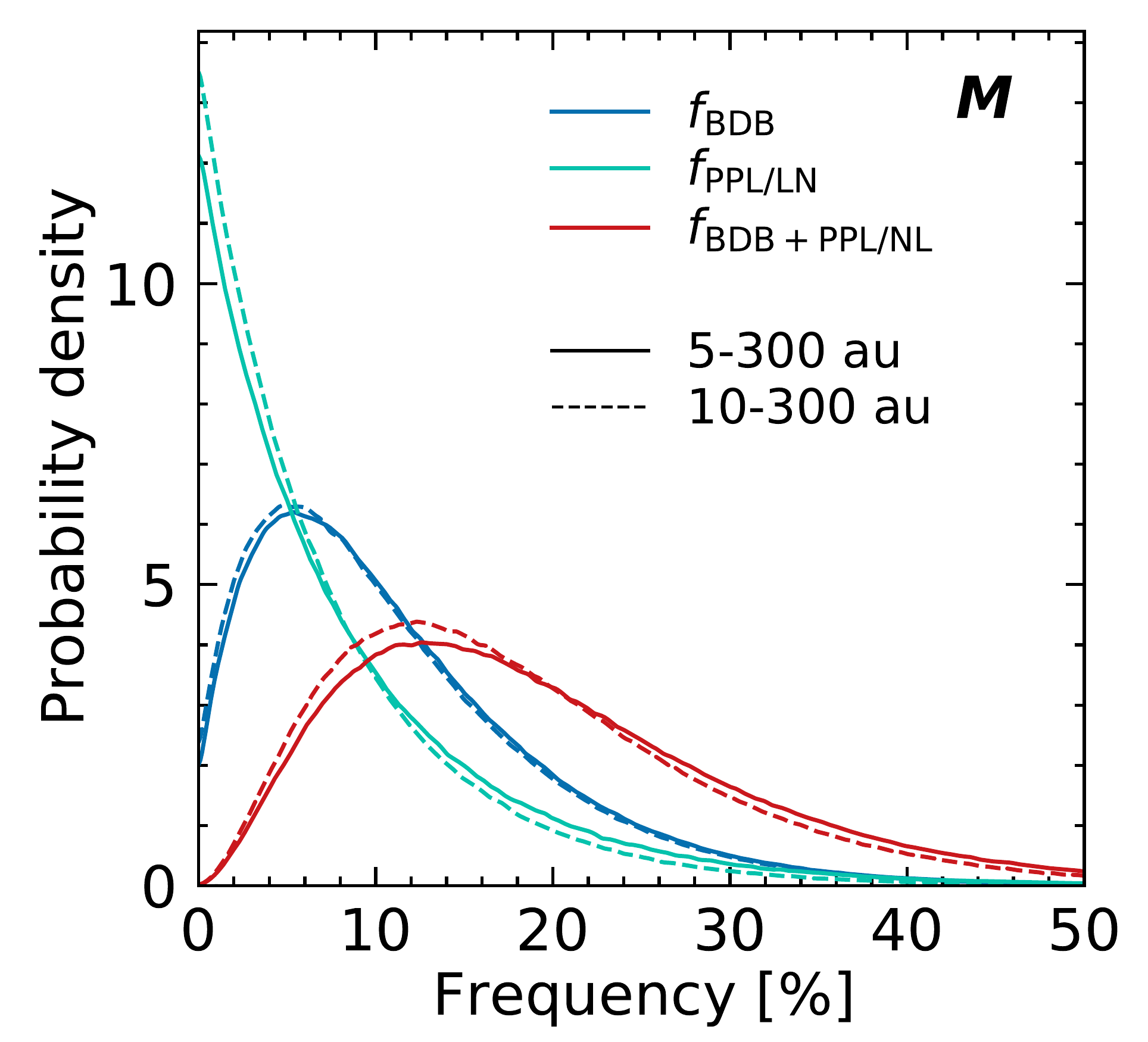}
    \caption{Probability density functions of the frequencies of substellar companions around BA (left) and M (right) stars based on the parametric model, computed for companions with semi-major axes in the range $a = 5$--300\,au (plain line) or $a = 10$--300\,au (dashed line), and using the BEX-COND-hot evolutionary tracks for the mass conversion of the detection limits. The same plot for FGK stars is shown in Fig.~\ref{fig:p-model_test_sma_FGK}.}
    \label{fig:p-model_test_sma_BA+M}
\end{figure*}

The effect of the inner limit of the semimajor axis range is not identical for all spectral types due to the different number of detections. In Fig.~\ref{fig:p-model_test_sma_FGK} we show that for FGK stars the effect of changing the lower limit from 5 to 10\,au is very weak, even though the detection around HIP\,107412 is removed.

Figure~\ref{fig:p-model_test_sma_BA+M} shows the effect of changing the semimajor axis lower limit for BA and M stars. For M stars, the effect is negligible because the only detection around an M star (CD\,-35\,2722) remains untouched since its semimajor axis is constrained in the range 76--216\,au. However, the effect is more significant for BA stars, for which the detection around $\beta$\,Pic must be removed because its semimajor axis is tightly constrained within 8.5--9.2\,au. The PPL/LN part of the model is the most affected, with a shift of the peak of the PDF from $\sim$15\% to $\sim$11\% when the cutoff changes from 5 to 10\,au. We highlight that a)~$\beta$\,Pic\,b is clearly in the mass range dominated by the PPL/LN part of the model (in contrast to HIP\,107412\,B), and b)~the 5--10\,au range is also where the peak of the PPL/LN distribution is expected. These two elements combined can easily explain that the effect on the PPL/LN part of the model is larger for the BA stars than for the FGK stars. However, despite the larger effect of the semimajor axis cutoff for the BA stars, our conclusions remain unchanged.

\clearpage
\section{Comparison of two asymmetric distributions}
\label{app:asym_distrib}

Let $X$ represent a random variable that has an asymmetric normal distribution. We call $X_0$ the mode of the distribution (i.e., value for which the distribution is maximum), and $\sigma_-$ and $\sigma_+$ the standard deviations toward values below and above~$X_0$, respectively. The PDF $\phi_X$ of~$X$ reads
\begin{equation}
    \phi_X(x) = \frac{\sqrt{2}}{\sqrt{\pi}\,\left(\sigma_{-} + \sigma_{+}\right)}    \displaystyle\left[\mathds{1}_{]-\infty,X_{0}]}\exp{\left(-\frac{(x-X_{0})^2}{2\,\sigma_{-}^2}\right)}+\displaystyle\mathds{1}_{]X_{0},+\infty[}\exp{\left(-\frac{(x-X_{0})^2}{2\,\sigma_+^2}\right)}\right],
    \label{eq:asy_distrib}
\end{equation}
where~$\mathds{1}_{[a,b]}$ is the identity function between~$a$ and $b,$ and it is zero elsewhere.

We assume that there are two independent measurements $x=X_0{}_{-\sigma_{X,-}}^{+\sigma_{X,+}}$ and $y=Y_0{}_{-\sigma_{Y,-}}^{+\sigma_{Y,+}}$ of the same measure. We then consider that $x$ and $y$ are realizations of the random variables~$X$ and~$Y$ that follow the asymmetric normal distributions~$\phi_X$ and~$\phi_Y$ defined by Eq.\,\ref{eq:asy_distrib}. We introduce the random variable~$Z=X-Y$ for which the probability density function $\phi_Z$ is
\begin{equation}
    \phi_Z(z) = \int_{-\infty}^{+\infty} \phi_X(t)\,\phi_Y(t+z)\,\mathrm{d}t
.\end{equation}
From~$\phi_Z$, we can calculate the probability $P(z_{\mathrm{threshold}})$ such that $|z|<z_{\mathrm{threshold}}$
\begin{equation}
    P(z_{\mathrm{threshold}}) = \int_{-z_{\mathrm{threshold}}}^{z_{\mathrm{threshold}}} \phi_Z(t)\,\mathrm{d}t
\end{equation}
and find the particular value $z_{95}$ such that 
\begin{equation}
    P(z_{95})=0.95
.\end{equation}
This value~$z_{95}$ depends on the modes and standard deviations of~$\phi_X$ and~$\phi_Y$.

As $x$ and $y$ are two independent measurements of the same measurand, we test the null hypothesis $x = y$. If the mode~$z_\mathrm{mode}$ of~$\phi_Z$ is such that $|z_\mathrm{mode}|<z_{95}$, then we accept the null hypothesis with a~$5\,\%$ risk.

For each line of Tab.\,\ref{tab:frequencies_comparison}, we compare one SHINE measurement to another. We calculate~$\phi_Z$ from the two asymmetric normal distributions, and then $z_{\mathrm{mode}}$ and $z_{95}$. If~$|z_\mathrm{mode}|<z_{95}$, we conclude that the two measurements are compatible with a~$5\,\%$ risk. In the other case, we conclude that the two measurements are not compatible with a~$5\,\%$ risk.

\end{document}